\documentclass[journal,10pt,a4paper,twocolumn,twoside]{IEEEtran}

\usepackage{graphicx}

\usepackage{subcaption}
\usepackage{amsmath}
\usepackage{booktabs}

\usepackage{xcolor}
\usepackage{amssymb}
\usepackage{amsmath,cases}
\usepackage{dsfont}
\usepackage{physics}

\usepackage{hyperref}
\usepackage{cite}
\usepackage{newpxtext,newpxmath}
\usepackage{acronym}

\definecolor{ocre}{HTML}{800000}
\definecolor{green}{RGB}{0, 128, 0}
\definecolor{sky}{HTML}{C6D9F1}
\definecolor{skybox}{HTML}{5F86B3}

\usepackage{tikz}
\usetikzlibrary{quantikz}
\usetikzlibrary{decorations.text, arrows.meta,calc,shadows.blur,shadings,intersections}
\usetikzlibrary{trees}

\begin{document}
	\title{Physical Layer Aspects of  Quantum Communications: A Survey}
	\author{Seid Koudia, Leonardo Oleynik, Mert Bayraktar, Junaid ur Rehman, and Symeon Chatzinotas,~\IEEEmembership{Fellow,~IEEE}\\
	
	\thanks{
 The authors are with the Interdisciplinary Centre for Security, Reliability, and Trust (SnT), Luxembourg, L-1855 Luxembourg. 
		(emails: \{seid.koudia, leonardo.oleynik, mert.bayraktar, junaid.urrehman, symeon.chatzinotas\}@uni.lu)

 }
    }

	%\date{}
	\maketitle

\begin{abstract}
    Quantum communication systems support unique applications in the form of distributed quantum computing, distributed quantum sensing, and several cryptographic protocols. The main enabler in these communication systems is an efficient infrastructure that is capable to transport unknown quantum states with high rate and fidelity. This feat requires a new approach to communication system design which efficiently exploits the available physical layer resources, while respecting the limitations and principles of quantum information. Despite the fundamental differences between the classic and quantum worlds, there exist universal communication concepts that may proven beneficial in quantum communication systems as well. In this survey, the distinctive aspects of physical layer quantum communications are highlighted in a attempt to draw commonalities and divergences between classic and quantum communications. More specifically, we begin by overviewing the quantum channels and use cases over diverse optical propagation media, shedding light on the concepts of crosstalk and interference. Subsequently, we survey quantum sources, detectors, channels and modulation techniques.  More importantly, we discuss and analyze spatial multiplexing techniques, such as coherent control, multiplexing, diversity and MIMO. Finally, we identify synergies between the two communication technologies and grand open challenges that can be pivotal in the development of next-generation quantum communication systems. 
\end{abstract}

\section{Introduction}

\Ac{QC} have recently emerged as the cornerstone of Quantum Internet, which envisages a network of interconnected quantum sensors and processors, working together to deliver high-performing applications, which outperform the classical counterparts in terms of sensing accuracy or computational capabilities \cite{WEH:18:Sci, CZW:22:CST, WJZ:22:LPR, KuP:24:IEEE_Network, AuR:24:OJCS}. Even though quantum phenomena can be observed across a variety of particles, in the \ac{QC} context photons have been established as the preferred option, either transmitted through wired media such as fiber optics or through wireless media, namely free space optics \cite{OFV:09:NPho, WSL:20:NPho, SP:19:APR}. Nevertheless, independently of the selected medium, the main challenge in \ac{QC} nowadays is the low rate of error-free communication, which creates an insurmountable bottleneck against enabling Quantum Internet applications. 

It is worth noting that classical communications also suffered from similar limitations in their early days \cite{RAB:02:COMMAG}. As technology evolved, the classical communication rates have increased tremendously by efficiently using all available degrees of freedom offered by nature, namely intensity, phase, polarization, frequency, and space \cite{SA:14:TIT, CJ:09:TIT}. In this direction, one of the most important milestones has been the multiplexing of complex signals across interfering channels and their effective demultiplexing at the receiver side, i.e., \ac{MIMO} communication. Depending on whether redundant or novel source info was multiplexed, \ac{MIMO} can either decrease the error rate or improve the data rate of the communication channel, an effect which is widely known as \ac{DMT} \cite{ZT:03:TIT}. 

\begin{figure*}[t]
	\centering
	\newcommand{\two}[4]{
	\draw[arrow,transform canvas={yshift=#1/2}] (#2) -- (#3) node [midway, above] {${#4}_1$};
	\draw[arrow,transform canvas={yshift=-#1/2}] (#2) -- (#3) node [midway, above] {${#4}_2$};
}

\newcommand{\three}[4]{
	\draw[arrow,transform canvas={yshift=#1}] (#2) -- (#3) node [midway, above] {${#4}_1$};
	\draw[arrow] (#2) -- (#3) node [midway, above] {${#4}_2$};
	\draw[arrow,transform canvas={yshift=-#1}] (#2) -- (#3) node [midway, above] {${#4}_3$};
}

\begin{tikzpicture}[
	% Environment Cfg
	font=\footnotesize, 
	text centered,
	rounded corners=0.08cm,
	%	text width=2.5cm, 
	% Styles
	conn/.style ={
		draw=black,
		text width=1cm, 
		text=white, minimum height=3cm,
		thick,
		fill=ocre!90,
		blur shadow, 
	},
	stan/.style ={
		draw=skybox,
		text width=1.5cm, 
		text=black, minimum height=3cm,
		thick,
		fill=sky,
		blur shadow, 
	},
	arrow/.style ={
		->,
		shorten >=1pt,
		semithick
	},
	Neutral/.style ={
		draw=black!50,
		thick,
		fill=black!10,
		blur shadow, 
	}
	]
	\def\x{1}
	\def\y{30}
	%for labels like "Alice's LAB"
	%Label/.style 
	
	%distinguish between classical and quantum data

	\node[stan] (A) at (0,0) {Quantum Pre Processing};
	\node[stan] (A1) [left=\x of A] {Quantum Circuit};
	\node[conn] (B) [right=\x of A]  {DV to CV};
	\node[stan] (C) [right=\x of B]  {Quantum Channel};
	
	\node[conn] (D) [right=\x of C]  {CV to DV};
	\node[stan] (E) [right=\x of D]  {Quantum Post Processing};
	\node[stan] (F) [right=\x of E]  {Quantum Circuit};
	
	%			\coordinate (G) [below=5cm of F];
	%			\draw [post] (F) -- (G) node [left] {KPI};

	\two{\y}{-2*\x,0}{A}{a}
	\three{\y}{A}{B}{{a}}
	\three{\y}{B}{C}{{a}}
	\three{\y}{C}{D}{{b}}
	\three{\y}{D}{E}{{b}}
	\two{\y}{E}{F}{b}
	
	\draw[arrow,<-] (A1) -++ (0,2) node[above] {Input};
	\node[stan] (A1) [left=\x of A] {Quantum Circuit};
		
	\draw[arrow,->] (F) -++ (0,-2) node[below] {Output};
%	\node[stan] (F) [right=\x of E]  {Quantum Circuit};

	\begin{scope}[on background layer]
		\node [dashed, draw=blue!50,fit=(A1) (A) (B) (C),inner sep=10pt ] {}; 
		\node [dashed, draw=green!50,fit=(C) (D) (E) (F),inner sep=15pt ] {};
	\end{scope}

\end{tikzpicture}
        \caption{The general structure of a quantum \ac{MIMO} channel. Preprocessing and post-processing are required to take full advantage of the higher dimensions offered by the \ac{MIMO} setup. \ac{DV} to \ac{CV} and \ac{CV} to \ac{DV} blocks might be required only in some select scenarios, e.g., when connecting \ac{DV} quantum nodes while transmitting \ac{CV} modes over the channel.}
	\label{fig:general_MIMO}
\end{figure*}
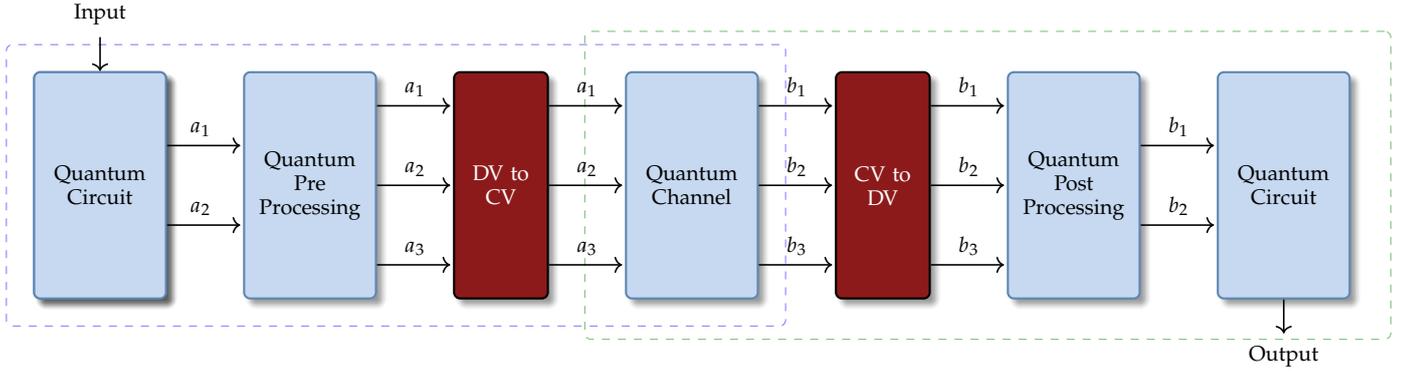

Following this train of thought, it is natural to pose the research question of whether it would be possible to use a similar rationale to improve the performance of quantum communication channels. As a first step in this direction, this paragraph overviews the classical vs quantum differences, when preparing, transmitting, propagating, receiving, and measuring information. To motivate the discussion, a general structure of possible quantum \ac{MIMO} communication setup is shown in Fig.~\ref{fig:general_MIMO}.
\subsection{Preparation \& Transmission}
A common assumption in the classic world is that information can be replicated with minimal effort and as a result introducing redundancy in terms of repeating information is trivial \cite{JC:13:EC_Book}. On the other hand, quantum mechanics forbid the cloning of quantum states \cite{WZ:82:Nat}. While this might not be an issue when quantum states can be repeatedly prepared at the transmitter based on classical info, it is certainly problematic when quantum sources are employed e.g., quantum computers or sensors. A relevant concept in this direction is quantum cloning, which effectively creates a larger number of low quality output quantum states based on a smaller number of input quantum states \cite{Wer:98:PRA, GM:97:PRL}. Unfortunately, the output states are not identical to the input and in addition they are entangled, ensuring this way that this method cannot be straightforwardly used to generate redundancy \cite{BEM:98:PRL, SIG:05:RMP}. The second relevant approach is \ac{QEC}, which finds applications in both computing and communication models \cite{Bra:98:Nat}. We refer the reader to \cite{Rof:19:CP, La:20:Book} for an overview.  
The second common assumption in the classical information is that the intensity or power of the prepared signal is an analog degree of freedom that can be increased arbitrarily (under public safety constraints) to compensate for the energy that is lost to the environment. However, in the quantum world, the information is encoded in particles (e.g. photons, wavepackets) whose energy levels are dictated by discrete quanta and should not be scaled arbitrarily to avoid the transition into a classical state. For higher energy pulses, distinguishing between different photon number states becomes increasingly difficult due to technological limitations and classical detection remains the only possibility \cite{ALS:10:LPR}. 

\subsection{Propagation}
The most important process towards controlling and mitigating  propagation impairments is channel estimation. In the classic world, multidimensional communication channels can be mathematically represented in the form of complex (intensity and phase) matrices, which capture all possible combinations of input and output dimensions \cite{LDM:02:JSAC, CEP:02:TB, AB:01:WCMC}. Assuming a sufficiently long coherence time, during which the matrix remains fixed, each complex coefficient is measured by preparing known sequences of information (pilots), propagating them and measuring the outcome at the receiver \cite{NZG:17:IET_Com, SCC:20:TCOM, MAK:17:TWC}. On the other hand, quantum systems are inherently discrete and probabilistic. For example, the propagation environment can impair quantum communications by probabilistically injecting (creation) or absorbing (annihilation) a discrete number of photons \cite{JC:22:JoPA}. This effect creates a large combinatorial set of probabilities, which is very challenging to estimate (quantum process tomography) without excessive pilot overhead \cite{MRL:08:PRA, RSM:11:NJP}. To be mathematically more exact, quantum channels or processes are modelled through operators over often infinite-dimensional Hilbert spaces. Similarly to their classic counterparts, quantum channels can very well be time-varying, so it remains an open question whether channel resources should be devoted for quantum process tomography or simpler (but less accurate) input-output relationships should be used for quantum channel models \cite{uAD:24:TCOM, EFC:21:npjQI, EFd:23:PRR}. For example, by focusing on specific observables at the receiver and not the full quantum state, classical or semi-classical channel models could be employed \cite{NPR:17:NP}. It is worth mentioning that blind feedback-based approaches have also shown promise in quantum channels. Blind refers to the fact that there are no specific pilots or states propagated with the purpose of learning channel coefficients. Feedback-based means that this is usually an iterative process where (classic or quantum) feedback is regularly sent from the receiver to the transmitter, in order to guide the next preparation phase \cite{XLM:12:Conf, MFC:20:Conf}. Such approaches fall under the umbrella of \ac{SCAPE} \cite{uS:22:Conf, uAD:24:TCOM}. 

\subsection{Reception \& Measurement}
\begin{figure}
    \centering
    \scriptsize
    \tikzstyle{every node}=[thick, anchor=west]
    \begin{tikzpicture}[scale=0.95,
        level distance=2cm, 
        grow via three points={one child at (0.3,-0.55) and
            two children at (0.3,-0.5) and (0.3,-1.15)},
        edge from parent path={([xshift=0.0mm] \tikzparentnode.south west) |- (\tikzchildnode.west)},
        growth parent anchor=south west,
        edge from parent/.style = {draw, -latex}
    ]
        \node {}
            child { node {\scriptsize I. Introduction}
                child[xshift=0.1cm] { node {\scriptsize I-A. Preparation and Transmission}}
                child[xshift=0.1cm] { node {\scriptsize I-B. Propagation}}
                child[xshift=0.1cm] { node {\scriptsize I-C. Reception \& Measurement}}
                child[xshift=0.1cm] { node {\scriptsize \begin{tabular}{@{}l@{}}I-D. Multidimensional Quantum \\ Communications \& Degrees of Freedom\end{tabular}}}
                child[xshift=0.1cm] { node {\scriptsize I-E. Reception \& Measurement}}
            }
            child[missing] {}
            child[missing] {}
            child[missing] {}                
            child[missing] {}
            child[missing] {}                
            child { node {\scriptsize II. Preliminaries}
                child[xshift=0.1cm] { node {\scriptsize II-A. CV vs DV Systems}}
                child[xshift=0.1cm] { node {\scriptsize \begin{tabular}{@{}l@{}}I-D. Probabilistic Cross-Talk (Classical Interference)  \\ vs Interference (Quantum) and Use Case\end{tabular}}}
                child[xshift=0.1cm] { node {\scriptsize II-C. Crosstalk in Free Space Optics}}
                child[xshift=0.1cm] { node {\scriptsize II-D. Crosstalk in Quantum Circuits}}
            }
            child[missing] {}
            child[missing] {}                
            child[missing] {}                
            child[missing] {}
            child { node {\scriptsize III. Quantum Information in Different Media}
                child[xshift=0.1cm] { node {\scriptsize III-A. Fiber Optics}}
                child[xshift=0.1cm] { node {\scriptsize III-B. Free Space Optics}}
                child[xshift=0.1cm] { node {\scriptsize III-C. Integrated Photonics}}
            }
            child[missing] {}                
            child[missing] {}                
            child[missing] {}
            child { node {\scriptsize IV. Quantum Communication}
                child[xshift=0.1cm] { node {\scriptsize IV-A. Quantum Sources}}
                child[xshift=0.1cm] { node {\scriptsize IV-B. Optical Channels}}
                child[xshift=0.1cm] { node {\scriptsize IV-C. Detection}}
                child[xshift=0.1cm] { node {\scriptsize IV-D. Quantum Modulation Techniques}}
            }
            child[missing] {}                
            child[missing] {}                
            child[missing] {}
            child[missing] {}
            child { node {\scriptsize V. Interacting Quantum Channels}
                child[xshift=0.1cm] { node {\scriptsize V-A. Quantum Coherent Control}}
                child[xshift=0.1cm] { node {\scriptsize V-B. Multiplexing and Diversity}}
                child[xshift=0.1cm] { node {\scriptsize V-C. Quantum MIMO}}
            }
            child[missing] {}                
            child[missing] {}                
            child[missing] {}
            child { node {\scriptsize\begin{tabular}{@{}l@{}} VI. Connecting DV Quantum Nodes with CV \\Quantum Links\end{tabular}}
                child[xshift=0.1cm] { node {\scriptsize VI-A. Cat Codes}}
                child[xshift=0.1cm] { node {\scriptsize VI-B. GKP Codes}}
                child[xshift=0.1cm] { node {\scriptsize VI-C. Truncation of Infinite-Dimensional Hilbert Space}}
            }
            child[missing] {}                
            child[missing] {}                
            child[missing] {}
            child { node {\scriptsize VII. Open Challenges}
                child[xshift=0.1cm] { node {\scriptsize\begin{tabular}{@{}l@{}} VII-A. Interference of Photons: The Hong–Ou–Mandel\\ Effect\end{tabular}}}
                child[xshift=0.1cm] { node {\scriptsize VII-B. Physical Phenomenon Beyond Losses}}
                child[xshift=0.1cm] { node {\scriptsize\begin{tabular}{@{}l@{}} VII-C. Performance Assessment of Quantum\\ Communication Systems\end{tabular}}}
                child[xshift=0.1cm] { node {\scriptsize VII-D. Selective Fading}}
                child[xshift=0.1cm] { node {\scriptsize VII-E. Diversity and Multiplexing in Quantum MIMO}}
                child[xshift=0.1cm] { node {\scriptsize VII-F. Interacting Quantum Channels}}
                child[xshift=0.1cm] { node {\scriptsize VII-G. The Interplay Between DV and CV Systems}}
            }
            child[missing] {}
            child[missing] {}
            child[missing] {}
            child[missing] {}
            child[missing] {}
            child[missing] {}
            child[missing] {}
            child { node {\scriptsize VIII. Conclusions}
            }
            child[missing] {};
    \end{tikzpicture}
    \caption{Paper Structure}
    \label{paper_structure}
    \hrulefill
\end{figure}
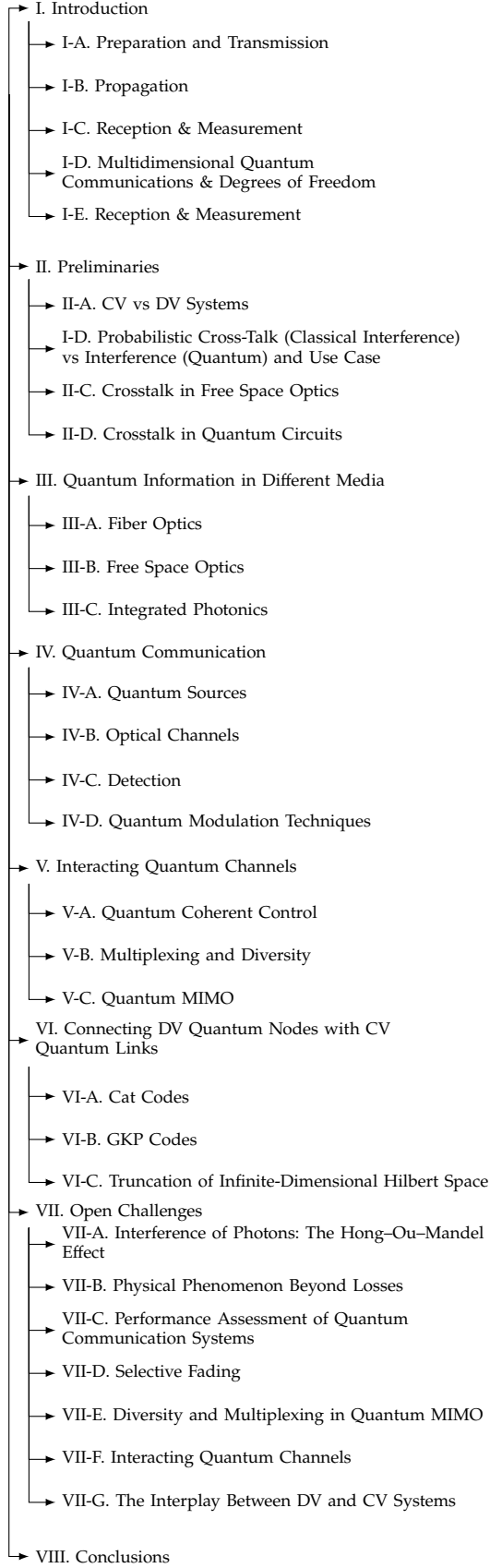
The concept of measurement is fundamental in \ac{QC} systems, because it signifies the process which extracts classical information from a quantum state \cite{NC:10:book, Wat:18:book}. Measurement does not necessarily happen right after the reception, for example when the received quantum state is fed directly to a quantum computer for further processing. However, eventually a post-processed quantum state has to be measured to extract information that can be perceived by humans or used for a classical computer interface. Compared to classic world, the performance metrics are not surprising, namely the quantum communication rate (qudits/sec) and the quantum state fidelity---a similarity metric between the transmitted and received state. Nevertheless, the actual process of measurement differs fundamentally. In the classic world, received information can be stored and measured an arbitrary number of times with the same outcome. In the quantum world, every time a measurement is performed, it unavoidably alters/evolves the quantum state. More importantly, unless the correct measurement operator is applied, there is a chance that no useful classical information will be extracted. In the special case, where the prepared quantum states were encoded based on classic information (e.g. output of digital sensors), the classical performance metrics can be also used, namely classic communication rate (bits/sec) and error probability/rate.   

\subsection{Multidimensional Quantum Communications \& Degrees of Freedom}
Before proceeding with this discussion, it is crucial to define the concept of a channel use or slot. In the classic world, a channel slot refers to an event where the communication channel is occupied to propagate information without, in theory, affecting the adjacent slots. The most obvious examples are time or frequency slots, where the orthogonality with adjacent slots is determined by pulse shaping. Another example is spatial dimensions, where channel slots can be seen as spatial directions, achieved by creating orthogonal directive beams across the propagation environment. Independently of the \ac{DoF} used to create multiple channel slots, the effect of orthogonality is achieved through pre-processing at the transmitter and/or post-processing at the receiver. The aforementioned \acp{DoF} have been already successfully exploited in real-life and are widely deployed in commercial services that are used on a daily basis (e.g. digital communications over copper, fiber, wireless). However, the extension of these paradigms to the quantum domain are not straightforward. Confining transmissions into a constrained time, frequency or spatial DoF highly depends on our capability to design and implement ideal sources. In addition, the interaction of light quanta over the propagation medium differs considerably depending on the employed type of quantum states (discrete variables vs continuous variable states). Finally, demultiplexing and detecting quantum states at the receiver side requires a fundamentally different approach, due to the no-cloning theorem. In the rest of this manuscript, we elaborate on these intricate points and we point to promising system models, transceiver architectures and open research challenges in this domain. 

The main contributions of this manuscript include:
\begin{itemize}
    \item We review the characteristics of optical beam propagation in free space from classical optics point-of-view and find its applications for quantum communications in \ac{FSO}. More concretely, we highlight the use of classical techniques, such as adaptive optics, and identify the situations where these can be used for quantum communications as well. 
    \item We draw parallels between the classical and quantum communication systems with the main aim of identifying situations where classically developed methods have a potential to be implemented in quantum communication systems for future developments. The main hypothesis here is that we do not need to develop wireless quantum communication systems from scratch but some of the already developed (classical) techniques can be repurposed with little effort for wireless quantum communication systems of the future. Some of the identified techniques in this article include channel modeling based on physical phenomenon, \ac{MIMO} for classical wireless communications, and interference/channel cross-talk management. 
    \item We identify fundamental models for the essential components of an end-to-end wireless quantum communication systems that need to be defined for effective utilization of classical communication techniques in quantum domain.
    \item We review promising techniques in quantum communication systems that may have a large role/impact in the quantum communication systems of future. These include the quantum equivalents of classically developed techniques, e.g., \ac{MIMO} channels and \ac{WDM}, as well as some uniquely quantum methods, e.g., quantum coherent control. 
\end{itemize}

The paper structure is summarized in Fig.~\ref{paper_structure} and is organized as follows. In Sec.~\ref{sec:prel} we define some preliminaries and make a distinction between some terms that are used both in classical and quantum communications with different meaning. In Sec.~\ref{sec:FSOBP} we different mediums that act as a communication channel for quantum information and the basic modeling of flow of information in these mediums. In Sec.~\ref{sec:FSQC}, we review the basic components of quantum communications, i.e., sources, channel models, and detectors. The techniques to combine quantum channels are discussed in Sec.~\ref{sec:CQC}. Open challenges are discussed in Sec.~\ref{sec:OQ} followed by conclusions in Sec.~\ref{sec:conc}.

\begin{table}
    \centering
    \caption{Table of Acronyms}\label{tab:acronyms}
    \begin{tabular}{p{0.45\textwidth}}
        \hline\\
        \textbf{Acronym \hspace{1em} Definition} \\[1ex]
        \hline
        \begin{acronym}[Acronym+-]
            \acro{MIMO}{multiple-input and multiple-output}
            \acro{WDM}{wavelength-division-multiplexing}

            \acro{QC}{quantum communications}
            \acro{DMT}{diversity-multiplexing trade-off}
            \acro{DOF}{degree of freedom}
            \acro{SCAPE}{simultaneous communication and parameter estimation}
            \acro{DoF}{degree of freedom}
            \acro{CV}{continuous variable}
            \acro{DV}{discrete variable}
            \acro{FSO}{free space optics}
            \acro{KPI}{key performance indicator}
            \acro{MCF}{multicore fiber}
            \acro{QKD}{quantum key distribution}
            \acro{RF}{radio frequency}
            \acro{SKR}{secret key rate}
            \acro{QWDM}{quantum wavelength division multiplexing}
            \acro{SMF}{single-mode fiber}
            \acro{WCP}{weak coherent pulse}
            \acro{SPP}{single-photon pulse}
            \acro{HOM}{Hong-Ou-Mandel}
            \acro{BS}{beam spliter}
            \acro{MDI}{measure-device independent}
            \acro{PAC}{probably approximately correctly}
            \acro{RRDPS}{round-robin differential-phase shift}
            \acro{SLM}{spatial light modulator}
            \acro{OAM}{orbital angular momentum}
            \acro{QEC}{quantum error correction}
            \acro{SPA}{signal-processing approach}
            \acro{QML}{quantum machine learning}
        \end{acronym}\\[-3ex]\hline
    \end{tabular}
\end{table}

\section{Preliminaries}\label{sec:prel}
In this section, we set some notation and define some key terminology that we use throughout the manuscript. A list of acronyms used throughout the manuscript are provided in the Table~\ref{tab:acronyms}.
\subsection{CV vs DV Systems}
Quantum information processing systems can be characterized into one of the two broad categories, i.e., discrete- and continuous-variable quantum system. The term discrete or continuous refers to the spectra of measurable/observable quantities. Encoding and processing based on the number of photons in a pulse is discrete variable processing, whereas encoding and processing in field variables such as amplitude or phase results into a continuous-variable system \cite{ALS:10:LPR}. 

A quantum system is said to be a \ac{DV} system when its associated observables have a discrete spectrum. That is, when the measurement outcomes result into one of the finitely many outcomes. In \ac{DV} quantum information systems, the basic unit of information is a quantum bit (qubit), which is a system with a two-level \ac{DOF}. A qubit can be expressed as a superposition of two basis states\footnote{A qudit is the qubit generalization to $d$ orthogonal states quantum superposition rather than two states}, e.g., 
\begin{align}
    \ket{\psi} = \alpha \ket{0} + \beta\ket{1},
    \label{eq:qubit_ket}
\end{align}
where $\left\{ \ket{0}, \ket{1}\right\}$ is an orthonormal basis; and $\alpha, \beta \in \mathds{C}$ normalize to $\left| \alpha \right|^2 + \left| \beta \right|^2 = 1$. In \eqref{eq:qubit_ket}, the implicit assumption is that the qubit is in a pure state, i.e., in a well-defined quantum state with no statistical mixing. A more general representation of a qubit is the density operator form that takes into account the possible classical mixing of multiple pure states. A density matrix $\rho$ is a convex combination of one or more pure states, i.e., 
\begin{align}
    \rho = \sum_{i = 1}^{n} p_i \ket{\psi_i}\bra{\psi_i},
\end{align}
where the pure state $\ket{\psi_i}$ appears in the ensemble with probability $p_i$. A quantum state that is not pure is called a mixed state. Examples of \ac{DV} systems include spin 1/2 particles such as electrons, the two lowest energy states of semiconductor quantum dots, or two polarization states of a single photon \cite{WPG:12:RMP}.

A \ac{CV} quantum system is defined over an infinite-dimensional Hilbert space and the associated observables have continuous spectra \cite{WPG:12:RMP}. Examples of \ac{CV} quantum systems include quantized modes of bosonic systems such as the different \acp{DOF} of electromagnetic field, vibrational modes of solids, and nuclear spins in a quantum dots \cite{WPG:12:RMP}. Similar to the \ac{DV} case, a \ac{CV} quantum system in pure state can be described as the superposition of eigenstates, i.e., 
\begin{align}
    \ket{\psi} = \int \braket{x}{\phi}\ket{x}\,\mathrm{d}x,
\end{align}
where $\braket{x}{\phi}$ is the wave function of the state in the continuous basis $x$ \cite{ALS:10:LPR}. 

The best known example of continuous variable is the quantized harmonic oscillator that is described by \acp{CV} quadratures such as position and momentum \cite{WPG:12:RMP}. The Hamiltonian $\hat{H}_k$ of a quantum harmonic oscillator can be expressed in terms of annihilation and creation operators\footnote{We are considering normalized creation and annihilation operators leading to quadratures of the same dimension}
\begin{align}
    \hat{H}_k = \left( \hat{a}_k^{\dagger}\hat{a}_k + \frac{1}{2}\right),
\end{align}
This Hamiltonian can be represented in terms of position $\hat{x}_k$ and momentum $\hat{p}_k$ operators
\begin{align}
    \hat{H}_k = \frac{1}{2} \left( \hat{p}_k^2 + \hat{x}_k^2\right),
\end{align}
where 
\begin{align}
    \hat{a}_k &= \frac{1}{\sqrt{2}}\left( \hat{x}_k + \dot{\iota} \hat{p}_k\right),\\
    \hat{a}_k^{\dagger} &= \frac{1}{\sqrt{2}}\left(  \hat{x} - \dot{\iota} \hat{p}_k\right)
\end{align}
and the commutation relation $\left[\hat{x}_k, \hat{p}_{k'}\right] = \dot{\iota}\delta_{k, k'}$ holds. 

While working with \ac{DV} quantum systems,  the density operator formalism is preferred, in \ac{CV} systems other representations are often more convenient. The Wigner function, for example, is suitable to describe the effects on quantum observable which may arise from quantum theory and statistical effect. The Wigner function behaves like classical probability and enabling calculations of mean and variances of the quadratures \cite{Bv:05:RMP}. In contrast to classical probability, Wigner function can become negative.\footnote{The negativity of the Wigner function is considered as a measure of a genuinely quantum non-Gaussianity \cite{genoni2010quantifying}. } For a given \ac{CV} quantum state with density operator $\rho$, the Wigner function can be obtained \cite{Bv:05:RMP, Wig:32:PR}
\begin{align}
    W\left(x, p\right) = \frac{2}{\pi}\int e^{\dot{\iota}4yp}\bra{x - y}\rho \ket{x + y}\, \mathrm{d}y,
\end{align}
where the integration is from $-\infty$ to $+\infty$. Gaussian states are the ones whose Wigner representation is Gaussian and can be completely specified with the first two moments \cite{WPG:12:RMP}.

Indeed, any \ac{CV} state can be expanded in the Fock space basis representation, comprising photon number states \(\{\ket{n}\}_{n=0}^{\infty}\). Different representations of \ac{CV} systems can be used depending on the context. For instance, when studying the entanglement of Gaussian states or their behavior under different Gaussian transformations, the Wigner or the Glauber-Sudarshan representations are appropriate. Conversely, when orthogonality is an important consideration, the Fock representation is more suitable as it provides orthogonality through its inherent orthonormal basis.

\subsection{Probabilistic cross-talk (classical interference) vs interference (quantum) and Use Cases}

The term ``interference'' is used in both wireless communication systems and quantum information systems, though with different meanings. In quantum information systems, interference refers to the interaction of matter waves. This interaction can occur between matter waves associated with different (sets of) physical particles or, more commonly, through the self-interference of a matter wave associated with a single (set of) physical particle. Designing quantum information processing protocols involves careful tracking of \emph{constructive} and \emph{destructive} interference as a result of this interaction, both of which can be desirable. The most well-known experiment for demonstrating quantum interference is Young's double-slit experiment where the wavefunctions of single photons interact with themselves to give rise to an interference pattern on the screen. The most well-known application/use case is Grover's quantum algorithm for database search. In Grover's algorithm, a marked element is to be searched in a database. Through a carefully designed quantum routine, the probability of measurement outcome corresponding to the marked element (other elements) is increased (decreased) by utilizing quantum interference. 

In classical communication systems, interference generally refers to the disruptive modification of a signal caused by the presence of another signal in its vicinity. Interference in classical communication systems can be of different types, e.g., co-channel interference, inter-symbol interference, adjacent-channel interference, multi-user interference and spatial interference. Similar to quantum interference, interference in classical communication systems can be among different electromagnetic waves (e.g., in co-channel and adjacent channel interference) or self-interaction of an electromagnetic wave (e.g., inter-symbol interference, temporal and spectral dispersion through the propagation channel). 

In contrast to quantum interference, classical interference is almost always undesirable and degrades the performance of the communication system, unless intentionally manipulated at the trasceivers. In the same context, untreated interference affecting optical systems (free-space or fiber) has been so far modelled as an additional source of noise under the term crosstalk.

To avoid ambiguity in this paper, we reserve the term interference for the quantum interference discussed in the previous paragraph and reserve the term cross-talk for the interaction of different signals regardless of their classical or quantum nature.

\subsection{Crosstalk in Free Space Optics}
Classical optical communication techniques play a crucial role in characterizing and mitigating the effects of turbulence on quantum communication systems. Turbulence in the atmosphere can cause random phase and intensity variations in propagating light beams, leading to unwanted distortions in the received field. These distortions affect most of the degrees of freedom of the light, including intensity, phase, and polarization, which are critical in quantum communication. Classical beam propagation methods can be adapted and enhanced to improve the performance of quantum communication systems in turbulent environments. These methods allow for calculating the received field and intensity expressions not only in free space but also in turbulent environments. By modeling the effects of turbulence on the transmitted quantum states, it becomes possible to predict and quantify the level of distortion experienced by the quantum signals. Moreover, classical techniques enable the calculation of the transmissivity of the beam, taking into account the impact of turbulence. By accurately assessing the transmissivity, researchers can determine the key rate of the quantum communication system, which is essential for evaluating its overall performance. In essence, by leveraging classical optical communication methodologies, researchers can develop strategies to mitigate the effects of turbulence on quantum communication systems. This includes understanding how turbulence affects the propagation of quantum states, modeling its impact on key parameters such as transmissivity, and ultimately optimizing the system to achieve reliable and efficient communication even in turbulent environments.

\subsection{Crosstalk in Quantum Circuits}
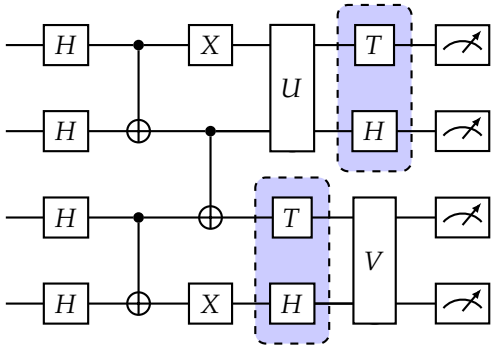
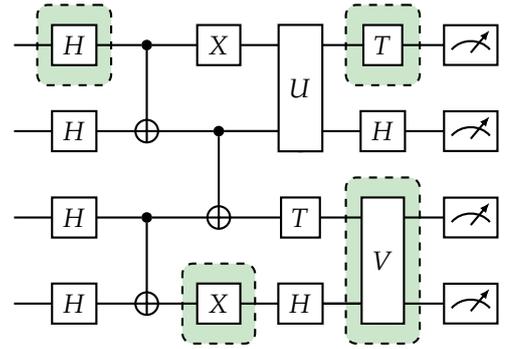
\begin{figure*}[t]
    \begin{minipage}[c] {0.49\textwidth}
        \centering
\begin{quantikz}
    & \gate{H} & \ctrl{1} & \gate{X} & \gate[2]{U}  & \gate{T}\gategroup[2, steps=1, style={dashed, rounded corners, fill=blue!20, inner xsep=2pt},
background]{{\sc }} & \meter{}\\
    & \gate{H} & \targ{} & \ctrl{1} & \targ{} & \gate{H} & \meter{}\\
    & \gate{H} & \ctrl{1} & \targ{} & \gate{T}\gategroup[2, steps=1, style={dashed, rounded corners, fill=blue!20, inner xsep=2pt},
background]{{\sc }} & \gate[2]{V} & \meter{}\\
    & \gate{H} & \targ{} & \gate{X} & \gate{H} & \targ{} & \meter{}
\end{quantikz}
\subcaption{Markovian cross-talk acting on different gates that are spacelike separated.}
\label{markovian}
\end{minipage}
\hspace{0.02\textwidth}
\begin{minipage}[c] {0.49\textwidth}
        \centering
\begin{quantikz}
    & \gate{H}\gategroup[1, steps=1, style={dashed, rounded corners, fill=green!20, inner xsep=2pt},
background]{{\sc }} & \ctrl{1} & \gate{X} & \gate[2]{U}  & \gate{T}\gategroup[1, steps=1, style={dashed, rounded corners, fill=green!20, inner xsep=2pt},
background]{{\sc }} & \meter{}\\
    & \gate{H} & \targ{} & \ctrl{1} & \targ{} & \gate{H} & \meter{}\\
    & \gate{H} & \ctrl{1} & \targ{} & \gate{T} & \gate[2]{V}\gategroup[2, steps=1, style={dashed, rounded corners, fill=green!20, inner xsep=2pt},
background]{{\sc }} & \meter{}\\
    & \gate{H} & \targ{} & \gate{X}\gategroup[1, steps=1, style={dashed, rounded corners, fill=green!20, inner xsep=2pt},
background]{{\sc }} & \gate{H} & \targ{} & \meter{}
\end{quantikz}
\subcaption{Non-Markovian cross-talk acting on different gates that are timelike separated.}
\label{nonmarkovian}
\end{minipage}
\caption{An illustration of different types of cross-talk that can affect quantum circuits}
\end{figure*}

Cross talk in quantum circuits refers to the undesired interaction among qubits, leading to errors and reduced fidelity in quantum computations. In an ideal quantum system, qubits should interact only with intended control signals or with other qubits in a controlled manner during the execution of quantum gates. However, due to physical proximity and the complex nature of quantum hardware, qubits can influence each other. This unwanted interaction can manifest as cross talk, which can introduce noise, causes decoherence, and results in computational errors. As a matter of fact, we distinguish two types of cross talk, Markovian and non-Markovian cross talks. Markovian cross talk is a memoryless process that afffects the computational qubits in a spacelike way, creating undesirable correlations which affects the readout after measurements as illustrated in Fig.~\ref{markovian}. Differentely, non-Markovian cross talk is a process that affects the computational qubits in a timelike manner, creating a memory of the individual systems of their past evolution as illustrated in Fig.~\ref{nonmarkovian}. quantum 
Cross talk is particularly problematic in scalable quantum computing, where maintaining high coherence and precise control over numerous qubits is crucial. Mitigating cross talk involves careful design and isolation of qubits, improving error correction techniques, and developing more sophisticated control mechanisms to ensure that qubits only interact as intended. Addressing cross talk is essential for advancing the reliability and efficiency of quantum circuits, paving the way for more robust quantum computing applications. In \cite{harper2024crosstalk} the authors designed a cross talk model and benchmarked it with the IBM QPU , in order to investigate the presence of cross talk noise and characterize it. The influence  of cross talk on different gates and circuits and the efficiency of the designed model were addressed  by a tomography on the IBM hardware, showing that the latter is significantly influenced by cross talk effects. A different approach was taken by the authors of \cite{seo2021mitigation} who considered cross talk effects arising from noise in a measurement readout stage where errors appear not only on individual qubits but also on multiple ones collectively. The authors have presented an error mitigation protocol consisting of a preproccessing stage, where local quantum gates are applied before the measurement setup, followed by a post-processing stage treating measurement outcomes to recover noiseless data. A similar approach using post-processing techniques for readout error mitigation in NISQ quantum devices was established in \cite{bravyi2021mitigating}. The authors considered a correlated noise model
based on Continuous Time Markov Processes to simulate the cross talk effect. The proposed error mitigation methods
were demonstrated experimentally for measurements of up to 20 qubits. Indeed, in a real world, cross talk effects in NISQ devices might follow non-markovian models. The authors of \cite{white2020demonstration} provided a way for characterizing non-markovian errors (arising from crosstalk or from other sources) and experimentally testing it on the IBM hardware, using the process tensor formalism. In the same spirit, using tensor-network quantum processes \cite{torlai2023quantum}, non-markovian effects in quantum processors can be characterized and mitigated. This has been reported in \cite{filippov2023scalable,mangini2024tensor}.

\section{Quantum Information in Different Media}\label{sec:FSOBP}
In this section, we review different media that acts as a communication channel for quantum communication systems, i.e., fibre optics and free-space.

\subsection{Fibre Optics}
\subsubsection{Fiber Optics MIMO}
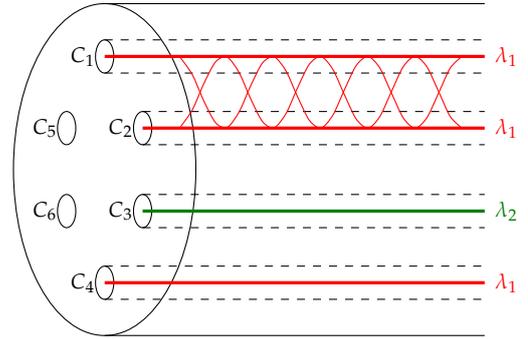
\begin{figure}[t!]
    \centering

    \usetikzlibrary{shapes.geometric}

\newcommand{\coreRed}[5]{
\coordinate (#1) at (#2,#3);
\node[#4] at (#1) {#5};
\draw \boundellipse{#1}{\dx/10}{\dy/10};
\draw[dashed] (#2,#3+\dy/10) -- (\dl,#3+\dy/10);
\draw[very thick,red] (#1) -- (\dl,#3) node [right] {$\lambda_1$};
\draw[dashed] (#2,#3-\dy/10) -- (\dl,#3-\dy/10);
}

\newcommand{\coreGreen}[5]{
	\coordinate (#1) at (#2,#3);
	\node[#4] at (#1) {#5};
	\draw \boundellipse{#1}{\dx/10}{\dy/10};
	\draw[dashed] (#2,#3+\dy/10) -- (\dl,#3+\dy/10);
	\draw[very thick,green] (#1) -- (\dl,#3) node [right] {$\lambda_2$};
	\draw[dashed] (#2,#3-\dy/10) -- (\dl,#3-\dy/10);
}

\newcommand{\coreL}[5]{
	\coordinate (#1) at (#2,#3);
	\draw \boundellipse{#1}{\dx/10}{\dy/10}; %factors of the main ellipse
	\node[#4]  at (#1) {#5}; 
}

\newcommand{\boundellipse}[3]% center, xdim, ydim
{(#1) ellipse (#2 and #3)
}

\begin{tikzpicture}[
	% Environment Cfg
	font=\footnotesize, 
	text centered,
%	rounded corners=0.08cm,
	%	text width=2.5cm, 
	% Styles
	conn/.style ={
		draw=black,
		text width=1cm, 
		text=white, minimum height=4cm,
		thick,
		fill=ocre!90,
		blur shadow, 
	},
	stan/.style ={
		draw=skybox,
		text width=2cm, 
		text=black, minimum height=4cm,
		thick,
		fill=sky,
		blur shadow, 
	},
	arrow/.style ={
		-{Latex[length=3pt]},
		semithick
	},
	Neutral/.style ={
		draw=black!50,
				text width=2cm, minimum height=2cm,
		thick,
%		fill=black!10,
%		blur shadow, 
	}
	]

	\def\fr{5}
	
	%fiber coordinate
	\def\dl{5}
	
	\def\dx{1.2}
	\def\dy{2.2}

	\def\xa{0}
	\def\ya{1.5}
	
	\def\xb{.5}
	\def\yb{.55}

	\draw \boundellipse{0,0}{\dx}{\dy};
	\draw (0,\dy) -- (\dl,\dy);
	\draw (0,-\dy) -- (\dl,-\dy);

	\coreRed{A}{\xa}{\ya}{left}{$C_1$}

	\coreRed{B}{\xb}{\yb}{left}{$C_2$}
	\coreGreen{C}{\xb}{-\yb}{left}{$C_3$}

	\coreRed{D}{\xa}{-\ya}{left}{$C_4$}
	
	\coreL{E}{-\xb}{\yb}{left}{$C_5$}
	\coreL{F}{-\xb}{-\yb}{left}{$C_6$}

	%cross-Talk		
	\draw[shift={(A)}, scale=1, domain=1:\dl-.3, smooth, variable=\x, red] plot ({\x}, {(\yb-\ya)*sin(\x r*\fr)/2+(\yb-\ya)/2});
	\draw[shift={(A)}, scale=1, domain=1:\dl-.3, smooth, variable=\x, red] plot ({\x}, {(\yb-\ya)*sin(-\x r*\fr)/2+(\yb-\ya)/2});

%	\draw \boundellipse{0,1}{.1}{.2};
%	
%	\draw \boundellipse{5,0}{1}{2};
%	

\end{tikzpicture}
   
    \caption{Crosstalk in six-core fiber. Signals transmitted through neighboring cores at the same wavelength can crosstalk with each other, as illustrated with cores $C_1$ and $C_2$.}
    
    \label{fig:MCF}
\end{figure}

\Acp{MCF} whose cross-section is shown in Fig.~\ref{fig:MCF} have garnered significant research attention recently due to their promising advantages in telecommunications and data transmission. These fibers offer the potential for achieving higher data rates while also providing space efficiency, fault tolerance, and reduced power consumption compared to traditional single-mode fibers. However, challenges such as crosstalk between the cores have been identified alongside these advantages. Crosstalk can occur due to interference between the cores, potentially impacting signal quality and transmission efficiency. Despite this challenge, researchers have been actively working on improving \ac{MCF} performance. Early studies measured inter-core cross-talk (XT) at around -40dB/km\cite{Rottondi2019}, but advancements have improved this to approximately -60dB/km\cite{Hayashi2019}. These improvements are influenced by the coupling methods used. Both non-coupled and coupled MCFs have been thoroughly analyzed. The impulse response width is also presented as a function of the operating wavelength and core diameter\cite{SAKAMOTO20178}. The influence of crosstalk on the signal-to-noise ratio(SNR) is critical. SNR penalty in \ac{MCF} is derived and proved experimentally considering the crosstalk effect \cite{9006884}. 
\subsubsection{Quantum Fiber Optics}
Additionally, researchers have investigated crosstalk power caused by forward Raman noise and forward Raman scattering noise resulting from inter-core crosstalk. Proposals have been made to model and evaluate QKD performance in multi-core and few-mode fibers, considering factors like dark current and inter-core crosstalk as background noise sources \cite{Moghaddam2021}. In these studies, authors have analyzed the feasibility of QKD transmission in multi-core fibers with quantum channels alone or both quantum and classical channels. Such research contributes to understanding the potential and limitations of MCFs in practical applications, paving the way for their future deployment in advanced telecommunications systems\cite{Urena:19}.One notable area of focus is \ac{QKD}, where MCFs have become a trending topic. For instance, a 52km \ac{QKD} link was successfully established using a 4-core MCF, achieving a transmission rate of 51.5kbps\cite{Mujtaba2024}. Furthermore, CV-QKD is less affected by crosstalk overall. However, when the CV-QKD wavelength overlaps with a classical channel, the excess noise from crosstalk rises sharply. Nevertheless, a secret key rate of 94.2 Gbps is still anticipated\cite{Eriksson2019}. 

\subsubsection{Crosstalk Mitigation}
The mitigation of crosstalk in \acp{MCF} has recently become a trending topic.  \Ac{MCF} design, core effective area, and propagation configurations are some of the crucial aspects that should be considered to mitigate the crosstalk. Different \ac{MCF} designs with trench or moat-assisted cores are evaluated, highlighting that larger trench/moat volumes significantly reduce inter-core crosstalk and radiation leakage loss. However, these designs pose manufacturing challenges and hydrogen sensitivity issues. The study shows that bidirectional transmission can further reduce crosstalk compared to co-propagating transmission. Optimal inter-core distances for minimizing crosstalk and maximizing SNR are identified, with specific values for different core effective areas. The results indicate that, despite potential higher SNR for larger effective area designs, smaller effective area designs may offer higher overall capacity due to accommodating more uncoupled cores \cite{TANDON2023129483}. In another study, key strategies include bi-directional transmission, which can suppress crosstalk by about 20 dB per core pair, and specialized core designs like trench-assisted (TA) and step-index (SI) MCFs that lower inter-core interference through careful geometric and refractive index adjustments. The proposed approaches also involve advanced routing and spectrum allocation algorithms, such as core prioritization and spectrum-splitting, to optimize spectral resource use and mitigate crosstalk.  New wavelength-dependent crosstalk models are presented, and their effectiveness is evaluated through simulations, highlighting that TA-MCFs with bi-directional transmission significantly enhance network capacity and link efficiency compared to traditional methods\cite{8336681}. Orthogonal filtering techniques suppress inter-core crosstalk (IC-XT) in weakly-coupled multicore fiber (WC-MCF) transmission systems. This method simplifies the reception of signals in WC-\ac{MCF} systems by eliminating the need for \ac{MIMO} digital signal processing, enabling individual lane extraction. The study demonstrates effective IC-XT mitigation using a pair of orthogonal filters in a 2-core MCF, achieving a maximum Q factor improvement of 6 dB for 600-Gbps 64-QAM signal transmission over 100 km. The research employs a transmission model based on coupled nonlinear Schrödinger equations to simulate and verify the proposed method's efficacy. Additionally, the impact of spatial mode dispersion on filter orthogonality performance is analyzed, highlighting the proposed system's considerable tolerance to IC-XT and its potential to enhance design flexibility for ultra-high-density WC-MCFs \cite{9893163}. Another method is designing and fabricating ultra-low crosstalk and low-loss MCF, focusing on reducing crosstalk in multicore fibers through various methods. It is achieved significantly low crosstalk and attenuation levels by employing trench-assisted pure-silica cores arranged hexagonally and utilizing fiber bends. Using an approximation model based on a coupled-mode theory with the equivalent index model, a relationship is derived among crosstalk, fiber parameters, and fiber bend. The mean crosstalk is estimated to be less than -30 dB even after 10,000 km of propagation. This study highlights the importance of managing fiber bends and using trench-assisted cores to achieve these results\cite{Hayashi:11}. The other methods are the development and advantages of trench-assisted multi-core fiber (TA-MCF) for reducing crosstalk in high-density fiber optic systems. TA-\ac{MCF} is proposed as an alternative to traditional step-index multi-core fibers (S-MCF), highlighting its ability to significantly reduce crosstalk through an index trench structure surrounding each core. This trench structure minimizes the overlap of electric fields between adjacent cores, thereby reducing crosstalk. The simulation results demonstrate that TA-MCFs can achieve crosstalk values up to 20 dB lower than S-MCFs over a 100 km distance. Additionally, the study includes the fabrication and testing of two types of TA-MCFs, which confirmed the simulation results, showing excellent crosstalk reduction, bending loss, and cutoff wavelength characteristics. The findings suggest that TA-MCFs are promising candidates for future high-capacity, long-haul optical communication systems\cite{5875695}. A hybrid analytic-numerical method reduces crosstalk in multicore fibers by optimizing the amplitude and frequency of random core radius fluctuations along the propagation direction. This method employs coupled mode theory and first-order perturbation theory to analytically determine noise amplitude parameters that minimize power transfer between inner and outer cores, validated by numerical experiments using finite difference beam propagation methods. The study reveals that moderate-noise regions effectively suppress crosstalk, with noise amplitude and fluctuation frequency parameters tailored to achieve optimal suppression. The proposed approach is generalizable to various optical systems described by coupled mode theory, facilitating efficient design and prediction of crosstalk suppression in multicore fibers and similar structures \cite{RN110}.

\subsection{Free space optics}
\subsubsection{MIMO FSO}
\begin{figure}[t!]
    \centering

    \usetikzlibrary{patterns}

\newcommand{\BS}[4]{
	\draw[stan] (#2-\l/2,#3-\l/2) rectangle (#2+\l/2,#3+\l/2);
	\draw[thick,skybox] (#2-\l/2+0.04,#3-\l/2+0.04) -- (#2+\l/2-0.04,#3+\l/2-0.04);
	\node[]  [below=\l*.6 cm of #1]  {#4};		
}

\newcommand{\BSa}[4]{
	\draw[stan] (#2-\l/2,#3-\l/2) rectangle (#2+\l/2,#3+\l/2);
	\draw[thick,skybox] (#2-\l/2+0.04,#3-\l/2+0.04) -- (#2+\l/2-0.04,#3+\l/2-0.04);
	\node[]  [above=\l*.6 cm of #1]  {#4};		
}

\newcommand{\mirror}[4]{
	\draw[thick,decoration={
		markings,
		mark=between positions 0.015 and 0.98 step 0.1072 with {\draw (0,0)--(60:3pt);}
	}] (#2-\l/2+0.04,#3-\l/2+0.04) -- (#2+\l/2-0.04,#3+\l/2-0.04);
	\node[]  [above=\l*.6 cm of #1]  {#4};		
}

\newcommand{\mirrorR}[4]{
	\draw[thick,decoration={
		markings,
		mark=between positions 0.015 and 0.98 step 0.1072 with {\draw (0,0)--(60+180:3pt);}
	}] (#2-\l/2+0.04,#3-\l/2+0.04) -- (#2+\l/2-0.04,#3+\l/2-0.04);
	\node[]  [below=\l*.6 cm of #1]  {#4};		
}

\begin{tikzpicture}[
	% Environment Cfg
	font=\footnotesize, 
	text centered,
%	rounded corners=0.08cm,
	%	text width=2.5cm, 
	% Styles
	conn/.style ={
		draw=black,
		text width=1cm, 
		text=white, minimum height=4cm,
		thick,
		fill=ocre!90,
		blur shadow, 
	},
	stan/.style ={
		draw=skybox,
		text width=2cm, 
		text=black, minimum height=1cm,
		thick,
		fill=sky,
%		blur shadow, 
	},
	arrow/.style ={
		{Latex[length=3pt]}-,
		semithick,
	},
	Neutral/.style ={
		draw=black!50,
		text width=2cm, minimum height=2cm,
		thick,
		%		fill=black!10,
		%		blur shadow, 
	}
	]
	\def\h{3.5} 
	\def\l{1}    %base of triangles
	\def\s{.5} %separation between TXs

	\def\xa{1}
	\def\ya{1}
	\coordinate (A1) at (\xa,\ya);
	\node[below] at (\xa,\ya) {$T_1$};

	\coordinate (m) at (\xa+\s/2,\ya+\s);

	%Triangle given h, l and coordinate
	 \coordinate (B) at (\xa-\l/2, \ya-\h);
	 \coordinate (C) at (\xa+\l/2, \ya-\h);
%	 \draw[fill=blue,fill opacity=.1,pattern=crosshatch] (A1) -- (B) -- (C) -- cycle;
 	\draw[draw=none, fill=green,fill opacity=.2] (A1) -- (B) -- (C) -- cycle;
%	 	\draw[pattern=crosshatch, pattern color=red] (A1) -- (B) -- (C) -- cycle;
	
	\def\xb{\xa+\s}
	\coordinate (A2) at (\xb,\ya);
	\node[below] at (\xb,\ya) {$T_2$};

	%Triangle given h, l and coordinate
	\coordinate (B) at (\xb-\l/2, \ya-\h);
	\coordinate (C) at (\xb+\l/2, \ya-\h);
	
	\draw[draw=none, fill=ocre,fill opacity=.2] (A2) -- (B) -- (C) -- cycle;
%	\draw[pattern=crosshatch, pattern color=blue] (A2) -- (B) -- (C) -- cycle;
	
	\coordinate[below=9*\h/10 of m] (CT);
	\draw[arrow,thin] (CT) --++ (-\h/2,\h/5) node[above] {Crosstalk};

	%Channel: a node with pattern
	
%	\node[pattern=crosshatch, pattern color=green!30,	text width=2cm, minimum height=1cm,rounded corners=0.08cm] (AB) [below=1.5 cm of m] {Quantum Channel}; %h/2

	%TRANSMITTER
	
	%panels's dimensions
	\def\xp{1.1*\s}
	\def\yp{3*\s}
	
	\draw[fill=black,rounded corners=0.08cm] (A1) -- ++ (\s,0) -- ++ (0,2*\s) -- ++ (-\s,0)  -- cycle ;
	
	\coordinate[above=\s of A2] (S3);
	\coordinate[right=\s/2 of S3] (S4);
	\draw (S3) -- (S4);	
	\draw[pattern=grid, pattern color=skybox] (S4)  -- ++ (0,\xp/2) --++ (\yp, 0) --++ (0,-\xp)--++ (-\yp,0) -- cycle;

	\coordinate[above=\s of A1] (S5);	
	\coordinate[left=\s/2 of S5] (S6);
	\draw (S5) -- (S6);
	\draw[pattern=grid, pattern color=skybox] (S6)  -- ++ (0,\xp/2) --++ (-\yp, 0) --++ (0,-\xp)--++ (\yp,0) -- cycle;

	%RECEIVER 

	%%receiver	
	\def\r{\s/2}
	\coordinate (AA1) at (\xa-\r,\ya-\h);
	\node[below=\r] at (\xa,\ya-\h) {$R_1$};
	
	\coordinate (AA2) at (\xb-\r,\ya-\h);
	\node[below=\r] at (\xb,\ya-\h) {$R_2$};
	
	\draw[fill=black] (AA1) arc[start angle=180, end angle=360, radius=\r];
	\draw[thin] (AA1) --++(\s/2,\s/4) --(\xa+\r,\ya-\h);
	
	\draw[fill=black] (AA2) arc[start angle=180, end angle=360, radius=\r];
	\draw[thin] (AA2) --++(\s/2,\s/4) --(\xb+\r,\ya-\h);

	% clearly show the optical array
	% write:
	%% crosstalk on the left (arrow)
	%% quantum channel on the right as a clave

	%labels
%	\node[right=1.5cm of m] (Tx) {Transmitter};
%	\node[below=of Tx] 	
	
	  \draw[decorate,decoration={brace,amplitude=10pt}] (\xa+4*\s,\ya) -- ++(0,-\h) node[midway,above,right=10pt,text width=2cm] {Quantum Channel};

\end{tikzpicture}
   
    \caption{Crosstalk in a \ac{MIMO} \ac{FSO} system operating in atmospheric turbulence. Two optical beams (depicted in green and red), transmitted by $T_1$ and $T_2$, crosstalk each other before reaching the receiver's detectors, $R_1$ and $R_2$.}
    
    \label{fig:MIMOFSO}
\end{figure}
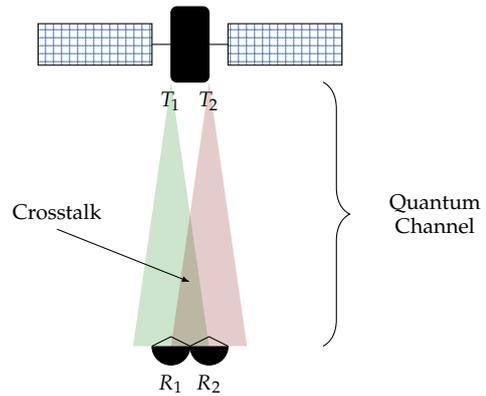

The efficacy of \ac{FSO} and free space \ac{QKD} is intricately tied to atmospheric turbulence. The unpredictable refractive index variations, random intensity fluctuations (scintillation), and erratic beam movements across the transverse receiver plane (beam wander) significantly impact communication performance. These atmospheric effects can be mitigated utilizing \ac{MIMO} configuration as given in \ref{fig:MIMOFSO}. It is presented that the scintillation index of \ac{MIMO} \ac{FSO} systems is mitigated by up to 1/3 as compared to the single-input single-output(SISO) configuration; therefore, it brings \(10^{-2}\) times less BER \cite{2016OptEn..55k1607G}. Aperture average scintillation of the FSO system is also inversely proportional to \ac{MIMO} complexity as it is in pointlike scintillation \cite{Gokce:16}.   Similarly, the BER of multiple-input single-output (MISO) decreases as the number of transmitters increases \cite{2016WRCM...26..642G}. Furthermore, array-type beams are more resistive against turbulence \cite{2023AnP...53500232N}.     

In addition to the mentioned \acp{DoF}, the performance of \ac{FSO} systems is significantly influenced by spatial, spectral, and polarization DoF.
Spatial DoF is particularly affected by the divergence of the optical beam caused by the turbulent environment. Atmospheric turbulence induces fluctuations in the refractive index, leading to random variations in the beam's propagation direction and spreading, thereby impacting the spatial focusing of the transmitted signal.
Polarization DoF is another critical aspect influenced by the atmosphere, primarily through phenomena like polarization mode dispersion (PMD). PMD occurs due to variations in the refractive index along different polarization axes, causing differential delays and phase shifts in the optical signal. Consequently, this leads to polarization fading and degradation in signal quality, especially in high-data-rate FSO systems.
Moreover, the spectral DoF of FSO systems is affected by specific wavelengths within the optical spectrum being attenuated by atmospheric constituents such as aerosols and water vapor. These attenuation windows can lead to wavelength-dependent losses, spectral broadening, and selective fading, impacting the transmission of information over specific frequency bands.
Understanding and mitigating the effects on these DoFs are essential for designing robust FSO communication systems capable of maintaining reliable and high-quality links in diverse environmental conditions.

\subsubsection{Free Space QKD}
Scientists leverage the degrees of freedom inherent in classical light to amplify data transmission capacity within the quantum realm. Consequently, they employ classical light for characterizing quantum channels. Moreover, it is stated that broken quantum states cannot be recovered without quantum tomography. As a result, the concurrence variation in quantum channels aligns with both classical light behavior and theoretical expectations \cite{nadagano2017}. In pioneering work, the first atmospheric turbulence model for quantum light was introduced in \cite{Semenov2009}, defining the transmission coefficient's interplay with input mode through the Glauber-Sudarshan P function. Thermal fields exert a significant influence on the behavior of light. In the context of phenomena like the Hong-Ou-Mandel Effect, where precise control over photon interactions is essential, the impact of thermal turbulence becomes particularly pronounced. The theoretical visibility limit, typically estimated at around 33\%, reflects the maximum visibility achievable under ideal conditions. However, experimental constraints often impose a more stringent limit, typically around 28\%, owing to the practical challenges posed by the thermal field\cite{PhysRevLett.131.233601}. 
Subsequently, a model utilizing Huygens-Kirchhoff integration \cite{Vasylyev2016}. emerged, offering a versatile validation spanning from weak to strong turbulence. This model commences its analysis by characterizing the elliptic beam within the transmitter aperture while incorporating the Kolmogorov power spectrum to simulate turbulent refractive index fluctuations.
Adjusting the atmospheric coherence length is achieved by modulating the refractive index structure constant, reflecting turbulence strength. 
Furthermore, beam wander, denoting undesired beam deviations from the line of sight \cite{andrews}, is scrutinized, with emphasis placed on its impact on nonclassicality \cite{Vasylyev2012}.

Leveraging the Kolmogorov power spectrum, this research computes the scintillation index of the Laguerre-Gaussian beam along a slant path, while considering refractive index variations via the Hufnagel-Valley model, elucidating the relationship between attitude and turbulence strength \cite{Yuceer2012}.
With a similar approach, log-normal power spectrum and gamma-gamma turbulence power spectrums are utilized to calculate the key rate of continuous wave QKD in weak and strong turbulences, respectively.
The inverse relation between the propagation distance and the key rate is emphasized due to the accumulated turbulence along the propagation distance \cite{Wang2018}. Accordingly, it is shown in \cite{Chai2019} that there is a 20\% difference in link distance between the fiber-based and free space QKD systems in the targeted secret key rate of $10^{-4}$. 

The numerical approach random phase screen(RPS) is used to obtain the transmissivity of the satellite to ground optical link. Then, the inverse relation between the zenith angle and the key rate is presented, benefiting from the transmissivity\cite{Villaseñor2020}.

From an experimental standpoint, a QKD setup is established, where turbulence is induced through a \ac{SLM}.
Random phase fluctuations are modeled by leveraging the inverse Fourier Transform of a frequency domain delta-correlated zero-mean Gaussian random complex function, combined with Kolmogorov power spectral density multiplication  \cite{Zhang2016}. 
In one early free-space QKD experiment, a 210m link was established, employing twisted photons and yielding a mean QBER of 3.81\% \cite{Vallone2014}. 
Subsequently, a significant milestone was achieved with the establishment of a 144km free-space entanglement-based QKD link between La Palma and Tenerife, wherein the European Space Agency's ground station played a pivotal role as the receiver. This endeavor shed light on the phenomenon wherein noise escalates as the detection of photons from uncorrelated pairs increases\cite{Ursin2007}.
\subsubsection{Mitigation Techniques of Atmospheric Turbulence}
The research is carried out to mitigate distortions due to atmospheric turbulence. Adaptive optics is a common method to correct the receiver plane wavefront. 
Adaptive optics plays a crucial role in QKD systems, especially in scenarios affected by atmospheric turbulence. 
This technology typically comprises a wavefront sensor, deformable mirror, and control unit. 
The correction order is determined based on the aberrations detected and the actuators in the deformable mirror.
Research, such as \cite{Gruneisen2021}, has demonstrated that in slant-path QKD systems, open-loop adaptive optics can notably improve signal-to-noise ratio and reduce quantum bit error rates compared to closed-loop configurations.
Building on this, recent studies, like \cite{Marulanda_2024}, have extended the application of adaptive optics to satellite-to-ground QKD links, both discrete and continuous wave systems. 
These studies indicate a significant enhancement in key rates, especially noticeable up to the 10th order of correction, with continued benefits observed up to the 20th order.
Furthermore, investigations such as \cite{Wang2019} emphasize the importance of closed-loop control frequency in improving secret key rates in CV-QKD setups. 
In challenging environments like turbulent mediums, as explored in \cite{Mikhael2024}, adaptive optics effectively mitigates wavefront distortions, thus improving the reliability of QKD operations.
Additionally, integrating orbital angular momentum (OAM) with QKD, as seen in \cite{Tao:21}, showcases a direct relationship between the correction order provided by adaptive optics and the secret key rate.
Higher correction orders enhance key rates and contribute to reducing quantum bit error rates, thus bolstering the overall security and performance of QKD systems.
In summary, adaptive optics emerges as a vital tool in advancing the capabilities of QKD systems, enabling improved performance and resilience against environmental challenges like turbulence.

Another method to mitigate atmospheric effects in free-space optical communication is by altering the structure of the beams. Some non-traditional beams exhibit reduced scintillation, smaller beam size, and less beam wander, making them advantageous for free-space optical communication. Consequently, shaped beams are utilized in free-space \ac{QKD} applications. For instance, the propagation properties of a vector Bessel-Gauss beam with a specific polarization state are analyzed in \cite{Li:17}. Changes in polarization and beam evolution are plotted as functions of propagation distance. The experimental setup for generating higher-order beams, in addition to zero-order beams, includes lenses, mirrors, a polarizing beam splitter, half-wave plates, quarter-wave plates, and a phase spatial light modulator.
Similarly, vector Bessel beams are employed in quantum systems due to their self-healing feature. This property enhances the signal-to-noise ratio as the beam can recover itself after encountering an obstacle \cite{Otte2018}. When considering the self-healing property of multiple Bessel-Gaussian beams with \ac{OAM}, multiplexing is performed on the testbed. This study finds that the simulation results for beams with lower \ac{OAM} modes closely match the experimental results. Additionally, Laguerre-Gaussian beams are generated using a spatial light modulator and are also used for multiplexing \cite{Ahmed2016}. However, the quantum bit error rate (QBER) of Laguerre-Gaussian beams is three times higher than that of Bessel-Gaussian modes, as noted in \cite{Nape:18}. Furthermore, the \ac{QKD} rate of flat-topped beams surpasses that of Laguerre-Gaussian beams under weak and moderate turbulence conditions \cite{He:21}.

\subsection{Integrated photonics}
Integrated photonics has become a cornerstone technology for advancing quantum communication and computing, leveraging the miniaturization, scalability, and stability of photonic circuits \cite{WSL:20:NPho}. This technology enables the integration of various quantum components such as single-photon sources, single-photon detectors, and quantum gates on a single chip, facilitating complex quantum information processing and secure communication protocols. For quantum communications, integrated photonics has been effectively utilized in QKD systems, capitalizing on established telecommunication infrastructure. Pioneering work includes SiO2-based optical interferometers for time-bin QKD systems \cite{honjo2004differential}. The first fully integrated chip-to-chip QKD system combined an InP transmitter and a silicon oxynitride (SiOxNy) receiver, achieving state rates up to 1.76 GHz \cite{sibson2017chip}. Moreover, integrated photonic devices have demonstrated the feasibility of Measurement-Device-Independent QKD (MDI-QKD) through Hong-Ou-Mandel (HOM) interference, with high visibility between weak coherent states from different integrated sources \cite{semenenko2019interference,agnesi2019hong}.

Quantum photonic chips represent a cutting-edge technology in quantum computing and quantum information processing, utilizing the unique properties of photons such as coherence and entanglement. These chips integrate various optical components on a single substrate, providing advantages in scalability, stability, and processing speed. Key platforms in this field include Xanadu's X-Series \cite{arrazola2021quantum}, PsiQuantum's chip \cite{alexander2024manufacturable}, and Quandela's Ascella \cite{maring2024versatile}, each offering distinct technologies and features tailored for different applications.
Xanadu's X-Series chips leverage silicon photonics and continuous-variable (CV) quantum computing, supporting a wide range of quantum applications with high-fidelity operations and reconfigurable circuits. PsiQuantum's chip and quandela's Ascella  focuse on  quantum computing with single-photon sources and detectors integrated on silicon photonic chips, emphasizing fault tolerance and scalability. 
Advances in single-photon sources and detectors, such as spontaneous parametric down-conversion (SPDC) and superconducting nanowire single-photon detectors (SNSPDs), which will be discussed in Sec.~\ref{sec:FSQC} have significantly improved the efficiency and reliability of quantum operations. Waveguide integration in materials like silicon, silicon nitride, and lithium niobate allows for precise photon routing with minimal loss, enabling the creation of complex photonic circuits. Quantum interference and entanglement generation on-chip have achieved high levels of fidelity, essential for scalable quantum networks. \Ac{QEC} on photonic chips involves encoding logical qubits into multiple physical qubits using error-correcting codes, with techniques such as the surface code and bosonic codes under active investigation. Implementing fault-tolerant operations requires low error rates and efficient error detection and correction, which remain critical research areas.
Applications of photonic quantum chips span quantum computing, communication, and sensing. In quantum computing, these chips promise scalable systems with inherent parallelism and low decoherence rates, targeting optimization, simulation, and cryptography. Quantum communication benefits from integrated photonics for secure communication channels and robust quantum repeaters. Quantum sensing applications include precision measurements in various fields, with integrated photonic chips offering compact and robust sensor solutions. Challenges in scaling up the number of qubits, maintaining high fidelity, and ensuring efficient integration with classical systems persist. Advances in manufacturing precision and hybrid quantum-classical architectures are essential. Establishing standards and benchmarks for performance evaluation will facilitate technology comparison and guide future development.

\section{Quantum Communication}\label{sec:FSQC}

Due to ease of implementation and low losses at small distances, optical fibers are the preferred medium for implementing short quantum channels \cite{PPA:09:NJP, SFI:11:OE, SLB:11:NJP, TYZ:16:PRX, CLL:09:OE, XCW:09:CSB, CWL:10:OE, HHL:16:OL}. However, the photon losses in fiber optic grow exponentially with the length of the channel making this approach non scalable \cite{LCP:22:RMP}. The vision of quantum internet incorporates a global-scale network that necessarily includes quantum links going beyond a few hundred kilometres. In parallel, a large percentage of the current digital networks operate over wireless links, either due to the infeasibility of underground or underwater cabling or due to the inherent mobility of the end nodes.  Therefore, the free-space links appear as a necessary component of this future quantum internet. In this section, we survey key components to achieve optical quantum communications, i.e., quantum sources, optical quantum channels, and detectors.

\subsection{Quantum sources}

\subsubsection{DV Sources}

\begin{table*}[t]
\centering
\caption{Key Performance Indicators (KPIs) for Different Single-Photon Sources}
\begin{tabular}{@{}ll@{}}
\toprule
\textbf{KPI} & \textbf{Value and Source} \\ \midrule
Production rate & Millions of single photons per second \cite{pomarico2012mhz} \\
Generation probability & $10^{-6}$ (SPDC and FWM) \\
Visibility in two-photon interference & Exceeds 90\% (Diamond color centers) \cite{pompili2021realization} \\
Zero phonon line incidence (ZPL) & 4\% (Diamond color centers) \\
Branching ratio & 80\% (Diamond color centers) \\
Collection efficiency & Up to 20\% (PAHs) \cite{colautti20203d} \\
Photon generation rate & Just under 50 million photons/s (PAHs) \\
Intensity squeezing & Up to 2.2 dB (PAHs) \cite{chu2017single} \\
Brightness (at detector) & 53\% (Quantum dots) \cite{thomas2021bright} \\
Indistinguishability (raw) & About 91\% (Quantum dots) \cite{thomas2021bright} \\
Photon coincidences & One per year (six single-photon Fock states) \\
Single-photon purity ($g^{(2)}(0)$) & $0.00(3)$ (PAHs) \cite{rezai2018coherence} \\
\bottomrule
\end{tabular}
\label{sources}
\end{table*}

A single-photon source ideally produces a single photon with high probability when requested. The generation of single-photon states is essentially based on two possible approaches, respectively relying on nonlinear effects in optically pumped crystals or on emission from individual quantum emitters. The main difference between the two  lies in the probabilistic nature of generation processes for nonlinear effects, compared to the potential of deterministic processes. Hereafter, we will provide a literature review on single photon sources with more focus on deterministic ones relying on quantum emitters.

Some systems utilize nonlinear interactions between the electromagnetic field and engineered materials to generate correlated photon pairs, such as through spontaneous parametric down-conversion (SPDC) or four-wave mixing (FWM). These interactions involve one or two pump photons decaying into pairs of photons. In these setups, one photon of the pair often serves as a herald, meaning its detection confirms the presence of a single photon in the other branch. Modern sources can produce millions of single photons per second in controlled states \cite{pomarico2012mhz} with specific spectral properties and well-regulated spatial modes. However, due to the probabilistic nature of these nonlinear optical processes, achieving both a high probability of photon production and high purity (the ratio between single-photon generation probability and the probability of multiple-photon detection) simultaneously is challenging.

Probabilistic single-photon sources are still widely used due to fundamental characteristics such as simple integration into photonic circuits \cite{LZL:21:AQT, CXE:17:LSA, FGR:20:AQT}, tailored spectral properties, and high indistinguishability. By employing many low-probability but high-purity heralded single-photon sources, systems can increase the probability of successful single-photon generation. Multiplexing in these sources utilizes time, space, and/or frequency degrees of freedom to parallelize photon creation in different modes, with active switching based on heralding event detection \cite{meyer2020single}. This approach can achieve very high brightness (generation rate over pump power) \cite{kaneda2019high}, though still lower than deterministic emitters.

Single-atom emissions are utilized to generate single-excitation states in the electromagnetic field \cite{cohen1998nobel}. These emissions stem from isolated quantum emitters, which are two-level systems excited by short, intense laser pulses to ensure a single excitation per cycle. The resulting spontaneous emission produces a single-photon state, crucial for applications requiring well-defined modes of the field. Quantum emitters are commonly integrated with optical cavities, single-mode fibers, or waveguides for enhanced processing capabilities \cite{lounis2005single}. This approach is promising due to its capability to generate photons on demand, driving exploration across various platforms to minimize experimental losses. Key technologies include color-center-based sources, quantum dots, and molecular sources.

Point defects in solids, or color centers, in a transparent host matrix have gained interest for applications like single-photon sources and spin qubits due to their unique combination of single-photon emission and long spin state coherence. Advanced sources include color centers in diamonds, such as nitrogen vacancies and silicon centers. Nitrogen vacancy centers have reported visibility in two-photon interference processes exceeding 90\% \cite{pompili2021realization}, even with emission from distinct centers. The main challenge is the low emission probability in the zero phonon line (ZPL) spectral region, at only 4\%. Silicon centers show an 80\% branching ratio but suffer from low radiative decay efficiency. Additionally, extracting light from diamond, a high refractive index material, is difficult, resulting in low brightness.

Photonics structures crafted from diamond aim to improve collection efficiency but often at the expense of coherence properties. An alternative method involves implanting impurities into structured materials using ion or electron beams, creating cavities, photonic crystals, or waveguides. Diamond-based sources are complex to fabricate, with only a select few groups worldwide consistently achieving favorable characteristics \cite{lenzini2018diamond}. These sources primarily serve as spin-photon interfaces for applications in quantum communication and distributed quantum computing protocols, as discussed further in \cite{neumann2023organic}.

Sources including single molecules of polycyclic aromatic hydrocarbons (PAHs) in transparent matrices,  exhibit sub-Poissonian statistics of anti-bunching \cite{basche1992photon}. PAH molecules show near-unit quantum yield and high purity values (\( g^{(2)}(0) = 0.00(3) \)) \cite{rezai2018coherence}, making them competitive in the solid state. Fluorescence intermittency is due to a non-zero probability of decaying to a triplet state, weakly allowed by spin-orbit coupling. For example, dibenzoterrylene (DBT) molecules in anthracene (Ac) matrices have a triplet state decay probability of \( 10^{-7} \) at low temperatures. This source emits in the near-infrared at around 785 nm and is stable at room and cryogenic temperatures, with an off-time of less than 1\%. Combined with intelligent photonic engineering, it produces a regular flow of single photons \cite{chu2017single}, achieving intensity squeezing up to 2.2 dB at room temperature. Collection efficiency is typically the limiting factor, with simple photonic structures improving efficiency from a small percentage to 20\% in a single spatial mode in cryogenic experiments \cite{colautti20203d}. Near-unit collection efficiency has been demonstrated at room temperature with a planar antenna and high-numerical-aperture optics, producing nearly 50 million photons per second. Molecular sources do not require optical cavities for high photon flux and indistinguishability, though strong coupling with a single emitter has been shown with open Fabry-Perot cavities \cite{wang2019turning}.
Molecular systems excel in coherence time, intensity, frequency stability, and the ability to interfere with photons from distinct molecules \cite{duquennoy2022real}.

Quantum-dot sources, are notable for generating single photons on demand. These exciton systems, with highly confined electron-hole pairs, behave like artificial atoms. The best single-photon sources based on quantum dots use III-V semiconductors grown epitaxially. Growth techniques leverage strain at interfaces between layers with different bandgaps, like gallium arsenide and indium arsenide. III-V QDs have large optical dipoles, facilitating broad coupling with confined or guided optical modes, essential for bright sources. While tunability is possible during growth, quantum dots are typically selected after optical characterization due to control challenges. Significant advancements in fabrication and excitation techniques (cross-polarization, phonon-assisted excitation, SUPER scheme, rapid adiabatic passage), coupling to photonic structures (bullseye cavity, open cavity, waveguide, micropillar), and local charge control in p-n junctions \cite{ZHL:21:AQT} position quantum dots as promising single-photon sources. They achieve an overall brightness (at the detector) of 53\% at a 76.3 MHz pump rate and raw indistinguishability of about 91\% \cite{thomas2021bright}. Other studies have demonstrated the use of demultiplexed quantum dot sources for photonic-quantum computing \cite{loredo2017boson}\cite{somaschi2016near}. For instance, Boson Sampling with up to 14 detected photons in an interferometer with 60 modes (see Table.~\ref{sources} for a summary of single-photon sources \acp{KPI} and their values).

\subsubsection{CV Sources}

Continuous variable quantum states are typically generated using various optical components and nonlinear interactions. Some common sources include:
\begin{enumerate}
\item \textbf{Laser Sources}: Coherent states are produced by highly stable laser sources, where the amplitude and phase can be precisely controlled.

\item \textbf{Squeezed states}:
Squeezed states of light are crucial for continuous variable quantum information processing, offering reduced quantum noise in one quadrature while increasing noise in the conjugate quadrature. This property is essential for quantum-enhanced metrology, quantum communication, and quantum computation. Below is an overview of the state-of-the-art techniques for generating squeezed states, focusing on advancements in optical parametric oscillators (OPOs), waveguide sources, and microresonators.
\item \textbf{Optical Parametric Oscillators:} 
Optical parametric oscillators remain the most widely used and versatile source for generating squeezed states. In an OPO, a nonlinear crystal (such as periodically poled lithium niobate, PPLN) is placed inside an optical cavity. When pumped with a laser at frequency \(\omega_p\), the crystal mediates the down-conversion process, generating signal and idler photons at frequencies \(\omega_s\) and \(\omega_i\), respectively, satisfying \(\omega_p = \omega_s + \omega_i\). The Hamiltonian governing this interaction is:
\begin{equation}
\hat{H}_{\text{OPO}} = i\hbar \kappa (\hat{a}_s^{\dagger} \hat{a}_i^{\dagger} e^{-i\Delta t} - \hat{a}_s \hat{a}_i e^{i\Delta t}),
\end{equation}
where \(\kappa\) is the nonlinear coupling coefficient, and \(\Delta = \omega_p - \omega_s - \omega_i\) is the detuning.
High level of squeezing exceeding $15db$  have been achieved using PPLN crystals and loss-cavities \cite{vahlbruch2016detection}.  Moreover, the Development of chip-scale OPOs integrating PPLN waveguides with microresonators, has enabled compact and stable squeezed light sources \cite{gehring2017towards}. Other experiments have been carried in this direction, staying within the state-of-the-art levels of squeezing \cite{VQL:19:PRApp,ZMT:21:NC}.
\item \textbf{Cubic non-linearities:} Cubic nonlinearity offers an alternative to second-order nonlinear processes for generating squeezed light. Typically, third-order nonlinear processes are much weaker than second-order ones; however, they are present in amorphous materials, broadening the selection of nonlinear media. In this context, optical fibers made from silica or specially designed highly-nonlinear glasses are particularly promising. The low intensity of interaction can be offset by longer interaction lengths, as advanced fiber production technology ensures low loss over large distances. The first observation of squeezing in an optical fiber occurred in 1986 using continuous radiation and a specialized detection scheme. This scheme involved reflecting the output beam from an optical resonator to achieve significant dispersion for necessary phase corrections of the sideband frequencies, enabling the measurement of squeezing. Despite considerable efforts to minimize noise, the observed squeezing was minimal, measuring only $0.5$ dB \cite{shelby1986broad}. Many experiments, afterwards, have been carried in this regard using different setups like optical solitons \cite{rosenbluh1991squeezed,drummond1993quantum} or sagnac fiber interferometers \cite{fujiwara2009generation}. 
\end{enumerate}

\subsection{Optical Channels}

\subsubsection{DV Channel Models}

Quantum channel models in discrete variable systems are fundamental for understanding the dynamics and performance of quantum information processes. These models mathematically describe the various mechanisms through which quantum states undergo perturbations due to interactions with their surroundings. The theoretical basis of quantum channels originates from the unitary interactions between a quantum system and its environment.

Consider a quantum system \( S \) and its environment \( E \), initially separate, with the composite system described by the Hilbert space \( \mathcal{H}_S \otimes \mathcal{H}_E \). The evolution of the combined system is governed by a unitary operator \( U \) acting on this composite Hilbert space. The density operator \( \rho \) of the system evolves as:
\begin{equation}
\rho \rightarrow \rho' = \text{Tr}_E \left( U (\rho \otimes \rho_E) U^\dagger \right),
\end{equation}
where \( \text{Tr}_E \) denotes the partial trace over the environment's degrees of freedom and \( \rho_E \) is the initial state of the environment.

This partial tracing process represents an averaging over the environment's behavior, leading to reduced dynamics described by the quantum channel \( \mathcal{E} \). The action of the quantum channel on the system's density matrix \( \rho \) is typically expressed in the Kraus representation:
\begin{equation}
\mathcal{E}(\rho) = \sum_i K_i \rho K_i^\dagger,
\end{equation}
where \( K_i \) are the Kraus operators satisfying the completeness relation \( \sum_i K_i^\dagger K_i = I \).

Different types of unitary interactions between the system and the environment yield different types of quantum channels:
\begin{enumerate}
\item \textbf{Phase Damping Channel}: This channel arises from the interaction of a two-level system (qubit) with a reservoir of quantum harmonic oscillators through a linear coupling where the spin of the two-level system is a constant of motion. The Kraus operators for this channel are:
   \begin{align}
   &K_0 = \begin{pmatrix} \sqrt{1 - \lambda}  & 0 \\ 0 & \sqrt{1 - \lambda} \end{pmatrix}, \quad K_1 = \begin{pmatrix} 0 & 0 \\ 0 & \sqrt{\lambda} \end{pmatrix}, \nonumber\\
   & K_2 = \begin{pmatrix} \sqrt{\lambda} & 0 \\ 0 & 0 \end{pmatrix},
   \end{align}
   where \( \lambda \) is the phase damping parameter, which has a time dependence as $1-\lambda=e^{-\Gamma(t)}$, where $\Gamma(t)$ can be different when different approximations of the interaction of the system with the environment are considered. For instance, in the markov approximation, $\Gamma(t)=C\cdot t $ with $C$ a constant depending on the spectral properties of the environment.

\item \textbf{Amplitude Damping Channel}: This channel results from a linear interaction between a two-level system and a thermal bath, where the spin of the system is not a constant of motion. The Kraus operators for this channel are:
   \begin{equation}
   K_0 = \begin{pmatrix} 1 & 0 \\ 0 & \sqrt{1 - \gamma} \end{pmatrix}, \quad K_1 = \begin{pmatrix} 0 & \sqrt{\gamma} \\ 0 & 0 \end{pmatrix},
   \end{equation}
   where $\gamma$ is the amplitude damping parameter. This channel can be encountered in superconducting platform for quantum computing. This channel has a similar time dependence to the dephasing channel in the Markovian approximation given explicitly as $1-p=e^{-C\cdot t}$ with $C$ a constant depending on the spectral properties of the environment. 

Quantum channels such as the depolarizing channel, bit-flip channel, and phase-flip channel are analogs of classical noise models like the white noise and binary symmetric channel.

\item \textbf{Depolarizing Channel}: This channel represents a situation where the qubit is replaced by a maximally mixed state with probability \( p \). The action of the depolarizing channel is:
  \begin{equation}
  \mathcal{E}(\rho) = (1 - p)\rho + \frac{p}{3} (X \rho X + Y \rho Y + Z \rho Z),
  \end{equation}
  where \( X, Y, Z \) are the Pauli matrices. The depolarization channel can perfectly model free space communications where the information is encoded in the polarization degree of freedom of the beam. 

\item \textbf{Bit-Flip Channel}: This channel flips the state of a qubit with probability \( p \). Its action is:
  \begin{equation}
  \mathcal{E}(\rho) = (1 - p)\rho + p X \rho X.
  \end{equation}

\item \textbf{Phase-Flip Channel}: This channel flips the phase of a qubit with probability \( p \). Its action is:
  \begin{equation}
  \mathcal{E}(\rho) = (1 - p)\rho + p Z \rho Z.
  \end{equation}
\end{enumerate}
These channels play a critical role in the study of \ac{QEC} and the robustness of quantum information protocols under various noise conditions.

\subsubsection{CV Channel Models}
\label{subsec:CVCM}

Bosonic channel models cater for optical quantum communication both in an optical fibre or in a free-space link \cite{MSG:24:arXiv, Guh:04:thesis_MIT}. The standard model for bosonic quantum communication caters for two noisy processes: photon loss and dephasing. A single-mode pure-loss bosonic channel is modeled by a beamsplitter that has the transmittivity $\eta$. The signal from transmitter is injected into one of the two input ports of the beamsplitter where the other input is injected with a fixed environment state, typically a vacuum state for pure-loss channels or thermal state to model noise \cite{LPG:20:PRL, Pir:21:PRR}. Similarly, one of the two output ports is fed to the receiver whereas the other output is lost to the environment that models the lossy process. The transmittivity $\eta$ of the beamsplitter under different transmission conditions can be obtained by taking the physical channel conditions into account, e.g., haze, rain, and turbulence \cite{VSV:17:PRA, Pir:21:PRR}. After characterizing $\eta$ by taking all relevant factors into account, the input-output relation of quantum states traversing through this channel can be written as \cite{VSV:17:PRA}
\begin{align}
    P_{\mathrm{out}}\left( \alpha\right) = \int_{0}^{1} \frac{1}{\eta}\mathcal{P}\left( \eta\right)P_{\mathrm{in}}\left( \frac{\alpha}{\sqrt{\eta}} \right)\, \mathrm{d}\eta
\end{align}
where $P_{\mathrm{out}}$ and $P_{\mathrm{in}}$ are quasiprobability distribution that characterize the output and input states, respectively, and $\mathcal{P}\left( \eta\right)$ is the probability distribution of transmittance.

The decoherence of quantum state during transmission is modeled by dephasing that provides the mapping (written in Fock basis)
\begin{align}
    \rho = \sum_{n, m = 0}^{\infty} \rho_{n, m}\ket{n}\bra{m} \rightarrow \sum_{n, m = 0}^{\infty} \rho_{n, m} e^{-\frac{\gamma}{2}\left( n - m\right)^2}\ket{n}\bra{m},
\end{align}
where $\gamma \in \left[ 0, 1\right]$ models the dephasing strength \cite{MSG:24:arXiv}. The dephasing process can take into account several forms of decoherence occurred in the medium, e.g., temperature fluctuation, Kerr nonlinearities in the fibre, imprecision in the path length, or the lack of shared phase reference between the communicating parties \cite{LW:23:NP}. A general bosonic channel model that incorporates both loss and dephasing can be obtained by the order-independent composition of the two processes \cite{MSG:24:arXiv}.

These types of optical quantum channel models are studied in detail in the literature. For example, exact quantum and private capacities of bosonic dephasing channels have been obtained recently \cite{LW:23:NP}. Anther recent work took diffraction, atmosheric extinction, turbulence, pointing errors, and background noise into accout to find the analyze the composable finite-size security of coherent state protocols in fibre optics and free-space fading quantum channels \cite{Pir:21:PRR}. A similar model for bosonic Gaussian channels was considered in \cite{GGC:14:NP} to obtain classical capacity of optical quantum channels.

A relevant type of bosonic channel is the thermal-loss channel.%
This class of channels is particularly important in optical and microwave communication scenarios, where laser noise and background radiation \cite{IQCTMN:2017:x40,ODQSTE:2018x41} have a significant impact, respectively. However, thermal-loss channels are more challenging to model than dephasing channels. They are neither degradable nor anti-degradable \cite{OMBGC:2006x42,OMQGC:2007x43}, and only lower and upper bounds for the quantum and private capacities are known \cite{NPJ:20:NC}.

\subsection{Detection}

\subsubsection{Detection of DV Quantum Signals}

\begin{table}[t]
\centering
\caption{Key Performance Indicators (KPIs) for Different Single-Photon Detectors}
\label{tab:detectors}
%\begin{tabular}{@{}ll@{}}
\begin{tabular}{p{2cm} p{6cm}}
\toprule
\textbf{KPI} & \textbf{Value and Source} \\ \midrule
System efficiency & 50-60\% at 800 nm (Si-based SPAD) \cite{gulinatti2021custom} \\
                  & 25-35\% (InGaAs/InP SPAD) \cite{zhang2015advances} \\
                  & >90\% (SNSPD) \cite{verma2015high} \cite{zhang2017nbn} \cite{zhang2019saturating} \\

Operating bandwidth & Visible to near-infrared (SPAD) \cite{ceccarelli2021recent} \\
                    & Telecom wavelengths (SNSPD) \cite{verma2015high} \cite{zhang2017nbn} \cite{zhang2019saturating} \\

Timing jitter & 50-180 ps (Si-based SPAD) \cite{gulinatti2021custom} \\
              & 70-150 ps (SNSPD) \cite{verma2015high} \cite{zhang2017nbn} \cite{zhang2019saturating} \\
              & As low as 40 ps (Commercial SNSPD) \\

Maximum count rate & Up to 37 MHz (Si-based SPAD) \cite{gulinatti2021custom} \\
                   & Exceeding 10 MHz (SNSPD) \cite{verma2015high} \cite{zhang2017nbn} \cite{zhang2019saturating} \\

Dark count rate & Up to 25 Hz (Si-based SPAD) \cite{gulinatti2021custom} \\
                & Approximately 1-2 kHz (InGaAs/InP SPAD) \cite{zhang2015advances} \\
                & Up to 10 Hz (SNSPD) \cite{verma2015high} \cite{zhang2017nbn} \cite{zhang2019saturating} \\ \bottomrule
\end{tabular}
\end{table}
Single photon detectors are classified into two primary categories based on their response types: Threshold Detectors and Photon-Number Resolving Detectors. Threshold detectors function by establishing a threshold level for photon detection. When the incoming signal surpasses this threshold, the detector records an event, signifying the presence of at least one photon. However, these detectors cannot distinguish between one photon and multiple photons; they only provide a binary output indicating detection above the threshold. They are typically used in applications where the precise number of photons is not critical, such as in quantum key distribution and certain forms of quantum computation like boson sampling. In contrast, Photon-number resolving detectors measure the energy or charge associated with each detected photon. By analyzing this information, they can determine the number of photons in the incoming signal.

The key performance indicators (KPIs) for assessing a detector's performance include system efficiency (the overall probability that an incident photon is detected by the detector, accounting for quantum efficiency and losses such as single-mode fiber coupling), operating bandwidth (the range of incident light wavelengths within which the detector maintains high system efficiency), timing jitter (the temporal uncertainty in the detector's response, where lower time jitter corresponds to higher temporal resolution, allowing more accurate identification of photon arrival times), maximum count rate (the highest number of counts per second the detector can register, depending on the reset time after each detection event), and dark count rate (the number of counts per second due to thermal effects, not from actual photon incidence). A comparison of different detectors against these \acp{KPI} is provided in Table~\ref{tab:detectors}
.

\paragraph{Single-Photon Avalanche Diodes (SPAD)}
Single-photon avalanche diodes (SPADs) are a type of photon detector based on semiconductor technology \cite{ceccarelli2021recent}. They consist of a p-n junction to which an appropriate voltage (reverse-bias) is applied to remove charge carriers from the depletion region . In this way, the incidence and absorption of a single photon are sufficient to initiate an avalanche ionization process, resulting in a macroscopic current. A greater number of photons does not lead to a significant difference in the generated current, and therefore, the detector's response does not provide information about the number of incident photons. The detector remains "blind" until it is reset to return to its initial condition. Among the main advantages of detectors based on this technology is their operation at room temperature. This characteristic makes them more suitable for portable applications where cryogenic equipment, such as those based on superconducting materials, is not feasible, in addition to lower manufacturing costs.

The choice of material for constructing the detectors depends on the required wavelength range. According to current technology, detectors with better specifications (quantum efficiency, dark count rate) can be obtained in the visible wavelength range, with performance tending to be lower as they move into the near-infrared region, especially in the O-band (1260-1360 nm) and C-band (1530-1565 nm) of the telecom wavelength. Among the best results reported in the literature, Si-based detectors \cite{gulinatti2021custom} can achieve quantum efficiencies of approximately 50-60\% at a wavelength of 800 nm, with dark count rates up to 25 Hz, jitter of 50-180 ps, and maximum counting rates up to 37 MHz. In the telecom band, the choice of material mainly falls on InGaAs and InGaAs/InP detectors \cite{zhang2015advances}, with quantum efficiencies of about 25-35\%, but associated with higher dark count rates (approximately 1-2 kHz). SPAD detectors are also commercially available from various companies.

\paragraph{Superconducting Detectors}
The second class of single-photon detectors is based on superconduncting technology. Within this category, several types based on different architectures have been developed, including transition edge sensors (TES) \cite{ullom2015review}, superconducting tunnel junctions (STJ) \cite{kurakado1982possibility} and microwave kinetic inductance detectors (MKID) \cite{mazin2005microwave}. Of particular interest both for quantum computing and quantum communications due to recent technological advancements and their characteristics are superconducting nanowire single-photon detectors (SNSPD) also known as nanowire detectors \cite{natarajan2012superconducting}. 

These detectors utilize nanowires, approximately 100 nm in diameter, fabricated from ultra-thin film with thickness about 5-10nm, made of superconducting material. The detection mechanism is based on the creation of cooper pairs, followed by a transition of the material from a superconducting state to a non-superconducting state, generating a detectable macroscopic current \cite{you2020superconducting}. Subsequently a reset time is required to return the detector to its initial superconducting state and become active again. SNSPDs have been reported in the literature with high detection efficiency, low dark count levels, and high count rates. For instance, in the telecom wavelength range in the C band, several works have been reported in the literature \cite{verma2015high,zhang2017nbn,zhang2019saturating} on various materials such as NbN, WSi, and MoSI, with performance parameters including efficiency >90\%, dark count rates up to 10 Hz, timing jitter of 70-150 ps, and maximum count rates exceeding 10 MHz. Similarly, many companies have introduced detectors of this type to the market, with typical parameters ranging from system efficiency of 80-90\%, timing jitter as low as 40 ps, and operating wavelengths ranging from telecom wavelengths to visible/near-infrared. Typically, a compromise needs to be made between the wavelength band of operation and detector efficiency. Unlike SPAD detectors, superconducting detectors require the use of a cryogenic apparatus at low temperatures. Therefore, this type of technology currently has a higher cost impact and lower portability.

\subsubsection{Detection of CV Quantum Signals}

Continuous Variable (CV) quantum detection is a crucial aspect of quantum information processing, particularly in quantum communication and computation. It involves measuring quantum states that are described by continuous variables, such as the quadratures of the electromagnetic field. The primary detection techniques in CV quantum systems are homodyne and heterodyne detection for Gaussian states, and PNRD for non-Gaussian states.
\paragraph{Homodyne Detection:}
Homodyne detection is used to measure the quadrature components of a light field, typically for Gaussian states. It involves mixing the signal field with a strong local oscillator (LO) at a beam splitter and measuring the resulting interference. 
The output photocurrent $I(t)$ of a homodyne detection is proportional to the quadrature component:
\begin{equation}
    I(t)=\sqrt{\eta}\langle\hat{x}_\theta\rangle+\xi(t)
\end{equation}
where $\eta$ is the detection efficiency, $\theta$ is the phase of the local oscilator  and $\xi(t)$ and $\langle\hat{x}_\theta\rangle$ are the noise term and the expected value of the quadrature respectively. 
\paragraph{Heterodyne Detection:}
Heterodyne detection measures both quadratures simultaneously by mixing the signal with a local oscillator at a frequency offset. The signal is down-converted to an intermediate frequency, which can be processed to extract both quadratures. In heterodyne detection, the measured field $\hat{\beta}$ is a complex combination of the quadratures:
\begin{equation}
    \hat{\beta}=\frac{1}{\sqrt{2}}\big(\hat{X}+i\hat{P}\big)
\end{equation}
The heterodyne photocurrent yields information about both quadratures:
\begin{equation}
    I(t)=\sqrt{\eta}\big(\langle\hat{x}_\theta\rangle+i\langle\hat{p}_\theta\rangle\big)+\xi(t)
\end{equation}
with $\langle\hat{x}_\theta\rangle$ and $\langle\hat{p}_\theta\rangle$ are the expectation values of the quadratures where $\theta$ is the phase of the local oscillator and $\eta$ and $\xi(t)$ are respectively the detection efficiency and the noise term. 

\subsection{Quantum Modulation Techniques}
In qauntum communications, modulation techniques are essential for encoding information into quantum states. Here we discuss several quantum modulation techniques, that are summarized in Table.~\ref{classicalvsquantum}, and are relevant for different discussions throught the manuscript.

\subsubsection{Frequency encoding}

 Frequency encoding uses different frequencies of photons to represent qubits. This method is beneficial for multiplexing and integrating with classical communication channels. Different frequency modes $\ket{f_1}$ and $\ket{f_2}$ represent logical states $\ket{0}$ and $\ket{1}$. 

 \subsubsection{Time bin}
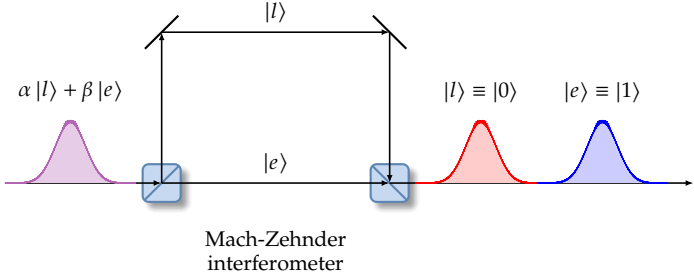
\begin{figure}
    \centering
\definecolor{ocre}{HTML}{800000}
\definecolor{green}{RGB}{0, 128, 0}
\definecolor{sky}{HTML}{C6D9F1}
\definecolor{skybox}{HTML}{5F86B3}

% Tikz Library
%\usetikzlibrary{3d, shapes.multipart}

% Styles
\tikzset{>=latex} % for LaTeX arrow head
\tikzset{axis/.style={black, thick,->}}

% Notation
%\usepackage{amsmath}

\newcommand{\BS}[4]{
	\draw[stan] (#2-\l/2,#3-\l/2) rectangle (#2+\l/2,#3+\l/2);
	\draw[thick,skybox] (#2-\l/2+0.03,#3-\l/2+0.03) --++ (\l-0.06, \l-0.06);
	%	\draw[thick,skybox] (#2-\s+0.03,#3+\s-0.03) --++ (\l-0.06,-\l+0.06);
	%	\node[]  [below=\l*.6 cm of #1]  {#4};		
}

\newcommand{\BSpd}[4]{
	\def\S{\l/2}
	\draw[stan] (#2-\l/2,#3-\l/2) rectangle (#2+\l/2,#3+\l/2);
	%	\draw[thick,skybox] (#2-\l/2+0.03,#3-\l/2+0.03) --++ (\l-0.06, \l-0.06);
	\draw[thick,skybox] (#2-\S+0.03,#3+\S-0.03) --++ (\l-0.06,-\l+0.06);
}

\newcommand{\mirror}[4]{
	\draw[thick] (#2-\l/2+0.03,#3-\l/2+0.03) --++ (\l-0.06, \l-0.06);
}

\newcommand{\mirrorpd}[4]{
	\def\S{\l/2}
	\draw[thick] (#2-\S+0.03,#3+\S-0.03) --++ (\l-0.06,-\l+0.06);
}

	\begin{tikzpicture}[
		% Environment Cfg
		font=\footnotesize, 
		text centered,
		rounded corners=0.08cm,
		%	text width=2.5cm, 
		% Styles
		stan/.style ={
			draw=skybox,
			text width=2cm, 
			text=black, minimum height=4cm,
			thick,
			fill=sky,
			blur shadow, 
		},
		arrow/.style ={
			-{Latex[length=3pt]},
			semithick
		},
		Neutral/.style ={
			draw=black!50,
			text width=2cm, minimum height=2cm,
			thick,
			%		fill=black!10,
			%		blur shadow, 
		}
		]
		\def\x{3}
		\def\s{\x}
		\def\y{2}
		\def\l{.5}
		%for labels like "Alice's LAB"
		%Label/.style 

		\coordinate (A) at (0,0);
		\BS{A}{0}{0}{}
		
		\coordinate[right=\x  cm  of A] (B);
		\BSpd{B}{\x}{0}{Mirror}
		
		\coordinate[above= \y cm of A] (C);
		\mirror{C}{0}{\y}{}
		
		\coordinate[right=\x cm  of C] (D);
		\mirrorpd{D}{\x}{\y}{}

		\draw[arrow] (A) -- (B) node[midway,above] {$\ket{e}$} node[midway,below=.5cm,text width=2cm] {Mach-Zehnder interferometer};
		\draw[arrow] (C) -- (D) node[midway,above] {$\ket{l}$};
		\draw[arrow] (A) -- (C);
		\draw[arrow] (D) -- (B);
		
		%		\coordinate[left=\s cm of A] (A1);
		%		\draw[arrow] (A1) -- (A) node [below=\l*.6 cm of A1] {Source} node [midway,above] {$\frac{\ket{0}+\ket{1}}{\sqrt{2}}$};
		%		\draw[stan] (-\s-\l/2,-\l/2) rectangle (-\s+\l/2,\l/2);
		%		
		
		\coordinate[left=2 cm of A] (A1);
		\draw[arrow] (A1) -- (A);
		
		\coordinate[right=4 cm of B] (B1);
		\draw[arrow] (B) -- (B1);
		
		\def\mu{2};
		\def\sigma{.5}
		
		\colorlet{mixedcolor}{red!50!blue} % 50% color1 and 50% color2
		
		\draw[shift={(A1)}, scale=.4,samples=100, smooth, domain=\mu-2:\mu+2, variable=\x, mixedcolor!60, fill=mixedcolor!20] plot ({\x}, {2 * exp(-(\x - \mu)^2 / (2 * \sigma * \sigma))} );
		
		\def\mu{3};
		\draw[shift={(B)}, scale=.4,samples=100, smooth, domain=\mu-2:\mu+2, variable=\x, red, fill=red!20] plot ({\x}, {2 * exp(-(\x - \mu)^2 / (2 * \sigma * \sigma))} );
		
		\def\mu{7};
		\draw[shift={(B)}, scale=.4,samples=100, smooth, domain=\mu-2:\mu+2, variable=\x, blue, fill=blue!20] plot ({\x}, {2 * exp(-(\x - \mu)^2 / (2 * \sigma * \sigma))} );

        \node[right=2*.4 of A1, above] (sup) {};
		\node[above=.7cm of sup] {$\alpha\ket{l} + \beta\ket{e}$};
		
		\node[right=3*.4 of B, above] (l) {};
		\node[above=.7cm of l] {$\ket{l}\equiv\ket{0}$};
		
		\node[right=7*.4 of B, above] (e) {};
		\node[above=.7cm of e] {$\ket{e}\equiv\ket{1}$};

	\end{tikzpicture}
	
	

    \caption{Time bin encoding}
    \label{timebin}
\end{figure}

 Time-bin encoding uses the arrival time of photons to encode quantum information. This technique is robust against decoherence over long distances. A Mach-Zehnder interferometer with a delay line can be used to create time-bin encoded states where the early arrival time $\ket{e}$ and the late arrival time $\ket{l}$ represent logical states $\ket{0}$ and $\ket{1}$ as illustrated in Fig.~\ref{timebin}
 
\subsubsection{Polarization encoding}

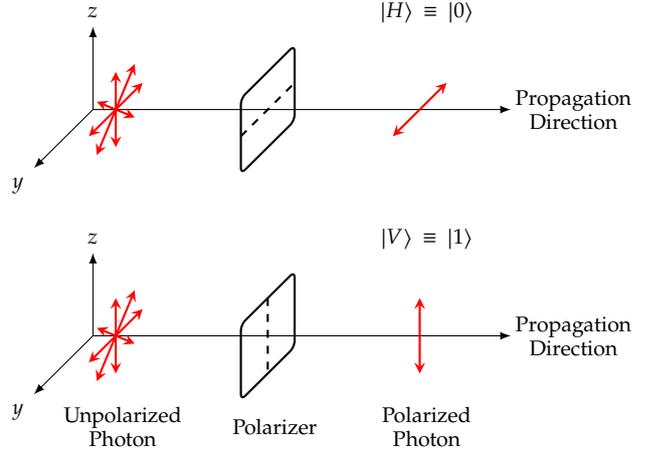
\begin{figure}
    \centering

% Tikz Library
\usetikzlibrary{3d, shapes.multipart}

% Styles
\tikzset{>=latex} % for LaTeX arrow head
\tikzset{axis/.style={black, thick,->}}

%% Draw in Polar Coordinates from (0,0) to (r,theta)
\newcommand{\cdraw}[2]{\draw[thick, -stealth, red] (0,0) -- ({#1*cos(#2)}, {#1*sin(#2)});}
\newcommand{\cdrawZ}[2]{\draw[thick, -stealth, red] (0,\Z) -- ({#1*cos(#2)}, {#1*sin(#2)+\Z});}

	\begin{tikzpicture}[
		x={(1cm,0cm)},y={(-0.707cm,-0.707cm)}, z={(0cm,1cm)}, 
		%		line cap=round, line join=round,
		% Environment Cfg
		font=\footnotesize, 
		text centered,
		rounded corners=0.08cm,
		%	text width=2.5cm, 
		% Styles
		stan/.style ={
			text width=2cm, 
			text=black, minimum height=4cm,
		},
		arrow/.style ={
			-{Latex[length=3pt]},
			semithick
		}
		]
		
		\def\d{.5}
		\def\s{.3}
		\def\Z{0}

		%		 Main Axes
		\draw[->] (0,0,\Z) -- (5.5,0,\Z) node[right,stan,xshift=-.3cm] {Propagation Direction};
		\draw[->] (0,0,\Z) -- (0,1.1,\Z) node[below left] {$y$};
		\draw[->] (0,0,\Z) -- (0,0,1.1+\Z) node[above] {$z$};

		\begin{scope}[canvas is yz plane at x=\s]
			\cdrawZ{\d}{0}
			\cdrawZ{\d}{180}
			\cdrawZ{\d}{90}
			\cdrawZ{\d}{270}
			\cdrawZ{\d}{45}
			\cdrawZ{\d}{45+180}
			
			\cdrawZ{\d}{-45}
			\cdrawZ{\d}{-45-180}
			
		\end{scope}

		\begin{scope}[canvas is yz plane at x=\s+2]
			\draw[thick] (\d,-\d+\Z) rectangle (-\d,\d+\Z);
			\draw[thick, dashed] (\d,\Z) -- (-\d,\Z);
		\end{scope}
		
		\begin{scope}[canvas is yz plane at x=\s+4]
			\cdrawZ{\d}{0}
			\cdrawZ{\d}{180}
		\end{scope}

		\begin{scope}[canvas is xy plane at z=2+\Z]
			\node[stan] at (\s+4+.8,1) {$\ket{H}\equiv \ket{0}$};
		\end{scope}

		\def\Z{-3}

		%		 Main Axes
		\draw[->] (0,0,\Z) -- (5.5,0,\Z) node[right,stan,xshift=-.3cm] {Propagation Direction};
		\draw[->] (0,0,\Z) -- (0,1.1,\Z) node[below left] {$y$};
		\draw[->] (0,0,\Z) -- (0,0,1.1+\Z) node[above] {$z$};

		\begin{scope}[canvas is yz plane at x=\s]
			\cdrawZ{\d}{0}
			\cdrawZ{\d}{180}
			\cdrawZ{\d}{90}
			\cdrawZ{\d}{270}
			\cdrawZ{\d}{45}
			\cdrawZ{\d}{45+180}
			
			\cdrawZ{\d}{-45}
			\cdrawZ{\d}{-45-180}
			
		\end{scope}

		\begin{scope}[canvas is yz plane at x=\s+2]
			\draw[thick] (\d,-\d+\Z) rectangle (-\d,\d+\Z);
			\draw[thick, dashed] (0,\Z+\d) -- (0,\Z-\d);
		\end{scope}
		
		\begin{scope}[canvas is yz plane at x=\s+4]
			\cdrawZ{\d}{90}
			\cdrawZ{\d}{90+180}
		\end{scope}

		\begin{scope}[canvas is xy plane at z=2+\Z]
			\node[stan] at (\s+4+.8,1) {$\ket{V}\equiv \ket{1}$};
		\end{scope}

				%LABELS
				\begin{scope}[canvas is xy plane at z=-.5+\Z]
						\node[stan] at (\s+.8,1) {Unpolarized Photon};
						\node[stan] at (\s+2+.8,1) {Polarizer};
						\node[stan] at (\s+4+.8,1) {Polarized Photon}; %Horizontally 
					\end{scope}

	\end{tikzpicture}
	
\caption{An illustration of polarization encoding}
    \label{polarization}
\end{figure}

Polarization encoding uses the polarization states of photons to represent qubits. This is one of the most common methods due to the ease of manipulating polarization states. Fig.~\ref{polarization} illustrates this type of encoding where for instance, the vertical $\ket{V}$ and the horizontal $\ket{H}$ polarizations can represent respectively the states $\ket{1}$ and $\ket{0}$. Other polarization states like the left and right circular polarizations might be used as well instead oif the horizontal and vertical polarizations.

\subsubsection{Orbital angular momentum}

OAM encoding uses the helical phase front of photons to encode information. Photons with OAM carry angular momentum along their propagation direction. Photons can have OAM values $\ket{l}$ where $l$ is an integer respresenting the number of twists in the phase front per wavelength. Spatial light modulators (SLMs) or q-plates can generate OAM 
 states. These states might be detected using mode sorters or forked gratings can detect OAM states. 

 \subsubsection{Spatial encoding of structured light}

 Spatial encoding uses the spatial mode structure of light beams to encode information. Structured light beams such as Hermite-Gaussian (HG) and Laguerre-Gaussian (LG) modes are used. These modes can carry information based on their spatial structure.  Spatial light modulators can be used to generate and manipulate these modes. This encoding is useful in high-dimensional quantum information processing.

\subsubsection{Path encoding}

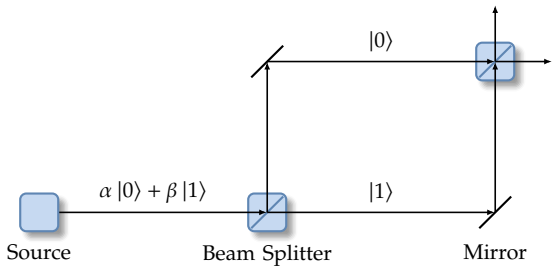
\begin{figure}
\begin{center}  
    \newcommand{\BS}[4]{
	\draw[stan] (#2-\l/2,#3-\l/2) rectangle (#2+\l/2,#3+\l/2);
	\draw[thick,skybox] (#2-\l/2+0.04,#3-\l/2+0.04) -- (#2+\l/2-0.04,#3+\l/2-0.04);
	\node[]  [below=\l*.6 cm of #1]  {#4};		
}

\newcommand{\BSa}[4]{
	\draw[stan] (#2-\l/2,#3-\l/2) rectangle (#2+\l/2,#3+\l/2);
	\draw[thick,skybox] (#2-\l/2+0.04,#3-\l/2+0.04) -- (#2+\l/2-0.04,#3+\l/2-0.04);
	\node[]  [above=\l*.6 cm of #1]  {#4};		
}

\newcommand{\mirror}[4]{
	\draw[thick,decoration={
		markings,
		mark=between positions 0.015 and 0.98 step 0.1072 with {\draw (0,0)--(60:3pt);}
	}] (#2-\l/2+0.04,#3-\l/2+0.04) -- (#2+\l/2-0.04,#3+\l/2-0.04);
	\node[]  [above=\l*.6 cm of #1]  {#4};		
}

\newcommand{\mirrorR}[4]{
	\draw[thick,decoration={
		markings,
		mark=between positions 0.015 and 0.98 step 0.1072 with {\draw (0,0)--(60+180:3pt);}
	}] (#2-\l/2+0.04,#3-\l/2+0.04) -- (#2+\l/2-0.04,#3+\l/2-0.04);
	\node[]  [below=\l*.6 cm of #1]  {#4};		
}

\begin{tikzpicture}[
	% Environment Cfg
	font=\footnotesize, 
	text centered,
	rounded corners=0.08cm,
	%	text width=2.5cm, 
	% Styles
	conn/.style ={
		draw=black,
		text width=1cm, 
		text=white, minimum height=4cm,
		thick,
		fill=ocre!90,
		blur shadow, 
	},
	stan/.style ={
		draw=skybox,
		text width=2cm, 
		text=black, minimum height=4cm,
		thick,
		fill=sky,
		blur shadow, 
	},
	arrow/.style ={
		-{Latex[length=3pt]},
		semithick
	},
	Neutral/.style ={
		draw=black!50,
				text width=2cm, minimum height=2cm,
		thick,
%		fill=black!10,
%		blur shadow, 
	}
	]
	\def\x{3}
	\def\s{\x}
	\def\y{2}
	\def\l{.5}
	%for labels like "Alice's LAB"
	%Label/.style 

	\coordinate (A) at (0,0);
	\BS{A}{0}{0}{Beam Splitter}
	
	\coordinate[right=\x cm  of A] (B);
	\mirrorR{B}{\x}{0}{Mirror}

	\coordinate[above= \y cm of A] (C);
	\mirror{C}{0}{\y}{}
	
	\coordinate[right=\x cm  of C] (D);
	\BSa{D}{\x}{\y}{}

	\draw[arrow] (A) -- (B) node[midway,above] {$\ket{1}$};
	\draw[arrow] (C) -- (D) node[midway,above] {$\ket{0}$};

	\draw[arrow] (A) -- (C);
	\draw[arrow] (B) -- (D);
	
	\coordinate[left=\s cm of A] (A1);
	\draw[arrow] (A1) -- (A) node [below=\l*.6 cm of A1] {Source} node [midway,above] {$\alpha\ket{0} + \beta\ket{1}$};
	\draw[stan] (-\s-\l/2,-\l/2) rectangle (-\s+\l/2,\l/2);

	\coordinate[right=\x/4 of D] (D1);
	\draw[arrow] (D) -- (D1) node [below=\l*.6 cm of D1] {};
	
	\coordinate[above=\x/4  of D] (D2);
	\draw[arrow] (D) -- (D2) node [below=\l*.6 cm of D1] {};

\end{tikzpicture}

    \caption{A Mach-Zehnder interferometer illustrating path encoding where the photon can be in a superposition of taking two different paths, $\ket{0}$ and $\ket{1}$ can represent the photon taking the left or right path, respectively. 
    }
    \label{machzender}
\end{center}
\end{figure}

Path encoding uses different spatial paths to represent quantum states. In this technique, the state of a qubit is determined by the path it takes. A simple Mach-Zehnder interferometer can be used to illustrate path encoding as in Fig.~\ref{machzender}. We highlight that path encoding is different from spatial modulation where the information unit is encoded on the selected antenna/path once at a time. Instead, in path encoding, many antennas transmit the information unit simultaneously through different paths in a quantum way.

\subsection{Modulation techniques and crosstalk }
Crosstalk is a significant challenge across various media, from free space and fiver, to waveguides in integrated photonics as was mentioned Sec.~\ref{sec:FSOBP}. Indeed, cross-talk highly depends on modulation and encoding techniques used in optical communication. Different from the \ac{MIMO} configuration, inter-channel, intra-channel, and inter symbol interference(ISI) are taken into account. Its impact varies depending on the specific method used and mitigating crosstalk often requires careful design considerations such as mode purity, synchronization, and the choice of modulation format. Each technique has specific strategies to minimize crosstalk, ensuring reliable and high-fidelity transmission of information. When it comes to modulation techniques, each has its unique vulnerabilities to cross-talk, which can degrade the performance of an optical communication system. For carrier-free signals, such as quadrature amplitude modulation (QAM), the crosstalk power is nearly constant under expected conditions in multi-core transmission systems \cite{rademacher2017crosstalk}\footnote{The expected comditions refers to  the modulation formats, symbol rates, and the nature of the transmitted signals. For instance,  Different modulation schemes, such as quadrature amplitude modulation (QAM) and on-off keying (OOK), have distinct characteristics that affect how they interact with crosstalk. QAM, for example, is noted for its constant power over time, which can make its crosstalk behavior more predictable compared to OOK, which has inherent fluctuations due to its binary nature. Moreover,   the symbol rate, or the speed at which symbols are transmitted, also influences crosstalk dynamics. Higher symbol rates can lead to different temporal and spectral characteristics of the crosstalk compared to lower symbol rates. Additionally, the relative delay between signals in different cores, known as inter-core skew, impacts the crosstalk. Variations in this skew can cause fluctuations in the crosstalk power, affecting the overall signal quality.}. In contrast, carrier-supported signals, such as on-off keying (OOK), always induce time-varying crosstalk powers due to the strong carrier component, necessitating a performance margin for these systems \cite{rademacher2017crosstalk}. Polarization states can experience crosstalk when different polarization modes couple. This can occur in systems using polarization multiplexing where maintaining the orthogonality of the polarization states is critical to minimize crosstalk. Crosstalk between polarization states can degrade the performance of the system and reduce the fidelity of the transmitted information. Crosstalk in dual polarization scenarios is influenced by random polarization rotations between phase-matching points. The crosstalk variance is affected by the inter-core skew and the frequency characteristics of the signal \cite{rademacher2017crosstalk}. OAM modes are susceptible to crosstalk, especially when transmitted through fibers. In \cite{cozzolino2019orbital} it has been reported that the separation of modes (5, 6, 7) in specially designed air-core fibers helps reduce crosstalk. However, lower order modes (<4) do not have sufficient separation, leading to higher crosstalk. The mode purity and extinction ratio are key indicators of crosstalk levels, with higher extinction ratios indicating lower crosstalk. In time-bin encoding, crosstalk can occur due to overlapping time bins, especially in systems with high symbol rates or poor synchronization. Proper time-bin separation and precise timing are essential to minimize crosstalk and ensure clear distinction between time bins \cite{kim2022quantum}. Path encoding, used in quantum key distribution (QKD) and other optical communication systems, can experience crosstalk when there is unwanted coupling between different paths. This is particularly challenging in multi-core fibers where inter-core coupling can induce crosstalk. The level of crosstalk is influenced by the inter-core skew and the modulation format, with higher order modulation formats showing lower crosstalk variations \cite{da2021path}. Path encoding, particularly in quantum photonic interconnects, can suffer from cross talk. In these systems, path-entangled states are generated on one chip and distributed to another chip by interconverting between path and polarization degrees of freedom using a two-dimensional grating coupler. Furthermore, cross talk can introduce noise and affect the fidelity of entanglement distribution \cite{wang2016chip}. 
\section{Interacting Quantum Channels }\label{sec:CQC}
In classical communication theory, noise models play a pivotal role in understanding the degradation of transmitted signals over communication links. Common noise models include additive white Gaussian noise (AWGN) which represents random fluctuations added to the received signal during detection. Other models encompass impairments on the propagation channel, such as thermal noise through the receiving aperture, multipath fading, and co-channel interference. To mitigate the effects of noise in modern classical communications, various techniques are employed. These include error-correcting codes like convolutional codes and Reed-Solomon codes, which introduce redundancy to the transmitted signal data, enabling receivers to detect and correct errors \cite{massoud2007digital}. These are usually blind to the exact physical impairment that generated the errors and are optimized for specific impairments such as the AWGN or erasure channels. Additionally, diversity techniques like spatial diversity, frequency diversity and time diversity are implemented to combat fading effects and enhance signal reliability \cite{goldsmith2005wireless}. Diversity techniques are normally aware of the underlying statistics of the heterogeneous channel streams/slots in order to apply intelligent forms of combining. Advanced modulation schemes such as quadrature amplitude modulations (QAM) and phase-shift keying (PSK) are also utilized to improve spectral efficiency and robustness against specific kinds of impairments e.g. phase noise, and non-linear channels. Furthermore, adaptive modulation and coding (AMC) techniques dynamically adjust modulation and coding parameters based on channel conditions, ensuring optimal performance under varying SNR conditions. Through a combination of these techniques, modern classical communication systems can achieve reliable and efficient data transmission over real-world communication links \cite{opticalcoms}.

Despite the huge efforts that the community is putting into the migration towards quantum communications, quantum communication systems are still susceptible to various sources of impairments, which can compromise the fidelity and security of transmitted quantum states. As such, noise mitigation techniques are indispensable in quantum communications to ensure reliable and secure transmission. Similar to noise mitigation techniques in classical communications, quantum communications systems employ strategies such as \ac{QEC} and quantum repeaters to combat noise and enhance the resilience of transmitted quantum states.  \Ac{QEC} techniques enable the detection and correction of errors that arise due to noise and decoherence, thereby preserving the integrity of quantum information. Quantum repeaters extend the range of quantum communication networks by mitigating the effects of signal attenuation and noise accumulation over long distances, thus enabling efficient long-distance quantum communication. These noise mitigation techniques are crucial for realizing the full potential of quantum communications and advancing towards practical applications such as secure quantum networks and quantum internet. We refer the reader to \cite{lidar2013quantum,cai2023quantum} for more details on \ac{QEC} and error mitigation (see Table.~\ref{classicalvsquantum} for a comparison between classical vs quantum techniques for noise mitigation and correction). 

Other more forward-looking techniques consider the employment of the genuinely quantum resource of quantum coherence to manipulate the propagation of the information between the transmitter and the receiver. In particular, the quantum coherent control of the causal orders between given channels and the coherent control of the path in which the information propagates have been investigated. Despite the differences and the similarities between the two schemes, they have proven to be very important for mitigating noise, when some or full knowledge about the communication channel is known to the transmitter \footnote{The indefinite causal paradigm of quantum channel requires in some cases only partial knowledge of the channel to provide an advantage, like its nature, without knowing the full behaviour of the channel. This is the example of the entanglement breaking channel. In some other cases full knowledge of the channel is needed, like the noise parameters of the channel.}. Using these quantum coherent control techniques, better transmission rates beyond the ultimate bounds posed by quantum Shannon theory can be achieved. 

Despite the different trials to come up with an efficient scheme to mitigate the impairments on information transfer in the quantum realm, communication strategies akeen to diversity, are still to be deeply investigated  due to the different nature of the interaction of information carriers with their environment causing a different effective behavior of the transmission lines on a quantum level \cite{wang2024exploiting}. Another problem is related to the nature of information to be transmitted. If the information is classical and is encoded in quantum states, it can be easily prepared in many copies and sent through different transmission lines. Instead, when the information is purely quantum and is being an outcome of a previous quantum process that might be not known, the information cannot be copied due to the no-cloning  theorem.

\begin{table*}[t]
\centering
\caption{Classical and Quantum Correspondence in Communication Systems}
\label{classicalvsquantum}
\begin{tabular}{p{4cm} p{6cm} p{6cm}}
\toprule
\textbf{Concept} & \textbf{Classical Communications} & \textbf{Quantum Communications} \\ \midrule
Forward Error Correction (FEC) & Classical Error Correction Codes (e.g., Hamming, Reed-Solomon, LDPC) & \ac{QEC} Codes (e.g., Stabilizer codes Codes, Surface Codes, Holographic codes, QLDPC) \\

Modulation & Amplitude, Frequency, Phase Modulation, spatial, time & path, frequency, polariziation, OAM, time, structured light \\

Adaptive Modulation and Coding (AMC) & Adaptive schemes based on channel conditions & Quantum Communications with Feedback\\

Quantum Coherent Control & None & Quantum Coherent Control Techniques, indefinite causal ordering \\ 
\bottomrule
\end{tabular}
\end{table*}

\subsection{Quantum Coherent Control}
\begin{figure}
    
    \newcommand{\BS}[4]{
	\draw[stan] (#2-\l/2,#3-\l/2) rectangle (#2+\l/2,#3+\l/2);
	\draw[thick,skybox] (#2-\l/2+0.04,#3-\l/2+0.04) -- (#2+\l/2-0.04,#3+\l/2-0.04);
	\node[]  [below=\l*.6 cm of #1]  {#4};		
}

\newcommand{\BSa}[4]{
	\draw[stan] (#2-\l/2,#3-\l/2) rectangle (#2+\l/2,#3+\l/2);
	\draw[thick,skybox] (#2-\l/2+0.04,#3-\l/2+0.04) -- (#2+\l/2-0.04,#3+\l/2-0.04);
	\node[]  [above=\l*.6 cm of #1]  {#4};		
}

\newcommand{\mirror}[4]{
	\draw[thick,decoration={
		markings,
		mark=between positions 0.015 and 0.98 step 0.1072 with {\draw (0,0)--(60:3pt);}
	}] (#2-\l/2+0.04,#3-\l/2+0.04) -- (#2+\l/2-0.04,#3+\l/2-0.04);
	\node[]  [above=\l*.6 cm of #1]  {#4};		
}

\newcommand{\mirrorR}[4]{
	\draw[thick,decoration={
		markings,
		mark=between positions 0.015 and 0.98 step 0.1072 with {\draw (0,0)--(60+180:3pt);}
	}] (#2-\l/2+0.04,#3-\l/2+0.04) -- (#2+\l/2-0.04,#3+\l/2-0.04);
	\node[]  [below=\l*.6 cm of #1]  {#4};		
}

\begin{tikzpicture}[
	% Environment Cfg
	font=\footnotesize, 
	text centered,
	rounded corners=0.08cm,
	%	text width=2.5cm, 
	% Styles
	conn/.style ={
		draw=black,
		text width=1cm, 
		text=white, minimum height=4cm,
		thick,
		fill=ocre!90,
		blur shadow, 
	},
	stan/.style ={
		draw=skybox,
%		text width=2cm, 
%		text=black, minimum height=2cm,
		thick,
		fill=sky,
		blur shadow, 
	},
	arrow/.style ={
		-{Latex[length=3pt]},
		semithick
	},
	Neutral/.style ={
		draw=black!50,
		text width=2cm, minimum height=2cm,
		thick,
		%		fill=black!10,
		%		blur shadow, 
	}
	]
	\def\x{3}
	\def\s{3}
	\def\y{2}
	\def\l{.5}
	%for labels like "Alice's LAB"
	%Label/.style 

	\coordinate (A) at (0,0);
	\BS{A}{0}{0}{PBS}
	
	\coordinate[right=\x cm  of A] (B);
	\mirrorR{B}{\x}{0}{Mirror}
	
	\coordinate[above= \y cm of A] (C);
	\mirror{C}{0}{\y}{}
	
	\coordinate[right=\x cm  of C] (D);
	\BSa{D}{\x}{\y}{}

	\draw[arrow] (A) -- (B) node[midway, stan,
	text width=\l cm, 
	minimum height=\l cm] {$\epsilon_1$};
	\draw[arrow] (C) -- (D) node[midway, stan,
	text width=\l cm, 
	minimum height=\l cm] {$\epsilon_0$};

	\draw[arrow] (A) -- (C);
	\draw[arrow] (B) -- (D);
	
	\coordinate[left=\s of A] (A1); 
	\draw[arrow] (A1) -- (A) node [below=\l*.6 cm of A1] {Source} node [midway,above] {$\dyad{+}\otimes \rho^{\text{t}}_\text{in}$};
	\draw[stan] (-\s-\l/2,-\l/2) rectangle (-\s+\l/2,\l/2);

	\coordinate[right=\x/4 of D] (D1);
%	\draw[arrow] (D) -- (D1); %node [below=\l*.6 cm of D1] {};
%	

	\coordinate[above=\x/4 of D] (D2);
	\draw[arrow] (D) -- (D2) node [above] {$\rho^{\text{ct}}_\text{out}$};

\end{tikzpicture}
    \caption{An illustration picture of the quantum coherent control process.}
    \label{fig:coherent control}
    
\end{figure}
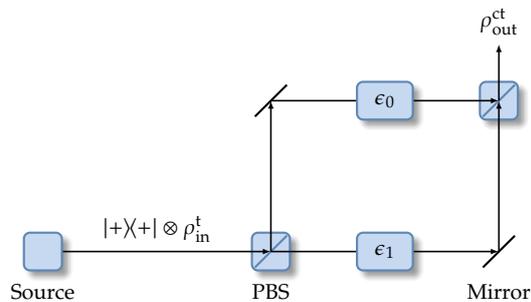

Quantum coherent control of quantum operations capitalizes on the distinctive attribute of quantum coherence to manipulate quantum operations or channels. Leveraging quantum coherence enables the simultaneous transmission of a single information carrier through multiple potential trajectories coherently, a capability absent in classical communications. This is inherently different from multipath propagation scenarios, where undesired side effects of information transfer resulting from reflection and diffraction, result in a constructive or destructive interference of the signal at a waveform level. The difference is mainly in the quantum nature of superposition between individual paths.
Employing such quantum coherent strategies for information transmission offers significant advantages in error and noise mitigation during transmission. For instance, pioneering research has showcased that coherent manipulation of the access point of the information carrier and its propagation path facilitates a scheme that exploits channel multiplexing in the path degree of freedom without signal multiplexing or redundancies, ultimately resulting in enhanced transmission rates \cite{Gisin_2005}. This innovation, termed "error filtration," has recently been extended to the domain of quantum coherent control of quantum channels \cite{Abbott_2020}. In this seminal study, a completely depolarizing channel, which inherently lacks the capacity to transmit any information due to its diminishing classical and quantum capacities, becomes viable for information transfer when a single qubit traverses two instances of the channel coherently. To illustrate, consider the scenario where the information carrier is a photon with its information encoded in its polarization degree of freedom. Utilizing a polarizing beam splitter (PBS), one can concurrently transmit the polarization qubit through both channels, with upward polarization traversing one completely depolarizing channel and downward polarization traversing the other, while maintaining coherence between the two polarization states. 
A more comprehensive depiction of coherent control of quantum channels using polarizing beam splitters (PBS) is elaborated in \cite{branciard2021coherent}, introducing a generalized version of PBS within a novel graphical framework inspired by optical configurations involving PBS. This approach aims to formally characterize quantum coherent control through a new descriptive formalism that extends beyond classical completely positive trace-preserving (CPTP) maps, which inadequately capture quantum coherent control. The motivation behind this endeavor is to provide a more nuanced and accurate representation of quantum coherent control, particularly in scenarios involving complex quantum channels and operations. 

In recent years, another framework gaining increasing attention in the realm of quantum coherent control is the concept of Quantum Switching, also known as quantum coherent control of the causal order of quantum channels. This area of study emerged as an extension of quantum information theory, aiming to transcend the limitations of conventional completely positive and trace-preserving maps \cite{Aharonov_2009}. The latter are applicable only under the assumption of a dynamical reduced dynamics on the information carrier, either initially correlated with the environment or during its transmission. The novel formalism expands upon this premise to encompass more exotic and theoretically plausible scenarios. Beginning with the notion of quantum combs \cite{perinotti_2009}, which may describe memory effects during transmission leading to potential information backflow from the environment to the system \footnote{The notion of a quantum comb is a generalization of the notion of a quantum channel. The latter does not exist only under the assumption that the system is initially not quantumlly correlated with the environment -- not in an entangled or discordant state-- as the reduced dynamics on the system cannot be recovered or the complete positivity of the channel is not guaranteed \cite{sudarchan,Milz_2021}. As such, the quantum comb comes as an extension of the notion of a quantum channel or a quantum map in these extreme scenarios.}, the concept of quantum supermaps emerges as higher-level quantum maps \cite{chiribella2008transforming}.Unlike conventional maps that act solely on the states of the information carrier, quantum supermaps operate on the configuration of quantum channels' placement, allowing for the exploration of genuinely quantum configurations that were previously inaccessible.

A comparison between the performance of the two schemes has been subject of investigation. One may be over-driven by the idea that the advantage of both schemes for mitigating noise comes from the constructive interference of the two signals sent through different legs or different orders in the case of quantum coherent control of paths and indefinite causal ordering respectively. This is only partially true! In indefinite causal ordering there is another additional effect coming from the commutation and anticommutation relations between the Kraus operators describing the individual channels. This has been reported in  \cite{loizeau2020channel}, where a comparison between the performance of indefinite causal ordering namely the quantum switch and quantum coherent control in randomly sampled channels. Another question of debate is the regime in which quantum coherent control of path of indefinite causal ordering can indeed provide an advantage over the ultimate limits of quantum Shannon theory of point-to-point communications. It has been observed that under some noise parameters of individual quantum channels these novel schemes can be advantageous, whereas in other regimes of the noise parameter, these schemes fail to provide any advantage. For instance, the quantum switch perfectly activates the capacity of a qubit entanglement breaking channel while if fails to do so for less drastic types of noise like the bit-flip and phase-flip noises. In \cite{koudia2021environment}, the authors proposed to think of the quantum switch as a particular instance of an environment assisted communication strategy which is optimal in the case of entanglement breaking channels, leading to perfect capacity activation, and fails to be so in the majority of cases where it might be even detrimental for information transmission.

\subsection{Multiplexing and Diversity}

Spatial multiplexing  involves the simultaneous transmission of multiple data streams through separate spatial paths within the propagation medium. Unlike traditional multiplexing techniques that allocate different time slots, frequencies, or wavelengths, spatial multiplexing exploits the spatial directions in the atmosphere or interfering cores in fiber cables to transmit multiple independent information streams concurrently. This is achieved by employing multiple transmitters and receivers strategically positioned to establish parallel communication links. By distributing information across distinct spatial paths, space multiplexing enhances spectral efficiency, increases energy efficiency, and mitigates the effects of atmospheric turbulence, crosstalk and other impairments. The key concept behind spatial multiplexing is that advanced joint pre/post-processing of information can transform energy leakage among spatial channels: from unwanted crosstalk treated as additional noise to useful signals which can enhance the communication performance. 

On the other hand, there is spatial diversity. Spatial diversity refers to the utilization of multiple spatially separated antennas or optical apertures to improve the reliability and robustness of communication links in the presence of propagation impairments. In \ac{FSO} systems, spatial diversity can be achieved by deploying multiple transmitters and receivers at different locations or orientations. By leveraging spatially separated paths, FSO systems can exploit variations in atmospheric conditions across different paths to enhance signal reception and mitigate the effects of fading and distortion caused by atmospheric turbulence and obstructions. Both methods, assume mutually quantum mechanically incoherent channels \footnote{By quantum mechanically incoherent channels we mean that the channels cannot propagate information simultaneously through both of them while having quantum coherence of the information carrier, making the channels placed in a parallel independent configuration used in multiplexing}.   Similarly, in fiber optic communication systems, spatial diversity can be employed by utilizing multiple optical fibers or deploying multiple spatially separated cores within a single fiber. This approach allows the system to distribute the optical signal across different paths, thereby reducing the impact of localized fiber impairments such as microbends, macrobends, and scattering losses. By taking advantage of spatially diverse paths, fiber optic systems can achieve improved signal integrity and robustness against physical layer impairments, ensuring more reliable and higher-quality transmission. This method also assumes that the channels are mutually quantum mechanically incoherent, allowing independent transmission paths to be used for mitigating the overall effect of signal degradation. 

We should highlight that quantum coherence between quantum channels should not be confused with phase coherence in optical systems. The latter is a property of a wave-like system where the phase relationship between different points or components of the system remains constant over time. This is crucial in classical waves such as optical systems. For instance, in laser beams, phase coherence ensures that the light waves are in step, producing a beam with a narrow frequency range and minimal dispersion. Maintaining phase coherence is essential for applications requiring high precision and stability, such as interferometry and coherent communication systems. Quantum coherence, on the other hand, is a fundamental concept in quantum mechanics. It describes the ability of a quantum system to exhibit superposition, where a particle can exist in multiple states simultaneously. Quantum coherence is characterized by the preservation of the relative phase between components of a quantum state. This property is crucial for phenomena such as quantum entanglement and quantum interference, which are the basis for quantum computing, quantum cryptography, and other quantum technologies. Quantum coherence is more delicate than classical phase coherence as it can be easily disrupted by interactions with the environment, leading to decoherence and loss of quantum information.

Quantum Coherent control in the counterpart, can harness the advantage of both schemes, namely, spatial multiplexing and spatial diversity by harnessing the genuinely  quantum superposition of different states during the transmission simultaneously \footnote{Quantum coherent control relies on the fact that the genuinely quantum coherence, which has no classical counterpart, stored in the information carrier, is mapped to the transmission channels putting them in a new quantum configuration that does not have a classical counterpart in its turn - it is not equivalent to sequential or parallel placement of channels.  }. One can easily notice that by increasing the number of coherent modes used for transmission, for instance with d-modes, a qudit codeword can be coherently transmitted, increasing the throughput. In the same time, without copying the information and transmitting it on different independent paths to the receiver, as in spatial diversity, quantum coherent control can indeed reduce the effect of noise arising from atmospheric effects in free space communications, as was proven by theory and experiment \cite{Giulia2021}.
Indeed, the signal from different spatial paths should be interfered at the receiver and coherent detection plays an important role to harness the full advantage of quantum coherent control. One can see this as an interferometric setup, where the interference of the signals at the end allows to project the global system into a constructive interference and destructive interference, leading to noise mitigation and filtration as was detailed in \cite{Gisin_2005}. One can of course tune the weights of each propagation path manually to achieve the desired constructive interference effect if a priori knowledge on the channels is given to optimize the transmission. 
Indeed, this scheme holds similarities with spatial multiplexing, in which the quantum state is encoded onto different spatial modes or paths of propagation, such as different optical fibers or waveguides. Each spatial mode carries a distinct component of the quantum state, and at the receiver, the components are combined or measured separately to retrieve the original quantum state.
This scenario has been investigated in different studies demonstrating that high-dimensional states exhibit greater resilience to noise, thereby suggesting a higher error tolerance in quantum communications.  Similarly, Quantum Coherent Control holds also similarities with path-based multiplexing, which involves encoding quantum information directly onto different paths of propagation, rather than using spatial modes. Each path corresponds to a distinct quantum state and the information is encoded in the choice of path taken by the quantum system which is decoded at the receiver by an interferometric detection cmprising single photon detectors. Path encoding shows promise due to its compatibility with photonic integrated circuits, facilitating the generation, manipulation and detection of path-encoded quantum states \cite{wang2016chip, solntsev2017path}. However, reliablely transmitting such states remains a major challenge. One potential solution is coupling each path to a \ac{SMF} each of which would be affected by varying random phase drifts due to temperature changes, bends, and mechanical stress could disrupt transmission, especially over long communication channels. Alternatively, coupling each path to different core of a multicore fiber (MCF) suppresses phase drifts among cores, enabling better transmission, particularly of superposition states. Previous experiments explored these fibers for high-dimensional quantum communication, but limitations in stability impacted achievable distances. Recently, this issue was addressed by actively stabilizing the system, achieving high-fidelity transmission of qudits of dimension $d=4$ and $d=7$ over a $2 KM$ long \ac{MCF}  \cite{da2021path}. For the knowledge of the authors no path encoding experiment has been carried in free space.

Although quantum coherent control shares similarities with spatial multiplexing and path encoding, there are still some differences between the two approaches. Notably, spatial encoding schemes may require additional resources, such as spatial light modulators, optical components for mode manipulation, or multiplexing/demultiplexing devices. Path encoding schemes may have simpler requirements in terms of routing and switching elements. Moreover,  in spatial encoding schemes, the receiver may need to perform spatial mode demultiplexing and spatial mode detection to retrieve the original quantum state. In path encoding schemes, the receiver typically measures the quantum state directly on each path, such as through photon counting or interferometric techniques. Additionally,  in spatial encoding, the quantum state is encoded onto different spatial modes, which can have higher dimensionality compared to path encoding. Each spatial mode can carry multiple degrees of freedom, such as different polarizations, spatial positions, or orbital angular momentum states. In path encoding, the quantum state is typically encoded onto discrete paths, which may have lower dimensionality compared to spatial modes.

\subsection{Quantum MIMO}

\begin{figure}[t!]
    \centering

    \includegraphics[width = 0.475\textwidth]{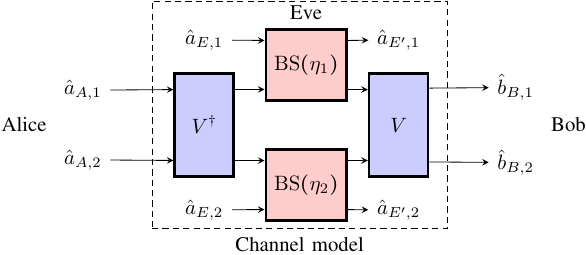}

    \caption{A $2\times 2$ \ac{MIMO} channel model for THz \ac{QKD} introduced in \cite{KDM:21:CL}. The interception of Eve is modelled by the two beamsplitters in the channel. Eve can arbitrarily change the input/output as well as the transmittivity of these beamsplitters.}
    \label{fig:QMIMO}
\end{figure}

\ac{MIMO} systems proved to bring a paradigm shift in wireless communication systems. In particular, the presence of multiple antennas at the transmitter and the receiver can either be used to increase the diversity or to increased the number of degrees of freedom in wireless communication systems. A seminal work  \cite{ZT:03:IEEE_T_IT} introduced a fundamental tradeoff between the diversity and multiplexing. Discussion on \ac{MIMO} channels with main focus on \ac{QKD} has also begun recently \cite{KDM:21:CL, KMC:23:TQE}. The main component of the channel model, again was a quantum optical circuit composed of beamsplitters where different inputs and outputs were controlled by different parties in a typical \ac{QKD} protocol, i.e., Alice, Bob, and Eve (cf.~Fig.~\ref{fig:QMIMO}). These channel models were utilized to analyze the performance of \ac{QKD} in Terahertz regime under different assumptions. 

One challenge in designing a \ac{MIMO} communication scheme with \emph{quantum information} is the no-cloning theorem \cite{WZ:82:Nat}. One fundamental operation to harvest spatial diversity gain from classical \ac{MIMO} communication is to transmit multiple copies of the signal. On the receiver end, the multiple copies of received signal are combined together to extract a higher quality of the received signal. Due to the no-cloning theorem, preparing multiple high-fidelity copies of a quantum signal is not possible, thus making it challenging to harness an advantage from \ac{MIMO} quantum channels. There are still a number of cases where transmitting multiple copies of a quantum state remain possible, e.g., when the underlying information is classical, e.g., in \ac{QKD} or when the transmitter has access/knowledge of the state preparation process. However, in the presence of multiple parallel channels (spatial or otherwise) it is possible to multiplex several streams of quantum signals together to harvest the multiplexing gain. 

\Ac{WDM} is a popular and effective method in optical communication for sending multiple data streams of different wavelengths on a single physical medium.\footnote{The frequency-division multiplexing is the same approach in the \ac{RF} communication.} A typical optical \ac{WDM} system consists of multiple optical sources of different wavelengths fed into an optical multiplexer that combines the input signals. The resulting output is fed into a \ac{SMF} for transmission. At the output, a demultiplexer splits the individual signals, which are then transmitted to the intended receiving ports. 

\Ac{WDM} remains to be an important approach for the co-existence of classical and quantum networks on the same infrastructure \cite{GNR:23:Conf, JQZ:24:Chip, ATB:21:conf, TAJ:21:conf}. Recent works have demonstrated the coexistence of classical and quantum signals on a single fiber using \ac{WDM} to achieve \ac{QKD} \ac{SKR} of more than 4 kbps at 65.5 km \cite{GNR:23:Conf}. Similarly, another recent experiment demonstrated the coexistence of a quantum channel and $8\times 200$ Gbps 16-QAM optical channels with launching power as high as -9dBm/channel \cite{ATB:21:conf}. Due to a strong disparity between the typical strengths of classical and quantum signals special care must be given to the channel spacing to ensure minimal cross-talk between the two \cite{BRS:16:SI, BRS:18:SR}. 

Another important application of \ac{WDM} is to multiplex multiple quantum signals in a \ac{SMF}. A recent work has proposed an all-\ac{QWDM} for the multiplexing of quantum signals \cite{BRS:23:PRA}. The work develops general models for the involved components in a complete \ac{QWDM} communication system. Specifically, the operations of quantum transmitters, a passive quantum wavelength distributor, and quantum receivers are analyzed in fully quantum terms. The model is adaptable to various scenarios; possible sources of quantum signal include \ac{WCP}, \ac{SPP}, or Poissonian mixed states; the wavelength distributor component is modeled as a unitary operator that admits various configurations including mux-fiber-demux configuration or a star coupler; finally the receiver also admits multiple configurations as a combination of demux or Filter followed by single-photon detectors or conjugate homodyne detectors. The analysis shows that \ac{QWDM} of coherent states gives a tensor product state at the output, i.e., there is no correlation between the states at the output of all quantum receivers. On the other hand, \acp{SPP} of different quantum receivers will turn out to be entangled in Lambdanet broadcasting communication system with a star coupler as the wavelength distributor. This unintended entanglement is a typical case of crosstalk between different signals and can be avoided by using an ideal router as a wavelength distributor of a router-based communication system.

\section{Connecting DV Quantum Nodes with CV Quantum Links}\label{CDVCV}

In the rapidly advancing fields of quantum communications and computing, one of the key challenges is to find efficient ways to encode and manipulate quantum information. Quantum harmonic oscillators provide a promising avenue for such tasks. Unlike classical oscillators, which can only occupy discrete energy levels, quantum harmonic oscillators can access a continuous spectrum of energy states, making them ideal candidates for encoding quantum information.

Encoding qudits in quantum harmonic oscillators offers several advantages. By utilizing the infinite dimensional Hilbert space of harmonic oscillators, we can achieve a higher information density compared to traditional qubit-based systems. This increased dimensionality allows for more robust error correction and potentially enables powerful quantum algorithms. The encoding process typically involves mapping the quantum states of the qudit onto energy eigenstates of the harmonic oscillator. This mapping can be achieved through various techniques such as superposition of Fock states or employing coherent states. Once encoded, the quantum information can be manipulated and processed using continuous-variable (CV) quantum gates and measurements acting as logical qudit gates and measurements.

The interconnection between discrete-variable (DV) quantum computing and continuous-variable (CV) quantum communications is particularly noteworthy. DV quantum computing typically involves qubits, which represent quantum information in discrete states. In contrast, CV quantum communications leverage quantum harmonic oscillators to encode information in continuous spectra. By integrating these approaches as shown in Figure.~\ref{fig:general_MIMO}, we can harness the strengths of both systems: the precise control and established error correction methods of DV systems and the high-dimensional encoding and flexibility of CV systems.

Furthermore, encoding qudits in quantum harmonic oscillators opens up new possibilities for quantum communications, quantum cryptography, and quantum computing. In quantum communications, the ability to encode information in a continuous spectrum allows for more efficient and secure transmission of data. Quantum cryptography benefits from the increased security offered by high-dimensional encoding, making eavesdropping significantly more challenging. In quantum computing, the hybrid approach of combining DV and CV systems can lead to the development of more versatile and powerful quantum algorithms, paving the way for advancements in various applications ranging from cryptography to complex simulations. This synergy represents a significant step forward in the quest for practical and scalable quantum technologies.

\subsection{Cat Codes}
 Cat states or Shroedinger states have been shown to be of paramount importance for quantum sensing \cite{he2023sensitive}, quantum communications \cite{bergmann2016quantum} and quantum computing \cite{vlastakis2013deterministically}.  Cat states are a coherent superposition of two coherent states with opposite displacement parameters. Although their seemingly simple structure, their generation still only probabilisitic in photonic platforms due to their non-Gaussian nature when only passive optics is used \cite{laghaout2013amplification}. By including non-Gaussianities through measurement setups \cite{winnel2024deterministic}, like PNRD measurmenets, or evolutions like cubic-gates, photon addition and subtraction, these state can be obtained in the optical domain \cite{ourjoumtsev2007generation, takase2021generation}. Many experimental studies have been devoted to the deterministic generation of these states based on Jaynes-Cummin interaction between light and single atoms in a cavity \cite{brune1992manipulation,buvzek1992schrodinger}. Notoriously, macro cat states of mechanical oscillators has been achieved recently in optomechanical systems \cite{liao2016generation,zeng2020macroscopic}. 
 Multi-component cat states (cat states with a coherent superposition of more than two coherent states) have shown promess towards loss tolerance, when encoding discrete variable degrees of freedom, making them suitable candidtaes for \ac{QEC} in a DV/CV interface \cite{bergmann2016quantum, chamberland2022building}.   

\subsection{GKP Codes}
GKP states were first introduced by Gottesman, Kitaev and Preskill as a method for encoding finite-dimensional quantum DV systems, namely qudits, into quantum harmonic oscillators \cite{gottesman2001encoding}. Recent theoretical advancements have expanded upon this seminal work, proposing and evaluating GKP state preparation using superconducing devices. After years of experimental advancement, GKP qudits have finally been demonstrated in superconducting microwave cavities \cite{campagne2020quantum} and in the harmonic motion of ions \cite{fluhmann2019encoding,kendell2024deterministic}. In the optical domain, however, preparing GKP states present significant challenges due to their non-Gaussian nature. The primary hurdle lies in the need for reliable and potent nonlinearities, which are not readily accessible in photonic platforms. One appraoch, Gaussian Boson Sampling \cite{takase2023gottesman} capitalizes on the fact that measurmenets can induce nonlinear effects. Here, Gaussian resource states are combined using passive liear optics and partially measured via PNRD detectors. While this method can yield high-quality optical GKP states, it does so only probabilistically by relying on the complex circuits of Gaussian Boson Sampling. To alleviate this experimental obstacle, alternative strategies have been proposed \cite{vasconcelos2010all}. By leveraging non-Gaussian resource states or non-Gaussian optical elements, a recurssive application of short linear circuits and homodyne measurements is sufficient for GKP state prparation \cite{weigand2018generating}. Another option involves combining photon-subtraction and homodyne-based elements to transform many mode Gaussian cluster states into non-Gaussian few mode states, which can be further processed into GKP states \cite{eaton2022measurement}. Approaches not relying on measurements, uses cubic phase gates as a resource for non-Gaussianity to generate GKP states \cite{yanagimoto2020engineering}. 
Stemming from the fact that GKP states are resilient for quantum communications over Gaussian lossy channels, discrete variable degrees of freedom encoded in GKP states is considered a promising approach towards fault tolerance quantum communications and computing. As a result,  quantum repeater protocols based on these states were designed achieving high end-to-end entanglement rates \cite{PhysRevResearch.3.033118}. Sequentially, quantum scheduling algorithm for GKP states-based quantum switches have been studied \cite{Azari2024QuantumSF}.

\subsection{Truncation of Infinite-Dimensional Hilbert Space}
Another practical method that connects \ac{DV} and \ac{CV} quantum system is the truncation of infinite-dimensional Hilbert space to a finite one. This method is typically used to make numerical calculations of \ac{CV} systems tractable. For example, the density matrix of a thermal noise in quantum systems is represented by infinite-dimensional continuum of coherent states. Only through appropriate truncation, it is possible to numerically treat the problems involving such noise \cite{Car:15:Book}. 
More concretely, in the case of pure noise, the density operator of corrupted signal can be defined in terms of number states
\begin{align}
	\rho \left( \gamma\right) = \sum_{m = 0}^{\infty}\sum_{n = 0}^{\infty} R_{m, n}\left( \gamma, N_{\mathrm{noise}}\right) \ket{m}\bra{n},
\end{align}
Where $\left|\gamma\right|^2$ is the mean number of signal photon, $N_{\mathrm{noise}}$ is the mean number of noise photons, and $R_{m, n}\left( \gamma, N_{\mathrm{noise}}\right)$ are the coefficients of the expansion. A finite representation of $\rho\left(\gamma\right)$ can be obtained by limiting the number of terms as
\begin{align}
    \rho\left(\gamma\right) \approx \sum_{h = 0}^{n-1}\sum_{k = 0}^{n-1} R_{h, k}\left( \gamma, N_{\mathrm{noise}}\right) \ket{h}\bra{k} := R\left(\gamma\right),
\end{align}
where $R\left(\gamma\right)$ is the $n \times n$ matrix approximating $\rho\left(\gamma\right)$. This representation can be made further compact by defining an accuracy $\epsilon$ and restricting $R\left(\gamma\right)$ to its eigenvalues larger than $\epsilon$ \cite{Car:15:Book}.

In practice, a similar truncation can be achieved by employing finite-dimensional projectors on the state \cite[Supplementary Note 1]{PLOB:17:NC}. Consider $m$ bosonic modes with the energy operator $H = \sum_{i = 1}^m N_i$, where $N_i$ is the number operator of mode $i$ and the energy-constrained states $\rho_E$, i.e., $\Tr{\rho_E H}\leq E$. Then there exists a finite-dimensional projector $P_d$ that projects these states to a $d$-dimensional support of the $m$-mode Hilbert space with the probability of success $\Tr{\rho_E P_d}\geq 1 - p$. Here $p = E/\sqrt[m]{d} - 1$. Further, this truncation ensures that the trace distance between the original and the truncated state is less than $\sqrt{p}$ \cite{PLOB:17:NC}. This truncation of infinite-dimensional Hilbert space can be extended to a truncation map (CPTP map), which can be extended to bipartite scenario where the map remains local to individual parties while boundin the trace distance between the original and the truncated state \cite{PLOB:17:NC}.

\section{Open Challenges}\label{sec:OQ}

 In this section, we delve into the open pivotal aspects of photonic-based quantum communications. We begin by examining the \ac{HOM} effect, a cornerstone quantum phenomenon where two indistinguishable photons entering a 50:50 beam splitter pair-off together, an outcome impossible to explain classically. Next, we address the challenges of optimizing unknown quantum communication systems through tomographic-based and signaling-processing methods. We discuss alternative strategies tailored to specific applications, highlighting how these approaches can effectively navigate the limitations of standard tomography. %
 We then address the unresolved issue of frequency-selective fading in the absence of a comprehensive quantum theory of atmospheric losses. Further, we discuss the fundamental challenges in adapting diversity and multiplexing techniques from classical \ac{MIMO} communications to the quantum domain. Finally, we touch upon the complexities of applying quantum coherent control and indefinite causal order in mitigating quantum noise, and examine the interplay between \ac{DV} and \ac{CV} systems within hybrid network architectures.

\subsection{Interference of Photons: The Hong–Ou–Mandel effect}

		The \ac{HOM} effect lies at the core of quantum photonics applications. This effect predicts that two indistinguishable incoming photons will pair off together after entering a 50:50 \ac{BS}. This is a strictly quantum-mechanical effect, since classically, the two photons could equally be detected separately. This phenomenon enables linear optics to be a feasible platform for quantum computing and lies at the heart of enhanced precision measurements and quantum communication applications such as teleportation, entanglement swapping, \Ac{MDI} \ac{QKD}, and others. In this section, we briefly introduce the \ac{HOM} effect for two and multiphoton systems, highlighting recent applications in \ac{QC}. For a through presentation of the topic check \cite{Survey:21:TPI}.

		\subsubsection{Two Photon Interference}
		
		Two-photon interference phenomenon is often introduced in a \ac{BS} scenario: assume that each input port of a 50:50 \ac{BS} receives a photon. If the photons are distinguishable (e.g.,if they have different polarization modes), the outcomes will be exactly as predicted in classical electromagnetism; that is, the photons will exit through the same or different ports. However, if the photons are indistinguishable, they will necessarily pair together; they will never be found in different ports. The latter case is often referred to as the \ac{HOM} effect \cite{HOM:87} and plays a central role in quantum information processing \cite{Survey:21:TPI}, particularly in the engineering and analysis of entangled states in quantum photonics.

		Two-photon interference can be utilized to perform a linear optical Bell state measurement and, therefore (given successful outcomes), to perform entanglement swapping. This idea was applied to trapped atoms \cite{Heralded:12:86} and transmon qubits \cite{RCRE:16:88}. In the first, a single trapped excited atom with two distinct photon polarizations is considered. Given two copies of such states, they showed that with a certain probability, entanglement between two separated atoms is achieved by interfering the emitted photons in a BS and measuring the outcomes. In other words, the authors in \cite{Heralded:12:86} were able to create entanglement between two separated atoms. Similarly, two transmon qubits were entangled in \cite{RCRE:16:88}.
		
		Two-photon interference combined with Bell state analysis plays a central role in \ac{MDI}-\ac{QKD} \cite{MDIQKD:12:x84}. As explained in \cite{Survey:21:TPI}, the protocol operates as follows: Alice and Bob each randomly prepare their individual photons in specific states and then send their photons to a third party, Charlie, who is not trusted. Upon receiving the photons from Alice and Bob, Charlie performs a Bell state measurement by directing each photon onto a \ac{BS} -- when the photons arrive simultaneously at the BS, two-photon interference is witnessed. Charlie then publicly announces the outcome of his Bell state measurement, which Alice and Bob use to establish a shared raw key. Subsequent classical post-processing steps, such as error correction and privacy amplification, are performed to generate a secure secret key between Alice and Bob. Although such measurement only identifies two of the four Bell states, this is enough for the security proof \cite{MDIQKD:12:x84}. 
		
		Another milestone in \ac{QKD} is the \Ac{RRDPS} protocol \cite{PQKDP:14:x102}. In practical \ac{QKD} schemes, by determining the amount of information leakage to a malicious attacker one can optimize the amount of post-processing required to establish a shared key. However, determining such quantity requires active monitoring of output quantities such as error rate \cite{Survey:21:TPI}, which decreases the overall protocol's efficiency. The \ac{RRDPS} protocol eliminates requirements by setting a bound on the information leaked to the attacker. This bound is determined solely by the parameters established during Alice’s generation stage. Moreover, the security of \ac{RRDPS}-based \ac{QKD} protocols is largely enhanced by considering large dimensions, requiring large and stable variable-delay interferometers. To overcome these challenges, a passive version of the \ac{RRDPS} was proposed \cite{PRRDPSQKD:15:x85}. In this scheme, the phase stability of the two-photon interference effect is exploited to avoid measuring the interference of a train of pulses at Bob's side.

\subsubsection{Multiphoton Interference with multimodes}

A first generalization of the two-photon \ac{HOM} effect occurs when three input photons enter a tritter \cite{Survey:21:TPI}, a three-mode analogue to a beam splitter. Similar to the two-photon \ac{HOM} effect, interference can be observed in a tritter if the photons are all indistinguishable. However, in contrast to the two-photon \ac{HOM} effect, indistinguishability alone is not a sufficient condition; the scattering behaviour of the photons in a tritter also depends on an additional collective phase \cite{IIPEBS:14:x146,CLTC:14:x147,PITME:15:x148} that affects the output probability when the photons are \textit{partially} distinguishable. This can alter the detection probability independently of the bi-photon transmission when all the photons are found in separate output ports \cite{Survey:21:TPI}. Moreover, the output probability distribution depends on the exact unitary transformation of the tritter and the input distribution. This complexity introduces additional features unique to multipartite interference \cite{17:DMPI:mi144}.

The above generalization can be extended further, increasing the complexity and opening a myriad of new applications. When two photons are sent into a multiport (e.g., multimode waveguides \cite{12:QWTIB:mi153,12:TPPE:mi154} or integrated waveguide structures \cite{09:IQP:mi156}), the transmission can be described as a complex bipartite quantum walk \cite{Survey:21:TPI}. This output distribution can be used to perform quantum simulations and investigate quantum correlation patterns. The interference of more than two photons offers even more powerful applications. In complex network structures, the interference of more than two photons has been utilized to enable the teleportation of high-dimensional quantum states \cite{19:QTHD:mi164}, generate multipartite entanglement \cite{13:LMEG:mi165}, and investigate the boson sampling problem \cite{11:CCLO:mi166}.

\subsection{Physical Phenomenon Beyond Losses}

An important work \cite{RN77} explores the innovative use of classical light to evaluate quantum channels, particularly those influenced by atmospheric turbulence. This study demonstrates that the orbital angular momentum (OAM) of light, which is conserved down to the single-photon level, can effectively analyze and correct quantum channels. Classical light, through non-separable states like vector beams, can replicate the behavior of quantum-entangled photons, enabling channel tomography without requiring a pre-existing quantum link. Experimental results confirm that the decay of classical entanglement mirrors that of quantum entanglement, validating the method. This approach facilitates real-time monitoring and correction of quantum channels, enhancing the robustness and efficiency of quantum communication systems by leveraging the conserved properties of OAM.

In addition, the authors present a new setup for quantum cryptography using Bell states based on energy-time entanglement, demonstrating its feasibility in laboratory experiments. The setup achieves bit rates of approximately 33 Hz and quantum bit error rates around 4\%, which is sufficiently low to ensure secure key distribution. Unlike schemes that rely on weak pulses containing mostly zero photons, our approach starts the transmission with a single photon. Additionally, using discrete energy-time states up to dimension 4 eliminates the need for rapid switching between non-commuting bases. Nature itself selects between the complementary properties of energy and time, reducing the effectiveness of photon number splitting attacks\cite{PhysRevLett.84.4737}. 

In terms of intensity modulation, Coherent One-Way(COW) QKD is used. It is demonstrated the effectiveness of a modulator-free QKD transmitter for the COW protocol. This system achieved estimated secure key rates ranging from 4.57 Mbit/s to 6.38 kbit/s over distances spanning from 7.5 km to 150 km. The absence of external modulators not only reduces the system's size but also simplifies its complexity. An exciting aspect of this work is its potential for integration into a multi-protocol network. Unlike current approaches, which often rely on bulky systems with numerous active components, our system allows for a single transmitter to seamlessly switch between protocols based on the client's requirements regarding bit-rate, distance, or security. Moreover, the versatility of this transmitter, offering both amplitude and phase modulation along with on-demand phase randomization, facilitates the easy adoption of newly developed protocols through firmware updates\cite{RN78}.

\subsection{Performance Assessment of Quantum Communication Systems}

Optimising an unknown quantum communication system presents a significant practical challenge. Unlike classical communications, where an unknown channel can be effectively characterised through full tomography, the same approach is not feasible in quantum communications. The standard approach (SA) to evaluate the performance of a quantum communication system primarily seeks to determine the channel's capacity by its full characterization. However, such characterization often requires a full tomographic reconstruction of the channel which is impractical as the complexity of the system increases --- $O(D^2/\epsilon)$ and $O(1/\epsilon^{2n})$ state copies are required for a tomography reconstruction with trace distance error $\epsilon$ of a $D$-dimensional \cite{Survey:CLQS:24} and energy-constrained $n$-mode \cite{Mele2024LearningQS} quantum state. This challenge becomes even more pronounced for time-varying channels, whose descriptions change more rapidly in time than tomography can be performed. Although some alternatives to speed up the channel characterization were proposed (cf. \cite{Survey:CLQS:24, QCB:20:survey-cer} for a comprehensive list), they lack generality, applying only to particular cases. The performance of a realistic quantum communication system cannot be efficiently evaluated in practice using channel characterization alone. 
However, there are viable alternatives tailored to specific applications and goals. In the following, we list them according to the amount of channel information they rely on --- ranging from full, partial, and none channel characterisation.

\subsubsection{Resource-efficient tomographic methods}

Resource-efficient tomographic methods offer a promising solution for structured systems and processes. Although standard tomography is excessively costly in the size of the system, as mentioned previously, some processes and states found in practical scenarios have properties or symmetries that can be exploited. For approximately pure or low-rank states, \textit{compressed sensing tomography} \cite{QSTCS:10:c53,FST:20:c56,QTPPCSP:15:c69} reduces the number of tomographic resources or increases the reliability of the recovered information using the same measurements found in full tomography. Similarly, if the system's Hamiltonian is local or has little entanglement, tomography reconstructions of large states can be substantially improved using tensor networks \cite{SRDM:2013:c10,EQST:10:Nature:c31,WTMPS:13:c66}. Recently, the Gaussianity of \ac{CV} states has been exploited to develop an efficient tomographic algorithm \cite{Mele2024LearningQS}. These methods leverage specific properties of the state or process to enable less resource-intensive tomography characterizations.

\subsubsection{Approximate evaluation through partial channel characterization}

For systems and processes where a partial characterization is sufficient to derive optimal conditions, \textit{probably approximately correctly} (\Ac{PAC}) learning techniques can be an advantageous alternative \cite{QCB:20:survey-cer}. For example, one might be interested in \ac{PAC} learning the expected outcomes of a local observable associated with a quantum state. Given specific conditions \cite{ELQS:19:c101}, \ac{PAC} learning is achievable with a measurement complexity that scales only linearly with the number of qubits, though it still requires exponential computational effort. A similar approach is possible if one is confident that the provided data is described by a parameterised Hamiltonian (or Liouvillian) model, whose parameters are, however unknown. In this case, learning techniques can recover the parameters associated with the corresponding model \cite{EOHL:12:c52,SRUPH:15:c63}.

Optimal communication conditions can be inferred through fidelity and trace distance estimations. Different from the above-mentioned tools, these methods are based on determining the overlap between distinct states and processes. This can be useful to assess the fidelity of an implemented state or process with respect to the ideal one or to compare different signal states, revealing optimal encoding strategies. Notably, importance sampling allows for the estimation of fidelity for imperfect preparations relative to ideal ones, with constant measurement complexity \cite{DFEP:2011:c46}. This concept can be extended to channels, as discussed in \cite{OSEAF:13:c98}. Moreover, such approaches can be efficiently implemented in noisy intermediate-scale quantum computers, as demonstrated in \cite{VEOSS:24:mine}.

Many characterization methods, including the ones mentioned above, rely on two strong assumptions: 1) reliable measurement devices, and 2) the quantum state or process can be accessed in independent and identically distributed (i.i.d.) experiments. While these assumptions are often reasonable in laboratory settings with substantial control over preparation, evolution, and measurement devices, they do not hold as well in communication scenarios. The second assumption, in particular, is challenging to meet in practical communication contexts, where the channel conditions can change more rapidly than the characterization protocol can accommodate.

\subsubsection{Signal-Processing Approach}

A \ac{SPA} circumvents the above challenges by framing the problem in terms of the received states rather than the channel's description, being agnostic to the latter. In such an approach, candidate states perform communication and encoding optimisation simultaneously. These states serve as probes and information carriers, updating the optimization and transmitting data. Mathematically, the primary objective of this framework is to achieve optimal discrimination of quantum signal states \cite{cariolaro2010theory:m10,2010_PQDTS:m9,vasquez2023quantum:m15,VEOSS:24:mine}. This task is simpler than tomographic-based methods, as it focuses on a quantum hypothesis testing scenario \cite[Section~2]{2015_QSD_survey:m21}, where the aim is to maximize the probability of success in identifying the transmitted state at the receiver. Rather than theoretically determining the channel's capacity, it seeks to evaluate its practical performance.

Given its simpler theoretical formulation, a \ac{SPA} might not be sufficient to predict the capacity of arbitrary channels. It is natural to expect that a simpler theoretical model, which is agnostic about the channel, will have less explanatory power, meaning that further assumptions must be imposed to identify the encoding scheme that achieves the channel's capacity. In fact, only in a few cases (e.g., two-state discrimination) optimal discrimination is known without further assumptions on the encoding scheme \cite[Section~9]{2015_QSD_survey:m21}. However, this limitation does not necessarily affect the channel's assessment since the same results are obtained once these assumptions are considered. The critical difference is that whereas in a characterization-based framework, one can derive the optimal encoding scheme conditions within its formalism, the same is not generally possible within the \ac{SPA}. However, this apparent advantage of tomographic-based methods has to be taken with a grain of salt because for practical channels (non-additive), determining the channel's capacity is generally intractable and, therefore, equally challenging.

A feedback-based \ac{SPA} can benefit from \ac{QML} techniques. 
By employing a variational algorithm to estimate the trace distance between the received states, the Authors in  \cite{VEOSS:24:mine} numerically determined the optimal encoding protocol for the amplitude damping and Pauli channels of a binary quantum communication system. Their simulations demonstrated the convergence and accuracy of the method with a few iterations, confirming that optimal conditions can be variationally determined with minimal computation. In their algorithm, the transmitter iteratively approximates the solution thanks to the receiver's feedback. In their case, the feedback is entirely classical; classical information is sent back through a classical channel. Nevertheless, other options could be explored, and a more general structure considering quantum outputs at the receiver's side and quantum feedback could unleash alternative \ac{QML} techniques.

\subsection{Frequency-selective fading}

Frequency-selective fading is a phenomenon in wireless communication where various frequency components of a transmitted signal experience different degrees of attenuation. As a result, certain frequencies within the signal's bandwidth are attenuated more than others. This effect is usually caused by multipath propagation, where every path has its own transmission coefficient. Therefore, a frequency-selective fading channel, as any other fading channel, is mainly characterized by its \textit{probability distribution of the transmission coefficient}, as explained previously in \autoref{subsec:CVCM}.

A consistent quantum theory of atmospheric losses has already been developed \cite{QLTA:19:fs12,Vasylyev2016}. As highlighted in \cite{17:HONEFLC}, this theory has considered a wide range of important free-space phenomenon that leads to channel fluctuations --- including weak turbulence caused mainly by beam wandering \cite{Vasylyev2012} as well as weak-to-moderate and strong turbulence dominated by beam-broadening and beam-shape deformation \cite{Vasylyev2016}. These findings have been validated in \ac{QKD} experiments \cite{Usenko_2012} and further explored in homodyne \cite{09:FFSQKD:fs16} and heterodyne \cite{16:FSQSUH:fs18} detection.

Despite the significant progress towards a quantum theory of atmospheric losses, much remains to be done regarding how the quantum features of light vary within these models and under what conditions quantum properties are preserved \cite{17:HONEFLC}. The first studies towards these questions focused on particular scenarios --- such as quantum teleportation \cite{12:QTED100:fs5}, tests of Bell inequalities \cite{10:ETTTA:fs19, Ursin2007}, and entanglement of Gaussian and non-Gaussian states \cite{17:HONEFLC} --- as reviewed in \cite{17:HONEFLC}. However, a universal understanding of quantum state dynamics subjected to atmospheric fluctuations has been scarcely addressed, with some exceptions as \cite{17:HONEFLC, 16:GETA:fs22}. In these works, different quantum effects are rigorously considered in the presence of an arbitrary fluctuating-loss medium.

\subsection{Diversity and Multiplexing in Quantum MIMO}
The concept of \ac{MIMO} communication appeared as a breakthrough in \ac{RF} wireless communication systems. The availability of multiple transmission chains can be exploited to either (a) improve the error rate of the communication (diversity gain) by transmitting multiple copies of the signal, or (b) increase the communication rate by superposing multiple data streams together (multiplexing gain) and separating them at the receiver. As it turns out these two types of gains are dual to each other and can be traded-off with each other \cite{ZT:03:IEEE_T_IT}. A direct generalization of classical diversity techniques to quantum communication seems challenging due to the no-cloning theorem \cite{WZ:82:Nat}. Similarly, quantum mechanics fundamentally makes it impossible to create a superposition of unknown quantum states, thus ruling out a direct generalization of classical multiplexing techniques to quantum ones \cite{OGH:16:PRL, Ban:20:PRA}. This severely restricts the efficient use of \ac{MIMO} systems to a select few cases to achieve either diversity or multiplexing gain. One possible scenario for multiplexing gain remains \ac{QKD} where the information to be encoded is classical and thus explicitly known to the sender \cite{KDM:21:CL, KMC:23:TQE}.
Then the fundamental question arises about the utilization of truly quantum \ac{MIMO} communication where the information is truly quantum and the transmitter has access to a single copy of the state to be transmitted. 

\subsection{Interacting Quantum Channels}
Despite significant advancements in genuinely quantum techniques to mitigate noise in quantum communications, such as quantum coherent control and indefinite causal ordering, these paradigms still face numerous challenges. Quantum coherent control, for instance, has not yet been thoroughly explored in the context of CV systems. While the coupling of discrete degrees of freedom into paths in quantum coherent control and the coupling of the control degree of freedom of an indefinite causal order in the quantum switch to the order of DV systems are well-understood, this understanding does not fully extend to Gaussian channels and CV states. Extending these quantum control strategies to the CV domain requires a paradigm shift.
Another challenge is expanding these quantum control strategies beyond the Single Input Single Output (SISO) framework, for which they were originally developed. To achieve this, new control schemes that account for the correlations between the control degrees of freedom and the correlations between channels during propagation need to be investigated. Additionally, these schemes have traditionally been developed for presumably uncorrelated channels, which is generally unrealistic, especially in scenarios involving fiber or free-space propagation where crosstalk is unavoidable. Investigating these schemes in the presence of channel correlations, particularly crosstalk, would represent a significant leap towards realistic application scenarios.
\subsection{The Interplay Between DV and CV Systems}
The interplay between DV and CV systems in quantum communications and computing using bosonic encodings presents several intriguing challenges. The first obstacle is the deterministic generation of non-Gaussian bosonic states that allow efficient interaction between DV and CV systems, and scaling these states to many modes to perfectly map multipartite entanglement from DV to CV bosonic states. Although efficient generation and mapping have been demonstrated in several microwave regime experiments suitable for superconducting platforms, these techniques have yet to be developed for the optical domain required for distributed photonic-based quantum computing.
Moreover, achieving fault-tolerant quantum networks requires the development of robust and efficient hybrid networks that can distribute entangled states over multi-hop nodes with arbitrary topologies. This necessitates the development of quantum repeaters based on bosonic codes, along with the required scheduling algorithms for entanglement distribution. In the realm of quantum computing, continuous-variable cluster states are known to enable fault-tolerant measurement-based quantum computing when used with the GKP encoding of a qubit into a bosonic mode \cite{walshe2021streamlined}. However, the distillation of magic states within this framework and the mapping of logical operations from DV to CV systems still require thorough investigation.

\section{Conclusions}\label{sec:conc}
We reviewed quantum communication technologies, focusing on free-space optical quantum communication and the potential opportunities arising from advancements in \ac{RF} wireless communication systems. To this end, we started with a component-wise review of quantum communication systems. That is, we reviewed (a) different types of quantum sources, (b) the atmospheric artefacts that degrade the quantum optical signals while transmission, and (c) different types of detectors for detecting and measuring quantum signals. In doing so, we also looked at several concepts in quantum communications that are similar to the concepts from mobile communication systems, most notable of which is \ac{MIMO} communications. We identified the opportunities and challenges in exploiting the innovative aspects of mobile communication for currently developing quantum communication systems.

\bibliographystyle{IEEEtran}
\bibliography{FSO}

\begin{thebibliography}{100}
\providecommand{\url}[1]{#1}
\csname url@samestyle\endcsname
\providecommand{\newblock}{\relax}
\providecommand{\bibinfo}[2]{#2}
\providecommand{\BIBentrySTDinterwordspacing}{\spaceskip=0pt\relax}
\providecommand{\BIBentryALTinterwordstretchfactor}{4}
\providecommand{\BIBentryALTinterwordspacing}{\spaceskip=\fontdimen2\font plus
\BIBentryALTinterwordstretchfactor\fontdimen3\font minus \fontdimen4\font\relax}
\providecommand{\BIBforeignlanguage}[2]{{%
\expandafter\ifx\csname l@#1\endcsname\relax
\typeout{** WARNING: IEEEtran.bst: No hyphenation pattern has been}%
\typeout{** loaded for the language `#1'. Using the pattern for}%
\typeout{** the default language instead.}%
\else
\language=\csname l@#1\endcsname
\fi
#2}}
\providecommand{\BIBdecl}{\relax}
\BIBdecl

\bibitem{WEH:18:Sci}
S.~Wehner, D.~Elkouss, and R.~Hanson, ``Quantum internet: {A} vision for the road ahead,'' \emph{Science}, vol. 362, no. 6412, p. eaam9288, Oct. 2018.

\bibitem{CZW:22:CST}
Y.~Cao, Y.~Zhao, Q.~Wang, J.~Zhang, S.~X. Ng, and L.~Hanzo, ``The evolution of quantum key distribution networks: On the road to the {Qinternet},'' \emph{IEEE Communications Surveys \& Tutorials}, vol.~24, no.~2, pp. 839--894, 2022.

\bibitem{WJZ:22:LPR}
S.-H. Wei, B.~Jing, X.-Y. Zhang, J.-Y. Liao, C.-Z. Yuan, B.-Y. Fan, C.~Lyu, D.-L. Zhou, Y.~Wang, G.-W. Deng, H.-Z. Song, D.~Oblak, G.-C. Guo, and Q.~Zhou, ``Towards real-world quantum networks: A review,'' \emph{Laser \& Photonics Reviews}, vol.~16, no.~3, p. 2100219, 2022.

\bibitem{KuP:24:IEEE_Network}
U.~Khalid, J.~ur~Rehman, S.~N. Paing, H.~Jung, T.~Q. Duong, and H.~Shin, ``Quantum network engineering in the {NISQ} age: Principles, missions, and challenges,'' \emph{IEEE Network}, vol.~38, no.~1, pp. 112--123, Jan. 2024.

\bibitem{AuR:24:OJCS}
H.~Al-Hraishawi, J.~ur~Rehman, M.~Razavi, and S.~Chatzinotas, ``Characterizing and utilizing the interplay between quantum technologies and non-terrestrial networks,'' \emph{IEEE Open Journal of the Communications Society}, vol.~5, pp. 1937--1957, 2024.

\bibitem{OFV:09:NPho}
J.~L. O'Brien, A.~Furusawa, and J.~Vučković, ``Photonic quantum technologies,'' \emph{Nature Photonics}, vol.~3, no.~12, pp. 687--695, Dec. 2009.

\bibitem{WSL:20:NPho}
J.~Wang, F.~Sciarrino, A.~Laing, and M.~G. Thompson, ``Integrated photonic quantum technologies,'' \emph{Nature Photonics}, vol.~14, no.~5, pp. 273--284, May 2020.

\bibitem{SP:19:APR}
S.~Slussarenko and G.~J. Pryde, ``{Photonic quantum information processing: A concise review},'' \emph{Applied Physics Reviews}, vol.~6, no.~4, p. 041303, 10 2019.

\bibitem{RAB:02:COMMAG}
T.~Rappaport, A.~Annamalai, R.~Buehrer, and W.~Tranter, ``Wireless communications: past events and a future perspective,'' \emph{IEEE Communications Magazine}, vol.~40, no.~5, pp. 148--161, May 2002.

\bibitem{SA:14:TIT}
I.~Shomorony and A.~S. Avestimehr, ``Degrees of freedom of two-hop wireless networks: Everyone gets the entire cake,'' \emph{IEEE Transactions on Information Theory}, vol.~60, no.~5, pp. 2417--2431, May 2014.

\bibitem{CJ:09:TIT}
V.~R. Cadambe and S.~A. Jafar, ``Degrees of freedom of wireless networks with relays, feedback, cooperation, and full duplex operation,'' \emph{IEEE Transactions on Information Theory}, vol.~55, no.~5, pp. 2334--2344, May 2009.

\bibitem{ZT:03:TIT}
L.~Zheng and D.~Tse, ``Diversity and multiplexing: a fundamental tradeoff in multiple-antenna channels,'' \emph{IEEE Transactions on Information Theory}, vol.~49, no.~5, pp. 1073--1096, May 2003.

\bibitem{JC:13:EC_Book}
G.~C.~C. Jr and J.~B. Cain, \emph{Error-Correction Coding for Digital Communications}.\hskip 1em plus 0.5em minus 0.4em\relax Springer Science \& Business Media, Jun. 2013.

\bibitem{WZ:82:Nat}
W.~K. Wootters and W.~H. Zurek, ``A single quantum cannot be cloned,'' \emph{Nature}, vol. 299, no. 5886, pp. 802--803, Oct. 1982.

\bibitem{Wer:98:PRA}
R.~F. Werner, ``Optimal cloning of pure states,'' \emph{Physical Review A}, vol.~58, no.~3, pp. 1827--1832, Sep. 1998.

\bibitem{GM:97:PRL}
N.~Gisin and S.~Massar, ``Optimal quantum cloning machines,'' \emph{Physical Review Letters}, vol.~79, no.~11, pp. 2153--2156, Sep. 1997.

\bibitem{BEM:98:PRL}
D.~Bruss, A.~Ekert, and C.~Macchiavello, ``Optimal universal quantum cloning and state estimation,'' \emph{Physical Review Letters}, vol.~81, no.~12, pp. 2598--2601, Sep. 1998.

\bibitem{SIG:05:RMP}
V.~Scarani, S.~Iblisdir, N.~Gisin, and A.~Acín, ``Quantum cloning,'' \emph{Reviews of Modern Physics}, vol.~77, no.~4, pp. 1225--1256, Nov. 2005.

\bibitem{Bra:98:Nat}
S.~L. Braunstein, ``Quantum error correction for communication with linear optics,'' \emph{Nature}, vol. 394, no. 6688, pp. 47--49, Jul. 1998.

\bibitem{Rof:19:CP}
J.~Roffe, ``Quantum error correction: an introductory guide,'' \emph{Contemporary Physics}, vol.~60, no.~3, pp. 226--245, Jul. 2019.

\bibitem{La:20:Book}
G.~G. La~Guardia, ``Quantum error-correcting codes,'' in \emph{Quantum Error Correction: Symmetric, Asymmetric, Synchronizable, and Convolutional Codes}, G.~G. La~Guardia, Ed.\hskip 1em plus 0.5em minus 0.4em\relax Cham: Springer International Publishing, 2020, pp. 25--41.

\bibitem{ALS:10:LPR}
U.~Andersen, G.~Leuchs, and C.~Silberhorn, ``Continuous-variable quantum information processing,'' \emph{Laser \& Photonics Reviews}, vol.~4, no.~3, pp. 337--354, 2010.

\bibitem{LDM:02:JSAC}
V.~Lottici, A.~D'Andrea, and U.~Mengali, ``Channel estimation for ultra-wideband communications,'' \emph{IEEE Journal on Selected Areas in Communications}, vol.~20, no.~9, pp. 1638--1645, Dec. 2002.

\bibitem{CEP:02:TB}
S.~Coleri, M.~Ergen, A.~Puri, and A.~Bahai, ``Channel estimation techniques based on pilot arrangement in {OFDM} systems,'' \emph{IEEE Transactions on Broadcasting}, vol.~48, no.~3, pp. 223--229, Sep. 2002.

\bibitem{AB:01:WCMC}
H.~Arslan and G.~E. Bottomley, ``Channel estimation in narrowband wireless communication systems,'' \emph{Wireless Communications and Mobile Computing}, vol.~1, no.~2, pp. 201--219, 2001.

\bibitem{NZG:17:IET_Com}
S.~Ni, J.~Zhao, and Y.~Gong, ``Optimal pilot design in massive {MIMO} systems based on channel estimation,'' \emph{IET Communications}, vol.~11, no.~7, pp. 975--984, 2017.

\bibitem{SCC:20:TCOM}
K.~Shen, H.~V. Cheng, X.~Chen, Y.~C. Eldar, and W.~Yu, ``Enhanced channel estimation in massive {MIMO} via coordinated pilot design,'' \emph{IEEE Transactions on Communications}, vol.~68, no.~11, pp. 6872--6885, Nov. 2020.

\bibitem{MAK:17:TWC}
R.~Mohammadian, A.~Amini, and B.~H. Khalaj, ``Deterministic pilot design for sparse channelestimation in {MISO}/multi-user {OFDM} systems,'' \emph{IEEE Transactions on Wireless Communications}, vol.~16, no.~1, pp. 129--140, Jan. 2017.

\bibitem{JC:22:JoPA}
R.~Jonsson and R.~D. Candia, ``Gaussian quantum estimation of the loss parameter in a thermal environment,'' \emph{Journal of Physics A: Mathematical and Theoretical}, vol.~55, no.~38, p. 385301, Aug. 2022.

\bibitem{MRL:08:PRA}
M.~Mohseni, A.~T. Rezakhani, and D.~A. Lidar, ``Quantum-process tomography: {Resource} analysis of different strategies,'' \emph{Physical Review A}, vol.~77, no.~3, p. 032322, Mar. 2008.

\bibitem{RSM:11:NJP}
S.~Rahimi-Keshari, A.~Scherer, A.~Mann, A.~T. Rezakhani, A.~I. Lvovsky, and B.~C. Sanders, ``Quantum process tomography with coherent states,'' \emph{New Journal of Physics}, vol.~13, no.~1, p. 013006, Jan. 2011.

\bibitem{uAD:24:TCOM}
J.~ur~Rehman, H.~Al-Hraishawi, T.~Q. Duong, S.~Chatzinotas, and H.~Shin, ``On estimating time-varying {Pauli} noise,'' \emph{IEEE Transactions on Communications}, vol.~72, no.~4, pp. 2079--2089, Apr. 2024.

\bibitem{EFC:21:npjQI}
J.~Etxezarreta~Martinez, P.~Fuentes, P.~Crespo, and J.~{Garcia-Frias}, ``Time-varying quantum channel models for superconducting qubits,'' \emph{npj Quantum Inf.}, vol.~7, no.~1, pp. 1--10, Jul. 2021.

\bibitem{EFd:23:PRR}
J.~Etxezarreta~Martinez, P.~Fuentes, A.~deMarti iOlius, J.~Garcia-Frias, J.~R. Fonollosa, and P.~M. Crespo, ``Multiqubit time-varying quantum channels for {NISQ}-era superconducting quantum processors,'' \emph{Physical Review Research}, vol.~5, no.~3, p. 033055, Jul. 2023.

\bibitem{NPR:17:NP}
B.~Ndagano, B.~Perez-Garcia, F.~S. Roux, M.~McLaren, C.~Rosales-Guzman, Y.~Zhang, O.~Mouane, R.~I. Hernandez-Aranda, T.~Konrad, and A.~Forbes, ``Characterizing quantum channels with non-separable states of classical light,'' \emph{Nature Physics}, vol.~13, no.~4, pp. 397--402, Apr. 2017.

\bibitem{XLM:12:Conf}
Y.~Xie, J.~Li, R.~Malaney, and J.~Yuan, ``Channel identification and its impact on quantum {LDPC} code performance,'' in \emph{2012 {Australian} {Communications} {Theory} {Workshop} ({AusCTW})}, Jan. 2012, pp. 140--144.

\bibitem{MFC:20:Conf}
J.~E. Martinez, P.~Fuentes, P.~M. Crespo, and J.~Garcia-Frías, ``Pauli channel online estimation protocol for quantum turbo codes,'' in \emph{2020 {IEEE} {International} {Conference} on {Quantum} {Computing} and {Engineering} ({QCE})}, Oct. 2020, pp. 102--108.

\bibitem{uS:22:Conf}
J.~ur~Rehman and H.~Shin, ``Simultaneous communication and parameter estimation of {Pauli} channels,'' in \emph{{ICC} 2022 - {IEEE} {International} {Conference} on {Communications}}, May 2022, pp. 648--653.

\bibitem{NC:10:book}
M.~A. Nielsen and I.~L. Chuang, \emph{Quantum Computation and Quantum Information: 10th Anniversary Edition}.\hskip 1em plus 0.5em minus 0.4em\relax Cambridge University Press, 2010.

\bibitem{Wat:18:book}
\BIBentryALTinterwordspacing
J.~Watrous, \emph{The Theory of Quantum Information}.\hskip 1em plus 0.5em minus 0.4em\relax Cambridge: Cambridge University Press, 2018. [Online]. Available: \url{https://www.cambridge.org/core/books/theory-of-quantum-information/AE4AA5638F808D2CFEB070C55431D897}
\BIBentrySTDinterwordspacing

\bibitem{WPG:12:RMP}
C.~Weedbrook, S.~Pirandola, R.~Garcia-Patron, N.~J. Cerf, T.~C. Ralph, J.~H. Shapiro, and S.~Lloyd, ``Gaussian quantum information,'' \emph{Reviews of Modern Physics}, vol.~84, no.~2, pp. 621--669, May 2012.

\bibitem{Bv:05:RMP}
S.~L. Braunstein and P.~van Loock, ``Quantum information with continuous variables,'' \emph{Rev. Mod. Phys.}, vol.~77, pp. 513--577, Jun 2005.

\bibitem{genoni2010quantifying}
M.~G. Genoni and M.~G. Paris, ``Quantifying non-gaussianity for quantum information,'' \emph{Physical Review A}, vol.~82, no.~5, p. 052341, 2010.

\bibitem{Wig:32:PR}
E.~Wigner, ``On the quantum correction for thermodynamic equilibrium,'' \emph{Phys. Rev.}, vol.~40, pp. 749--759, Jun 1932.

\bibitem{harper2024crosstalk}
B.~Harper, B.~Tonekaboni, B.~Goldozian, M.~Sevior, and M.~Usman, ``Crosstalk attacks and defence in a shared quantum computing environment,'' \emph{arXiv preprint arXiv:2402.02753}, 2024.

\bibitem{seo2021mitigation}
S.~Seo, J.~Seong, and J.~Bae, ``Mitigation of crosstalk errors in a quantum measurement and its applications,'' \emph{arXiv preprint arXiv:2112.10651}, 2021.

\bibitem{bravyi2021mitigating}
S.~Bravyi, S.~Sheldon, A.~Kandala, D.~C. Mckay, and J.~M. Gambetta, ``Mitigating measurement errors in multiqubit experiments,'' \emph{Physical Review A}, vol. 103, no.~4, p. 042605, 2021.

\bibitem{white2020demonstration}
G.~A. White, C.~D. Hill, F.~A. Pollock, L.~C. Hollenberg, and K.~Modi, ``Demonstration of non-markovian process characterisation and control on a quantum processor,'' \emph{Nature Communications}, vol.~11, no.~1, p. 6301, 2020.

\bibitem{torlai2023quantum}
G.~Torlai, C.~J. Wood, A.~Acharya, G.~Carleo, J.~Carrasquilla, and L.~Aolita, ``Quantum process tomography with unsupervised learning and tensor networks,'' \emph{Nature Communications}, vol.~14, no.~1, p. 2858, 2023.

\bibitem{filippov2023scalable}
S.~Filippov, M.~Leahy, M.~A. Rossi, and G.~Garc{\'\i}a-P{\'e}rez, ``Scalable tensor-network error mitigation for near-term quantum computing,'' \emph{arXiv preprint arXiv:2307.11740}, 2023.

\bibitem{mangini2024tensor}
S.~Mangini, M.~Cattaneo, D.~Cavalcanti, S.~Filippov, M.~A. Rossi, and G.~Garc{\'\i}a-P{\'e}rez, ``Tensor network noise characterization for near-term quantum computers,'' \emph{arXiv preprint arXiv:2402.08556}, 2024.

\bibitem{Rottondi2019}
C.~Rottondi, P.~Martelli, P.~Boffi, L.~Barletta, and M.~Tornatore, ``Crosstalk-aware core and spectrum assignment in a multicore optical link with flexible grid,'' \emph{IEEE Transactions on Communications}, vol.~67, no.~3, pp. 2144--2156, 2019.

\bibitem{Hayashi2019}
T.~Hayashi, T.~Nagashima, T.~Nakanishi, T.~Morishima, R.~Kawawada, A.~Mecozzi, and C.~Antonelli, ``Field-deployed multi-core fiber testbed,'' in \emph{2019 24th OptoElectronics and Communications Conference (OECC) and 2019 International Conference on Photonics in Switching and Computing (PSC)}, 2019, pp. 1--3.

\bibitem{SAKAMOTO20178}
\BIBentryALTinterwordspacing
T.~Sakamoto, T.~Mori, M.~Wada, T.~Yamamoto, F.~Yamamoto, and K.~Nakajima, ``Strongly-coupled multi-core fiber and its optical characteristics for mimo transmission systems,'' \emph{Optical Fiber Technology}, vol.~35, pp. 8--18, 2017, next Generation Multiplexing Schemes in Fiber-based Systems. [Online]. Available: \url{https://www.sciencedirect.com/science/article/pii/S1068520016300608}
\BIBentrySTDinterwordspacing

\bibitem{9006884}
R.~S. Luís, G.~Rademacher, B.~J. Puttnam, D.~Semrau, R.~I. Killey, P.~Bayvel, Y.~Awaji, and H.~Furukawa, ``Crosstalk impact on the performance of wideband multicore-fiber transmission systems,'' \emph{IEEE Journal of Selected Topics in Quantum Electronics}, vol.~26, no.~4, pp. 1--9, 2020.

\bibitem{Moghaddam2021}
E.~E. Moghaddam, H.~Beyranvand, and J.~A. Salehi, ``Resource allocation in space division multiplexed elastic optical networks secured with quantum key distribution,'' \emph{IEEE Journal on Selected Areas in Communications}, vol.~39, no.~9, pp. 2688--2700, 2021.

\bibitem{Urena:19}
\BIBentryALTinterwordspacing
M.~U. {n}a, I.~Gasulla, F.~J. Fraile, and J.~Capmany, ``Modeling optical fiber space division multiplexed quantum key distribution systems,'' \emph{Opt. Express}, vol.~27, no.~5, pp. 7047--7063, Mar 2019. [Online]. Available: \url{https://opg.optica.org/oe/abstract.cfm?URI=oe-27-5-7047}
\BIBentrySTDinterwordspacing

\bibitem{Mujtaba2024}
M.~Zahidy, D.~Ribezzo, C.~{De Lazzari}, I.~Vagniluca, N.~Biagi, R.~M{\"u}ller, T.~Occhipinti, L.~Oxenl{\o}we, M.~Galili, T.~Hayashi, D.~Cassioli, A.~Mecozzi, C.~Antonelli, A.~Zavatta, and D.~Bacco, ``\BIBforeignlanguage{English}{Practical high-dimensional quantum key distribution protocol over deployed multicore fiber},'' \emph{\BIBforeignlanguage{English}{Nature Communications}}, vol.~15, no.~1, 2024, publisher Copyright: {\textcopyright} The Author(s) 2024.

\bibitem{Eriksson2019}
T.~A. Eriksson, B.~J. Puttnam, G.~Rademacher, R.~S. Luís, M.~Takeoka, Y.~Awaji, M.~Sasaki, and N.~Wada, ``Inter-core crosstalk impact of classical channels on cv-qkd in multicore fiber transmission,'' in \emph{2019 Optical Fiber Communications Conference and Exhibition (OFC)}, 2019, pp. 1--3.

\bibitem{TANDON2023129483}
\BIBentryALTinterwordspacing
P.~Tandon and A.~Zakharian, ``Impact of multicore fiber (mcf) opticals, cross-talk, radiative leakage loss, splice loss and propagation configuration on the system transmission performance,'' \emph{Optics Communications}, vol. 539, p. 129483, 2023. [Online]. Available: \url{https://www.sciencedirect.com/science/article/pii/S0030401823002304}
\BIBentrySTDinterwordspacing

\bibitem{8336681}
H.~Yuan, M.~Furdek, A.~Muhammad, A.~Saljoghei, L.~Wosinska, and G.~Zervas, ``Space-division multiplexing in data center networks: on multi-core fiber solutions and crosstalk-suppressed resource allocation,'' \emph{Journal of Optical Communications and Networking}, vol.~10, no.~4, pp. 272--288, 2018.

\bibitem{9893163}
L.~Sun, G.~N. Liu, Y.~Cai, J.~Du, Z.~He, Z.~Li, C.~Lu, and G.~Shen, ``Mitigating the inter-core crosstalk of multicore fiber transmission by orthogonal filtering,'' \emph{IEEE Photonics Technology Letters}, vol.~34, no.~24, pp. 1373--1376, 2022.

\bibitem{Hayashi:11}
\BIBentryALTinterwordspacing
T.~Hayashi, T.~Taru, O.~Shimakawa, T.~Sasaki, and E.~Sasaoka, ``Design and fabrication of ultra-low crosstalk and low-loss multi-core fiber,'' \emph{Opt. Express}, vol.~19, no.~17, pp. 16\,576--16\,592, Aug 2011. [Online]. Available: \url{https://opg.optica.org/oe/abstract.cfm?URI=oe-19-17-16576}
\BIBentrySTDinterwordspacing

\bibitem{5875695}
K.~Takenaga, Y.~Arakawa, S.~Tanigawa, N.~Guan, S.~Matsuo, K.~Saitoh, and M.~Koshiba, ``Reduction of crosstalk by trench-assisted multi-core fiber,'' in \emph{2011 Optical Fiber Communication Conference and Exposition and the National Fiber Optic Engineers Conference}, 2011, pp. 1--3.

\bibitem{RN110}
\BIBentryALTinterwordspacing
B.~Jaramillo~Ávila, J.~M. Torres, R.~d.~J. León-Montiel, and B.~M. Rodríguez-Lara, ``Optimal crosstalk suppression in multicore fibers,'' \emph{Scientific Reports}, vol.~9, no.~1, p. 15737, 2019. [Online]. Available: \url{https://doi.org/10.1038/s41598-019-51854-x}
\BIBentrySTDinterwordspacing

\bibitem{2016OptEn..55k1607G}
M.~C. {G{\"o}k{\c{c}}e}, Y.~{Baykal}, and M.~{Uysal}, ``{Performance analysis of multiple-input multiple-output free-space optical systems with partially coherent Gaussian beams and finite-sized detectors},'' \emph{Optical Engineering}, vol.~55, p. 111607, Nov. 2016.

\bibitem{Gokce:16}
\BIBentryALTinterwordspacing
M.~C. G\"{o}k\c{c}e, Y.~Baykal, and M.~Uysal, ``Aperture averaging in multiple-input single-output free-space optical systems using partially coherent radial array beams,'' \emph{J. Opt. Soc. Am. A}, vol.~33, no.~6, pp. 1041--1048, Jun 2016. [Online]. Available: \url{https://opg.optica.org/josaa/abstract.cfm?URI=josaa-33-6-1041}
\BIBentrySTDinterwordspacing

\bibitem{2016WRCM...26..642G}
M.~C. {G{\"o}k{\c{c}}e}, Y.~{Baykal}, and M.~{Uysal}, ``{Bit error rate analysis of MISO FSO systems},'' \emph{Waves in Random and Complex Media}, vol.~26, no.~4, pp. 642--649, Oct. 2016.

\bibitem{2023AnP...53500232N}
H.~{Nabil}, A.~{Balhamri}, M.~{Bayraktar}, and A.~{Belafhal}, ``{Propagation Characteristics of a Partially Coherent Gaussian Schell‑model Array Vortex Beam in the Joint Turbulence Effect of a Jet Engine and Atmosphere},'' \emph{Annalen der Physik}, vol. 535, no.~10, p. 2300232, Oct. 2023.

\bibitem{nadagano2017}
B.~{Ndagano}, B.~{Perez-Garcia}, F.~S. {Roux}, M.~{McLaren}, C.~{Rosales-Guzman}, Y.~{Zhang}, O.~{Mouane}, R.~I. {Hernandez-Aranda}, T.~{Konrad}, and A.~{Forbes}, ``{Characterizing quantum channels with non-separable states of classical light},'' \emph{Nature Physics}, vol.~13, no.~4, pp. 397--402, Apr. 2017.

\bibitem{Semenov2009}
\BIBentryALTinterwordspacing
A.~A. Semenov and W.~Vogel, ``Quantum light in the turbulent atmosphere,'' \emph{Phys. Rev. A}, vol.~80, p. 021802, Aug 2009. [Online]. Available: \url{https://link.aps.org/doi/10.1103/PhysRevA.80.021802}
\BIBentrySTDinterwordspacing

\bibitem{PhysRevLett.131.233601}
\BIBentryALTinterwordspacing
A.~Khodadad~Kashi, L.~Caspani, and M.~Kues, ``Spectral hong-ou-mandel effect between a heralded single-photon state and a thermal field: Multiphoton contamination and the nonclassicality threshold,'' \emph{Phys. Rev. Lett.}, vol. 131, p. 233601, Dec 2023. [Online]. Available: \url{https://link.aps.org/doi/10.1103/PhysRevLett.131.233601}
\BIBentrySTDinterwordspacing

\bibitem{Vasylyev2016}
\BIBentryALTinterwordspacing
D.~Vasylyev, A.~A. Semenov, and W.~Vogel, ``Atmospheric quantum channels with weak and strong turbulence,'' \emph{Phys. Rev. Lett.}, vol. 117, p. 090501, Aug 2016. [Online]. Available: \url{https://link.aps.org/doi/10.1103/PhysRevLett.117.090501}
\BIBentrySTDinterwordspacing

\bibitem{andrews}
L.~C. Andrews and R.~L. Philips, \emph{Laser beam propagation through random media}.\hskip 1em plus 0.5em minus 0.4em\relax SPIE Press, 2005.

\bibitem{Vasylyev2012}
\BIBentryALTinterwordspacing
D.~Y. Vasylyev, A.~A. Semenov, and W.~Vogel, ``Toward global quantum communication: Beam wandering preserves nonclassicality,'' \emph{Phys. Rev. Lett.}, vol. 108, p. 220501, Jun 2012. [Online]. Available: \url{https://link.aps.org/doi/10.1103/PhysRevLett.108.220501}
\BIBentrySTDinterwordspacing

\bibitem{Yuceer2012}
\BIBentryALTinterwordspacing
M.~Yüceer and H.~T. Eyyuboglu, ``Laguerre-gaussian beam scintillation on slant paths,'' \emph{Appl. Phys. B}, vol. 109, pp. 311--316, 2012. [Online]. Available: \url{https://link.springer.com/article/10.1007/s00340-012-5186-3}
\BIBentrySTDinterwordspacing

\bibitem{Wang2018}
\BIBentryALTinterwordspacing
S.~Wang, P.~Huang, T.~Wang, and G.~Zeng, ``Atmospheric effects on continuous-variable quantum key distribution,'' \emph{New J. Phys.}, vol.~20, p. 083037, 2018. [Online]. Available: \url{https://iopscience.iop.org/article/10.1088/1367-2630/aad9c4}
\BIBentrySTDinterwordspacing

\bibitem{Chai2019}
\BIBentryALTinterwordspacing
G.~Chai, Z.~Cao, W.~Liu, S.~Wang, P.~Huang, and G.~Zeng, ``Parameter estimation of atmospheric continuous-variable quantum key distribution,'' \emph{Phys. Rev. A}, vol.~99, p. 032326, Mar 2019. [Online]. Available: \url{https://link.aps.org/doi/10.1103/PhysRevA.99.032326}
\BIBentrySTDinterwordspacing

\bibitem{Villaseñor2020}
E.~Villaseñor, R.~Malaney, K.~A. Mudge, and K.~J. Grant, ``Atmospheric effects on satellite-to-ground quantum key distribution using coherent states,'' in \emph{GLOBECOM 2020 - 2020 IEEE Global Communications Conference}, 2020, pp. 1--6.

\bibitem{Zhang2016}
\BIBentryALTinterwordspacing
Y.~Zhang, S.~Prabhakar, A.~H. Ibrahim, F.~S. Roux, A.~Forbes, and T.~Konrad, ``Experimentally observed decay of high-dimensional entanglement through turbulence,'' \emph{Phys. Rev. A}, vol.~94, p. 032310, Sep 2016. [Online]. Available: \url{https://link.aps.org/doi/10.1103/PhysRevA.94.032310}
\BIBentrySTDinterwordspacing

\bibitem{Vallone2014}
\BIBentryALTinterwordspacing
G.~Vallone, V.~D'Ambrosio, A.~Sponselli, S.~Slussarenko, L.~Marrucci, F.~Sciarrino, and P.~Villoresi, ``Free-space quantum key distribution by rotation-invariant twisted photons,'' \emph{Phys. Rev. Lett.}, vol. 113, p. 060503, Aug 2014. [Online]. Available: \url{https://link.aps.org/doi/10.1103/PhysRevLett.113.060503}
\BIBentrySTDinterwordspacing

\bibitem{Ursin2007}
R.~{Ursin}, F.~{Tiefenbacher}, T.~{Schmitt-Manderbach}, H.~{Weier}, T.~{Scheidl}, M.~{Lindenthal}, B.~{Blauensteiner}, T.~{Jennewein}, J.~{Perdigues}, P.~{Trojek}, B.~{{\"O}mer}, M.~{F{\"u}rst}, M.~{Meyenburg}, J.~{Rarity}, Z.~{Sodnik}, C.~{Barbieri}, H.~{Weinfurter}, and A.~{Zeilinger}, ``{Entanglement-based quantum communication over 144km},'' \emph{Nature Physics}, vol.~3, no.~7, pp. 481--486, Jul. 2007.

\bibitem{Gruneisen2021}
\BIBentryALTinterwordspacing
M.~T. Gruneisen, M.~L. Eickhoff, S.~C. Newey, K.~E. Stoltenberg, J.~F. Morris, M.~Bareian, M.~A. Harris, D.~W. Oesch, M.~D. Oliker, M.~B. Flanagan, B.~T. Kay, J.~D. Schiller, and R.~N. Lanning, ``Adaptive-optics-enabled quantum communication: A technique for daytime space-to-earth links,'' \emph{Phys. Rev. Appl.}, vol.~16, p. 014067, Jul 2021. [Online]. Available: \url{https://link.aps.org/doi/10.1103/PhysRevApplied.16.014067}
\BIBentrySTDinterwordspacing

\bibitem{Marulanda_2024}
\BIBentryALTinterwordspacing
V.~M. Acosta, D.~Dequal, M.~Schiavon, A.~Montmerle-Bonnefois, C.~B. Lim, J.-M. Conan, and E.~Diamanti, ``Analysis of satellite-to-ground quantum key distribution with adaptive optics,'' \emph{New Journal of Physics}, vol.~26, no.~2, p. 023039, feb 2024. [Online]. Available: \url{https://dx.doi.org/10.1088/1367-2630/ad231c}
\BIBentrySTDinterwordspacing

\bibitem{Wang2019}
Y.~{Wang}, X.~{Wu}, L.~{Zhang}, D.~{Huang}, Q.~{Liao}, and Y.~{Guo}, ``{Performance improvement of free-space continuous-variable quantum key distribution with an adaptive optics unit},'' \emph{Quantum Information Processing}, vol.~18, no.~8, p. 251, Aug. 2019.

\bibitem{Mikhael2024}
\BIBentryALTinterwordspacing
M.~Sayat, M.~Birch, O.~Thearle, M.~Copeland, E.~Jager, F.~Bennet, P.~K. Lam, N.~Rattenbury, and J.~Cater, ``{Experimental effects of turbulence on coherent states in a free-space channel using adaptive optics for continuous variable quantum key distribution},'' in \emph{Quantum Computing, Communication, and Simulation IV}, P.~R. Hemmer and A.~L. Migdall, Eds., vol. 12911, International Society for Optics and Photonics.\hskip 1em plus 0.5em minus 0.4em\relax SPIE, 2024, p. 1291107. [Online]. Available: \url{https://doi.org/10.1117/12.3000049}
\BIBentrySTDinterwordspacing

\bibitem{Tao:21}
\BIBentryALTinterwordspacing
Z.~Tao, Y.~Ren, A.~Abdukirim, S.~Liu, and R.~Rao, ``Mitigating the effect of atmospheric turbulence on orbital angular momentum-based quantum key distribution using real-time adaptive optics with phase unwrapping,'' \emph{Opt. Express}, vol.~29, no.~20, pp. 31\,078--31\,098, Sep 2021. [Online]. Available: \url{https://opg.optica.org/oe/abstract.cfm?URI=oe-29-20-31078}
\BIBentrySTDinterwordspacing

\bibitem{Li:17}
\BIBentryALTinterwordspacing
P.~Li, Y.~Zhang, S.~Liu, H.~Cheng, L.~Han, D.~Wu, and J.~Zhao, ``Generation and self-healing of vector bessel-gauss beams with variant state of polarizations upon propagation,'' \emph{Opt. Express}, vol.~25, no.~5, pp. 5821--5831, Mar 2017. [Online]. Available: \url{https://opg.optica.org/oe/abstract.cfm?URI=oe-25-5-5821}
\BIBentrySTDinterwordspacing

\bibitem{Otte2018}
\BIBentryALTinterwordspacing
E.~Otte, I.~Nape, C.~Rosales-Guzm\'an, A.~Vall\'es, C.~Denz, and A.~Forbes, ``Recovery of nonseparability in self-healing vector bessel beams,'' \emph{Phys. Rev. A}, vol.~98, p. 053818, Nov 2018. [Online]. Available: \url{https://link.aps.org/doi/10.1103/PhysRevA.98.053818}
\BIBentrySTDinterwordspacing

\bibitem{Ahmed2016}
N.~Ahmed, Z.~Zhao, L.~Li, H.~Huang, M.~P.~J. Lavery, P.~Liao, Y.~Yan, Z.~Wang, G.~Xie, Y.~Ren, A.~Almaiman, A.~J. Willner, S.~Ashrafi, A.~F. Molisch, M.~Tur, and A.~E. Willner, ``Mode-division-multiplexing of multiple bessel-gaussian beams carrying orbital-angular-momentum for obstruction-tolerant free-space optical and millimetre-wave communication links,'' \emph{Scientific reports}, vol.~6, p. 22082, March 2016.

\bibitem{Nape:18}
\BIBentryALTinterwordspacing
I.~Nape, E.~Otte, A.~Vall\'{e}s, C.~Rosales-Guzm\'{a}n, F.~Cardano, C.~Denz, and A.~Forbes, ``Self-healing high-dimensional quantum key distribution using hybrid spin-orbit bessel states,'' \emph{Opt. Express}, vol.~26, no.~21, pp. 26\,946--26\,960, Oct 2018. [Online]. Available: \url{https://opg.optica.org/oe/abstract.cfm?URI=oe-26-21-26946}
\BIBentrySTDinterwordspacing

\bibitem{He:21}
\BIBentryALTinterwordspacing
W.~He, S.~Guha, J.~H. Shapiro, and B.~A. Bash, ``Performance analysis of free-space quantum key distribution using multiple spatial modes,'' \emph{Opt. Express}, vol.~29, no.~13, pp. 19\,305--19\,318, Jun 2021. [Online]. Available: \url{https://opg.optica.org/oe/abstract.cfm?URI=oe-29-13-19305}
\BIBentrySTDinterwordspacing

\bibitem{honjo2004differential}
T.~Honjo, K.~Inoue, and H.~Takahashi, ``Differential-phase-shift quantum key distribution experiment with a planar light-wave circuit mach--zehnder interferometer,'' \emph{Optics letters}, vol.~29, no.~23, pp. 2797--2799, 2004.

\bibitem{sibson2017chip}
P.~Sibson, C.~Erven, M.~Godfrey, S.~Miki, T.~Yamashita, M.~Fujiwara, M.~Sasaki, H.~Terai, M.~G. Tanner, C.~M. Natarajan \emph{et~al.}, ``Chip-based quantum key distribution,'' \emph{Nature communications}, vol.~8, no.~1, p. 13984, 2017.

\bibitem{semenenko2019interference}
H.~Semenenko, P.~Sibson, M.~G. Thompson, and C.~Erven, ``Interference between independent photonic integrated devices for quantum key distribution,'' \emph{Optics letters}, vol.~44, no.~2, pp. 275--278, 2019.

\bibitem{agnesi2019hong}
C.~Agnesi, B.~Da~Lio, D.~Cozzolino, L.~Cardi, B.~B. Bakir, K.~Hassan, A.~Della~Frera, A.~Ruggeri, A.~Giudice, G.~Vallone \emph{et~al.}, ``Hong--ou--mandel interference between independent iii--v on silicon waveguide integrated lasers,'' \emph{Optics letters}, vol.~44, no.~2, pp. 271--274, 2019.

\bibitem{arrazola2021quantum}
J.~M. Arrazola, V.~Bergholm, K.~Br{\'a}dler, T.~R. Bromley, M.~J. Collins, I.~Dhand, A.~Fumagalli, T.~Gerrits, A.~Goussev, L.~G. Helt \emph{et~al.}, ``Quantum circuits with many photons on a programmable nanophotonic chip,'' \emph{Nature}, vol. 591, no. 7848, pp. 54--60, 2021.

\bibitem{alexander2024manufacturable}
K.~Alexander, A.~Bahgat, A.~Benyamini, D.~Black, D.~Bonneau, S.~Burgos, B.~Burridge, G.~Campbell, G.~Catalano, A.~Ceballos \emph{et~al.}, ``A manufacturable platform for photonic quantum computing,'' \emph{arXiv preprint arXiv:2404.17570}, 2024.

\bibitem{maring2024versatile}
N.~Maring, A.~Fyrillas, M.~Pont, E.~Ivanov, P.~Stepanov, N.~Margaria, W.~Hease, A.~Pishchagin, A.~Lema{\^\i}tre, I.~Sagnes \emph{et~al.}, ``A versatile single-photon-based quantum computing platform,'' \emph{Nature Photonics}, vol.~18, no.~6, pp. 603--609, 2024.

\bibitem{PPA:09:NJP}
M.~Peev, C.~Pacher, R.~Alléaume, C.~Barreiro, J.~Bouda, W.~Boxleitner, T.~Debuisschert, E.~Diamanti, M.~Dianati, J.~F. Dynes, S.~Fasel, S.~Fossier, M.~Fürst, J.-D. Gautier, O.~Gay, N.~Gisin, P.~Grangier, A.~Happe, Y.~Hasani, M.~Hentschel, H.~Hübel, G.~Humer, T.~Länger, M.~Legré, R.~Lieger, J.~Lodewyck, T.~Lorünser, N.~Lütkenhaus, A.~Marhold, T.~Matyus, O.~Maurhart, L.~Monat, S.~Nauerth, J.-B. Page, A.~Poppe, E.~Querasser, G.~Ribordy, S.~Robyr, L.~Salvail, A.~W. Sharpe, A.~J. Shields, D.~Stucki, M.~Suda, C.~Tamas, T.~Themel, R.~T. Thew, Y.~Thoma, A.~Treiber, P.~Trinkler, R.~Tualle-Brouri, F.~Vannel, N.~Walenta, H.~Weier, H.~Weinfurter, I.~Wimberger, Z.~L. Yuan, H.~Zbinden, and A.~Zeilinger, ``The {SECOQC} quantum key distribution network in {Vienna},'' \emph{New Journal of Physics}, vol.~11, no.~7, p. 075001, Jul. 2009.

\bibitem{SFI:11:OE}
M.~Sasaki, M.~Fujiwara, H.~Ishizuka, W.~Klaus, K.~Wakui, M.~Takeoka, S.~Miki, T.~Yamashita, Z.~Wang, A.~Tanaka, K.~Yoshino, Y.~Nambu, S.~Takahashi, A.~Tajima, A.~Tomita, T.~Domeki, T.~Hasegawa, Y.~Sakai, H.~Kobayashi, T.~Asai, K.~Shimizu, T.~Tokura, T.~Tsurumaru, M.~Matsui, T.~Honjo, K.~Tamaki, H.~Takesue, Y.~Tokura, J.~F. Dynes, A.~R. Dixon, A.~W. Sharpe, Z.~L. Yuan, A.~J. Shields, S.~Uchikoga, M.~Legré, S.~Robyr, P.~Trinkler, L.~Monat, J.-B. Page, G.~Ribordy, A.~Poppe, A.~Allacher, O.~Maurhart, T.~Länger, M.~Peev, and A.~Zeilinger, ``Field test of quantum key distribution in the {Tokyo} {QKD} {Network},'' \emph{Optics Express}, vol.~19, no.~11, pp. 10\,387--10\,409, May 2011.

\bibitem{SLB:11:NJP}
D.~Stucki, M.~Legré, F.~Buntschu, B.~Clausen, N.~Felber, N.~Gisin, L.~Henzen, P.~Junod, G.~Litzistorf, P.~Monbaron, L.~Monat, J.-B. Page, D.~Perroud, G.~Ribordy, A.~Rochas, S.~Robyr, J.~Tavares, R.~Thew, P.~Trinkler, S.~Ventura, R.~Voirol, N.~Walenta, and H.~Zbinden, ``Long-term performance of the {SwissQuantum} quantum key distribution network in a field environment,'' \emph{New Journal of Physics}, vol.~13, no.~12, p. 123001, Dec. 2011.

\bibitem{TYZ:16:PRX}
Y.-L. Tang, H.-L. Yin, Q.~Zhao, H.~Liu, X.-X. Sun, M.-Q. Huang, W.-J. Zhang, S.-J. Chen, L.~Zhang, L.-X. You, Z.~Wang, Y.~Liu, C.-Y. Lu, X.~Jiang, X.~Ma, Q.~Zhang, T.-Y. Chen, and J.-W. Pan, ``Measurement-device-independent quantum key distribution over untrustful metropolitan network,'' \emph{Physical Review X}, vol.~6, no.~1, p. 011024, Mar. 2016.

\bibitem{CLL:09:OE}
T.-Y. Chen, H.~Liang, Y.~Liu, W.-Q. Cai, L.~Ju, W.-Y. Liu, J.~Wang, H.~Yin, K.~Chen, Z.-B. Chen, C.-Z. Peng, and J.-W. Pan, ``Field test of a practical secure communication network with decoy-state quantum cryptography,'' \emph{Optics Express}, vol.~17, no.~8, pp. 6540--6549, Apr. 2009.

\bibitem{XCW:09:CSB}
F.~Xu, W.~Chen, S.~Wang, Z.~Yin, Y.~Zhang, Y.~Liu, Z.~Zhou, Y.~Zhao, H.~Li, D.~Liu, Z.~Han, and G.~Guo, ``Field experiment on a robust hierarchical metropolitan quantum cryptography network,'' \emph{Chinese Science Bulletin}, vol.~54, no.~17, pp. 2991--2997, Sep. 2009.

\bibitem{CWL:10:OE}
T.-Y. Chen, J.~Wang, H.~Liang, W.-Y. Liu, Y.~Liu, X.~Jiang, Y.~Wang, X.~Wan, W.-Q. Cai, L.~Ju, L.-K. Chen, L.-J. Wang, Y.~Gao, K.~Chen, C.-Z. Peng, Z.-B. Chen, and J.-W. Pan, ``Metropolitan all-pass and inter-city quantum communication network,'' \emph{Optics Express}, vol.~18, no.~26, pp. 27\,217--27\,225, Dec. 2010.

\bibitem{HHL:16:OL}
D.~Huang, P.~Huang, H.~Li, T.~Wang, Y.~Zhou, and G.~Zeng, ``Field demonstration of a continuous-variable quantum key distribution network,'' \emph{Optics Letters}, vol.~41, no.~15, pp. 3511--3514, Aug. 2016.

\bibitem{LCP:22:RMP}
C.-Y. Lu, Y.~Cao, C.-Z. Peng, and J.-W. Pan, ``Micius quantum experiments in space,'' \emph{Reviews of Modern Physics}, vol.~94, no.~3, p. 035001, Jul. 2022.

\bibitem{pomarico2012mhz}
E.~Pomarico, B.~Sanguinetti, T.~Guerreiro, R.~Thew, and H.~Zbinden, ``Mhz rate and efficient synchronous heralding of single photons at telecom wavelengths,'' \emph{Optics express}, vol.~20, no.~21, pp. 23\,846--23\,855, 2012.

\bibitem{pompili2021realization}
M.~Pompili, S.~L. Hermans, S.~Baier, H.~K. Beukers, P.~C. Humphreys, R.~N. Schouten, R.~F. Vermeulen, M.~J. Tiggelman, L.~dos Santos~Martins, B.~Dirkse \emph{et~al.}, ``Realization of a multinode quantum network of remote solid-state qubits,'' \emph{Science}, vol. 372, no. 6539, pp. 259--264, 2021.

\bibitem{colautti20203d}
M.~Colautti, P.~Lombardi, M.~Trapuzzano, F.~S. Piccioli, S.~Pazzagli, B.~Tiribilli, S.~Nocentini, F.~S. Cataliotti, D.~S. Wiersma, and C.~Toninelli, ``A 3d polymeric platform for photonic quantum technologies,'' \emph{Advanced Quantum Technologies}, vol.~3, no.~7, p. 2000004, 2020.

\bibitem{chu2017single}
X.-L. Chu, S.~G{\"o}tzinger, and V.~Sandoghdar, ``A single molecule as a high-fidelity photon gun for producing intensity-squeezed light,'' \emph{Nature Photonics}, vol.~11, no.~1, pp. 58--62, 2017.

\bibitem{thomas2021bright}
S.~Thomas, M.~Billard, N.~Coste, S.~Wein, H.~Ollivier, O.~Krebs, L.~Taza{\"\i}rt, A.~Harouri, A.~Lemaitre, I.~Sagnes \emph{et~al.}, ``Bright polarized single-photon source based on a linear dipole,'' \emph{Physical review letters}, vol. 126, no.~23, p. 233601, 2021.

\bibitem{rezai2018coherence}
M.~Rezai, J.~Wrachtrup, and I.~Gerhardt, ``Coherence properties of molecular single photons for quantum networks,'' \emph{Physical Review X}, vol.~8, no.~3, p. 031026, 2018.

\bibitem{LZL:21:AQT}
L.~Lu, X.~Zheng, Y.~Lu, S.~Zhu, and X.-S. Ma, ``Advances in chip-scale quantum photonic technologies,'' \emph{Advanced Quantum Technologies}, vol.~4, no.~12, p. 2100068, 2021.

\bibitem{CXE:17:LSA}
L.~Caspani, C.~Xiong, B.~J. Eggleton, D.~Bajoni, M.~Liscidini, M.~Galli, R.~Morandotti, and D.~J. Moss, ``Integrated sources of photon quantum states based on nonlinear optics,'' \emph{Light: Science \& Applications}, vol.~6, no.~11, p. e17100, Nov. 2017.

\bibitem{FGR:20:AQT}
L.-T. Feng, G.-C. Guo, and X.-F. Ren, ``Progress on integrated quantum photonic sources with silicon,'' \emph{Advanced Quantum Technologies}, vol.~3, no.~2, p. 1900058, 2020.

\bibitem{meyer2020single}
E.~Meyer-Scott, C.~Silberhorn, and A.~Migdall, ``Single-photon sources: Approaching the ideal through multiplexing,'' \emph{Review of Scientific Instruments}, vol.~91, no.~4, p. 041101, 2020.

\bibitem{kaneda2019high}
F.~Kaneda and P.~G. Kwiat, ``High-efficiency single-photon generation via large-scale active time multiplexing,'' \emph{Science advances}, vol.~5, no.~10, p. eaaw8586, 2019.

\bibitem{cohen1998nobel}
C.~N. Cohen-Tannoudji, ``Nobel lecture: Manipulating atoms with photons,'' \emph{Reviews of Modern Physics}, vol.~70, no.~3, p. 707, 1998.

\bibitem{lounis2005single}
B.~Lounis and M.~Orrit, ``Single-photon sources,'' \emph{Reports on Progress in Physics}, vol.~68, no.~5, p. 1129, 2005.

\bibitem{lenzini2018diamond}
F.~Lenzini, N.~Gruhler, N.~Walter, and W.~H. Pernice, ``Diamond as a platform for integrated quantum photonics,'' \emph{Advanced Quantum Technologies}, vol.~1, no.~3, p. 1800061, 2018.

\bibitem{neumann2023organic}
M.~Neumann, X.~Wei, L.~Morales-Inostroza, S.~Song, S.-G. Lee, K.~Watanabe, T.~Taniguchi, S.~Gootzinger, and Y.~H. Lee, ``Organic molecules as origin of visible-range single photon emission from hexagonal boron nitride and mica,'' \emph{ACS nano}, vol.~17, no.~12, pp. 11\,679--11\,691, 2023.

\bibitem{basche1992photon}
T.~Basch{\'e}, W.~Moerner, M.~Orrit, and H.~Talon, ``Photon antibunching in the fluorescence of a single dye molecule trapped in a solid,'' \emph{Physical review letters}, vol.~69, no.~10, p. 1516, 1992.

\bibitem{wang2019turning}
D.~Wang, H.~Kelkar, D.~Martin-Cano, D.~Rattenbacher, A.~Shkarin, T.~Utikal, S.~G{\"o}tzinger, and V.~Sandoghdar, ``Turning a molecule into a coherent two-level quantum system,'' \emph{Nature Physics}, vol.~15, no.~5, pp. 483--489, 2019.

\bibitem{duquennoy2022real}
R.~Duquennoy, M.~Colautti, R.~Emadi, P.~Majumder, P.~Lombardi, and C.~Toninelli, ``Real-time two-photon interference from distinct molecules on the same chip,'' \emph{Optica}, vol.~9, no.~7, pp. 731--737, 2022.

\bibitem{ZHL:21:AQT}
C.~Zhang, Y.-F. Huang, B.-H. Liu, C.-F. Li, and G.-C. Guo, ``Spontaneous parametric down-conversion sources for multiphoton experiments,'' \emph{Advanced Quantum Technologies}, vol.~4, no.~5, p. 2000132, 2021.

\bibitem{loredo2017boson}
J.~Loredo, M.~Broome, P.~Hilaire, O.~Gazzano, I.~Sagnes, A.~Lemaitre, M.~Almeida, P.~Senellart, and A.~White, ``Boson sampling with single-photon fock states from a bright solid-state source,'' \emph{Physical review letters}, vol. 118, no.~13, p. 130503, 2017.

\bibitem{somaschi2016near}
N.~Somaschi, V.~Giesz, L.~De~Santis, J.~Loredo, M.~P. Almeida, G.~Hornecker, S.~L. Portalupi, T.~Grange, C.~Anton, J.~Demory \emph{et~al.}, ``Near-optimal single-photon sources in the solid state,'' \emph{Nature Photonics}, vol.~10, no.~5, pp. 340--345, 2016.

\bibitem{vahlbruch2016detection}
H.~Vahlbruch, M.~Mehmet, K.~Danzmann, and R.~Schnabel, ``Detection of 15 db squeezed states of light and their application for the absolute calibration of photoelectric quantum efficiency,'' \emph{Physical review letters}, vol. 117, no.~11, p. 110801, 2016.

\bibitem{gehring2017towards}
T.~Gehring, U.~B. Hoff, T.~Iskhakov, and U.~L. Andersen, ``Towards an integrated squeezed light source,'' in \emph{Integrated Photonics: Materials, Devices, and Applications IV}, vol. 10249.\hskip 1em plus 0.5em minus 0.4em\relax SPIE, 2017, pp. 19--26.

\bibitem{VQL:19:PRApp}
Z.~Vernon, N.~Quesada, M.~Liscidini, B.~Morrison, M.~Menotti, K.~Tan, and J.~Sipe, ``Scalable squeezed-light source for continuous-variable quantum sampling,'' \emph{Physical Review Applied}, vol.~12, no.~6, p. 064024, Dec. 2019.

\bibitem{ZMT:21:NC}
Y.~Zhang, M.~Menotti, K.~Tan, V.~D. Vaidya, D.~H. Mahler, L.~G. Helt, L.~Zatti, M.~Liscidini, B.~Morrison, and Z.~Vernon, ``Squeezed light from a nanophotonic molecule,'' \emph{Nature Communications}, vol.~12, no.~1, p. 2233, Apr. 2021.

\bibitem{shelby1986broad}
R.~M. Shelby, M.~D. Levenson, S.~H. Perlmutter, R.~G. DeVoe, and D.~F. Walls, ``Broad-band parametric deamplification of quantum noise in an optical fiber,'' \emph{Physical review letters}, vol.~57, no.~6, p. 691, 1986.

\bibitem{rosenbluh1991squeezed}
M.~Rosenbluh and R.~M. Shelby, ``Squeezed optical solitons,'' \emph{Physical review letters}, vol.~66, no.~2, p. 153, 1991.

\bibitem{drummond1993quantum}
P.~Drummond, R.~Shelby, S.~Friberg, and Y.~Yamamoto, ``Quantum solitons in optical fibres,'' \emph{Nature}, vol. 365, no. 6444, pp. 307--313, 1993.

\bibitem{fujiwara2009generation}
Y.~Fujiwara, H.~Nakagome, K.~Hirosawa, and F.~Kannari, ``Generation of squeezed pulses with a sagnac loop fiber interferometer using a non-soliton femtosecond laser pulse at 800 nm,'' \emph{Optics Express}, vol.~17, no.~13, pp. 11\,197--11\,204, 2009.

\bibitem{MSG:24:arXiv}
F.~A. Mele, F.~Salek, V.~Giovannetti, and L.~Lami, ``Quantum communication on the bosonic loss-dephasing channel,'' \emph{arXiv}, Jan. 2024, arXiv:2401.15634 [quant-ph].

\bibitem{Guh:04:thesis_MIT}
\BIBentryALTinterwordspacing
S.~Guha, ``Classical capacity of the free-space quantum-optical channel,'' Thesis, Massachusetts Institute of Technology, 2004. [Online]. Available: \url{https://dspace.mit.edu/handle/1721.1/87908}
\BIBentrySTDinterwordspacing

\bibitem{LPG:20:PRL}
L.~Lami, M.~B. Plenio, V.~Giovannetti, and A.~S. Holevo, ``Bosonic quantum communication across arbitrarily high loss channels,'' \emph{Physical Review Letters}, vol. 125, no.~11, p. 110504, Sep. 2020.

\bibitem{Pir:21:PRR}
S.~Pirandola, ``Limits and security of free-space quantum communications,'' \emph{Physical Review Research}, vol.~3, no.~1, p. 013279, Mar. 2021.

\bibitem{VSV:17:PRA}
D.~Vasylyev, A.~A. Semenov, W.~Vogel, K.~Gunthner, A.~Thurn, O.~Bayraktar, and C.~Marquardt, ``Free-space quantum links under diverse weather conditions,'' \emph{Physical Review A}, vol.~96, no.~4, p. 043856, Oct. 2017.

\bibitem{LW:23:NP}
L.~Lami and M.~M. Wilde, ``Exact solution for the quantum and private capacities of bosonic dephasing channels,'' \emph{Nature Photonics}, vol.~17, no.~6, pp. 525--530, Jun. 2023.

\bibitem{GGC:14:NP}
V.~Giovannetti, R.~García-Patrón, N.~J. Cerf, and A.~S. Holevo, ``Ultimate classical communication rates of quantum optical channels,'' \emph{Nature Photonics}, vol.~8, no.~10, pp. 796--800, Oct. 2014.

\bibitem{IQCTMN:2017:x40}
\BIBentryALTinterwordspacing
Z.-L. Xiang, M.~Zhang, L.~Jiang, and P.~Rabl, ``Intracity quantum communication via thermal microwave networks,'' \emph{Phys. Rev. X}, vol.~7, p. 011035, Mar 2017. [Online]. Available: \url{https://link.aps.org/doi/10.1103/PhysRevX.7.011035}
\BIBentrySTDinterwordspacing

\bibitem{ODQSTE:2018x41}
\BIBentryALTinterwordspacing
C.~J. Axline, L.~D. Burkhart, W.~Pfaff, M.~Zhang, K.~Chou, P.~Campagne-Ibarcq, P.~Reinhold, L.~Frunzio, S.~M. Girvin, L.~Jiang, M.~H. Devoret, and R.~J. Schoelkopf, ``On-demand quantum state transfer and entanglement between remote microwave cavity memories,'' \emph{Nature Physics}, vol.~14, no.~7, pp. 705--710, 2018. [Online]. Available: \url{https://doi.org/10.1038/s41567-018-0115-y}
\BIBentrySTDinterwordspacing

\bibitem{OMBGC:2006x42}
\BIBentryALTinterwordspacing
F.~Caruso, V.~Giovannetti, and A.~S. Holevo, ``One-mode bosonic gaussian channels: a full weak-degradability classification,'' \emph{New Journal of Physics}, vol.~8, no.~12, p. 310, dec 2006. [Online]. Available: \url{https://dx.doi.org/10.1088/1367-2630/8/12/310}
\BIBentrySTDinterwordspacing

\bibitem{OMQGC:2007x43}
\BIBentryALTinterwordspacing
A.~S. Holevo, ``One-mode quantum gaussian channels: Structure and quantum capacity,'' \emph{Problems of Information Transmission}, vol.~43, no.~1, pp. 1--11, 2007. [Online]. Available: \url{https://doi.org/10.1134/S0032946007010012}
\BIBentrySTDinterwordspacing

\bibitem{NPJ:20:NC}
K.~Noh, S.~Pirandola, and L.~Jiang, ``Enhanced energy-constrained quantum communication over bosonic {Gaussian} channels,'' \emph{Nature Communications}, vol.~11, no.~1, p. 457, Jan. 2020.

\bibitem{gulinatti2021custom}
A.~Gulinatti, F.~Ceccarelli, M.~Ghioni, and I.~Rech, ``Custom silicon technology for spad-arrays with red-enhanced sensitivity and low timing jitter,'' \emph{Optics Express}, vol.~29, no.~3, pp. 4559--4581, 2021.

\bibitem{zhang2015advances}
J.~Zhang, M.~A. Itzler, H.~Zbinden, and J.-W. Pan, ``Advances in ingaas/inp single-photon detector systems for quantum communication,'' \emph{Light: Science \& Applications}, vol.~4, no.~5, pp. e286--e286, 2015.

\bibitem{verma2015high}
V.~B. Verma, B.~Korzh, F.~Bussieres, R.~D. Horansky, S.~D. Dyer, A.~E. Lita, I.~Vayshenker, F.~Marsili, M.~D. Shaw, H.~Zbinden \emph{et~al.}, ``High-efficiency superconducting nanowire single-photon detectors fabricated from mosi thin-films,'' \emph{Optics express}, vol.~23, no.~26, pp. 33\,792--33\,801, 2015.

\bibitem{zhang2017nbn}
W.~Zhang, L.~You, H.~Li, J.~Huang, C.~Lv, L.~Zhang, X.~Liu, J.~Wu, Z.~Wang, and X.~Xie, ``Nbn superconducting nanowire single photon detector with efficiency over 90\% at 1550 nm wavelength operational at compact cryocooler temperature,'' \emph{Science China Physics, Mechanics \& Astronomy}, vol.~60, pp. 1--10, 2017.

\bibitem{zhang2019saturating}
W.~Zhang, Q.~Jia, L.~You, X.~Ou, H.~Huang, L.~Zhang, H.~Li, Z.~Wang, and X.~Xie, ``Saturating intrinsic detection efficiency of superconducting nanowire single-photon detectors via defect engineering,'' \emph{Physical Review Applied}, vol.~12, no.~4, p. 044040, 2019.

\bibitem{ceccarelli2021recent}
F.~Ceccarelli, G.~Acconcia, A.~Gulinatti, M.~Ghioni, I.~Rech, and R.~Osellame, ``Recent advances and future perspectives of single-photon avalanche diodes for quantum photonics applications,'' \emph{Advanced Quantum Technologies}, vol.~4, no.~2, p. 2000102, 2021.

\bibitem{ullom2015review}
J.~N. Ullom and D.~A. Bennett, ``Review of superconducting transition-edge sensors for x-ray and gamma-ray spectroscopy,'' \emph{Superconductor Science and Technology}, vol.~28, no.~8, p. 084003, 2015.

\bibitem{kurakado1982possibility}
M.~Kurakado, ``Possibility of high resolution detectors using superconducting tunnel junctions,'' \emph{Nuclear Instruments and Methods in Physics Research}, vol. 196, no.~1, pp. 275--277, 1982.

\bibitem{mazin2005microwave}
B.~A. Mazin, \emph{Microwave kinetic inductance detectors}.\hskip 1em plus 0.5em minus 0.4em\relax California Institute of Technology, 2005.

\bibitem{natarajan2012superconducting}
C.~M. Natarajan, M.~G. Tanner, and R.~H. Hadfield, ``Superconducting nanowire single-photon detectors: physics and applications,'' \emph{Superconductor science and technology}, vol.~25, no.~6, p. 063001, 2012.

\bibitem{you2020superconducting}
L.~You, ``Superconducting nanowire single-photon detectors for quantum information,'' \emph{Nanophotonics}, vol.~9, no.~9, pp. 2673--2692, 2020.

\bibitem{rademacher2017crosstalk}
G.~Rademacher, R.~S. Lu{\'\i}s, B.~J. Puttnam, Y.~Awaji, and N.~Wada, ``Crosstalk dynamics in multi-core fibers,'' \emph{Optics express}, vol.~25, no.~10, pp. 12\,020--12\,028, 2017.

\bibitem{cozzolino2019orbital}
D.~Cozzolino, D.~Bacco, B.~Da~Lio, K.~Ingerslev, Y.~Ding, K.~Dalgaard, P.~Kristensen, M.~Galili, K.~Rottwitt, S.~Ramachandran \emph{et~al.}, ``Orbital angular momentum states enabling fiber-based high-dimensional quantum communication,'' \emph{Physical Review Applied}, vol.~11, no.~6, p. 064058, 2019.

\bibitem{kim2022quantum}
J.-H. Kim, J.-W. Chae, Y.-C. Jeong, and Y.-H. Kim, ``Quantum communication with time-bin entanglement over a wavelength-multiplexed fiber network,'' \emph{APL Photonics}, vol.~7, no.~1, 2022.

\bibitem{da2021path}
B.~Da~Lio, D.~Cozzolino, N.~Biagi, Y.~Ding, K.~Rottwitt, A.~Zavatta, D.~Bacco, and L.~K. Oxenl{\o}we, ``Path-encoded high-dimensional quantum communication over a 2-km multicore fiber,'' \emph{npj Quantum Information}, vol.~7, no.~1, p.~63, 2021.

\bibitem{wang2016chip}
J.~Wang, D.~Bonneau, M.~Villa, J.~W. Silverstone, R.~Santagati, S.~Miki, T.~Yamashita, M.~Fujiwara, M.~Sasaki, H.~Terai \emph{et~al.}, ``Chip-to-chip quantum photonic interconnect by path-polarization interconversion,'' \emph{Optica}, vol.~3, no.~4, pp. 407--413, 2016.

\bibitem{massoud2007digital}
\BIBentryALTinterwordspacing
P.~Massoud~Salehi and J.~Proakis, \emph{Digital Communications}.\hskip 1em plus 0.5em minus 0.4em\relax McGraw-Hill Education, 2007. [Online]. Available: \url{https://books.google.lu/books?id=HroiQAAACAAJ}
\BIBentrySTDinterwordspacing

\bibitem{goldsmith2005wireless}
\BIBentryALTinterwordspacing
A.~Goldsmith, \emph{Wireless Communications}, ser. Cambridge Core.\hskip 1em plus 0.5em minus 0.4em\relax Cambridge University Press, 2005. [Online]. Available: \url{https://books.google.lu/books?id=n-3ZZ9i0s-cC}
\BIBentrySTDinterwordspacing

\bibitem{opticalcoms}
H.~Kaushal and G.~Kaddoum, ``Optical communication in space: Challenges and mitigation techniques,'' \emph{IEEE Communications Surveys \& Tutorials}, vol.~19, no.~1, pp. 57--96, 2017.

\bibitem{lidar2013quantum}
D.~A. Lidar and T.~A. Brun, \emph{Quantum error correction}.\hskip 1em plus 0.5em minus 0.4em\relax Cambridge university press, 2013.

\bibitem{cai2023quantum}
Z.~Cai, R.~Babbush, S.~C. Benjamin, S.~Endo, W.~J. Huggins, Y.~Li, J.~R. McClean, and T.~E. O’Brien, ``Quantum error mitigation,'' \emph{Reviews of Modern Physics}, vol.~95, no.~4, p. 045005, 2023.

\bibitem{wang2024exploiting}
Z.~Wang, T.~C. Ralph, R.~Aguinaldo, and R.~Malaney, ``Exploiting spatial diversity in earth-to-satellite quantum-classical communications,'' \emph{arXiv preprint arXiv:2407.02224}, 2024.

\bibitem{Gisin_2005}
\BIBentryALTinterwordspacing
N.~Gisin, N.~Linden, S.~Massar, and S.~Popescu, ``Error filtration and entanglement purification for quantum communication,'' \emph{Physical Review A}, vol.~72, no.~1, Jul. 2005. [Online]. Available: \url{http://dx.doi.org/10.1103/PhysRevA.72.012338}
\BIBentrySTDinterwordspacing

\bibitem{Abbott_2020}
\BIBentryALTinterwordspacing
A.~A. Abbott, J.~Wechs, D.~Horsman, M.~Mhalla, and C.~Branciard, ``Communication through coherent control of quantum channels,'' \emph{Quantum}, vol.~4, p. 333, Sep. 2020. [Online]. Available: \url{http://dx.doi.org/10.22331/Q-2020-09-24-333}
\BIBentrySTDinterwordspacing

\bibitem{branciard2021coherent}
C.~Branciard, A.~Cl{\'e}ment, M.~Mhalla, and S.~Perdrix, ``Coherent control and distinguishability of quantum channels via pbs-diagrams,'' \emph{arXiv preprint arXiv:2103.02073}, 2021.

\bibitem{Aharonov_2009}
\BIBentryALTinterwordspacing
Y.~Aharonov, S.~Popescu, J.~Tollaksen, and L.~Vaidman, ``Multiple-time states and multiple-time measurements in quantum mechanics,'' \emph{Physical Review A}, vol.~79, no.~5, May 2009. [Online]. Available: \url{http://dx.doi.org/10.1103/PhysRevA.79.052110}
\BIBentrySTDinterwordspacing

\bibitem{perinotti_2009}
\BIBentryALTinterwordspacing
G.~Chiribella, G.~M. D'Ariano, and P.~Perinotti, ``Theoretical framework for quantum networks,'' \emph{Phys. Rev. A}, vol.~80, p. 022339, Aug 2009. [Online]. Available: \url{https://link.aps.org/doi/10.1103/PhysRevA.80.022339}
\BIBentrySTDinterwordspacing

\bibitem{sudarchan}
T.~F. Jordan, A.~Shaji, and E.~C.~G. Sudarshan, ``Dynamics of initially entangled open quantum systems,'' \emph{Phys. Rev. A}, vol.~70, p. 052110, Nov 2004.

\bibitem{Milz_2021}
S.~Milz and K.~Modi, ``Quantum stochastic processes and quantum non-markovian phenomena,'' \emph{PRX Quantum}, vol.~2, no.~3, Jul 2021.

\bibitem{chiribella2008transforming}
G.~Chiribella, G.~M. D'Ariano, and P.~Perinotti, ``Transforming quantum operations: Quantum supermaps,'' \emph{Europhysics Letters}, vol.~83, no.~3, p. 30004, 2008.

\bibitem{loizeau2020channel}
N.~Loizeau and A.~Grinbaum, ``Channel capacity enhancement with indefinite causal order,'' \emph{Physical Review A}, vol. 101, no.~1, p. 012340, 2020.

\bibitem{koudia2021environment}
S.~Koudia, A.~S. Cacciapuoti, and M.~Caleffi, ``From the environment-assisted paradigm to the quantum switch,'' in \emph{2021 IEEE Global Communications Conference (GLOBECOM)}.\hskip 1em plus 0.5em minus 0.4em\relax IEEE, 2021, pp. 1--6.

\bibitem{Giulia2021}
\BIBentryALTinterwordspacing
G.~Rubino, L.~A. Rozema, D.~Ebler, H.~Kristj\'ansson, S.~Salek, P.~Allard~Gu\'erin, A.~A. Abbott, C.~Branciard, i.~c.~v. Brukner, G.~Chiribella, and P.~Walther, ``Experimental quantum communication enhancement by superposing trajectories,'' \emph{Phys. Rev. Res.}, vol.~3, p. 013093, Jan 2021. [Online]. Available: \url{https://link.aps.org/doi/10.1103/PhysRevResearch.3.013093}
\BIBentrySTDinterwordspacing

\bibitem{solntsev2017path}
A.~S. Solntsev and A.~A. Sukhorukov, ``Path-entangled photon sources on nonlinear chips,'' \emph{Reviews in Physics}, vol.~2, pp. 19--31, 2017.

\bibitem{ZT:03:IEEE_T_IT}
{Lizhong Zheng} and D.~Tse, ``Diversity and multiplexing: a fundamental tradeoff in multiple-antenna channels,'' \emph{IEEE Transactions on Information Theory}, vol.~49, no.~5, pp. 1073--1096, May 2003.

\bibitem{KDM:21:CL}
N.~K. Kundu, S.~P. Dash, M.~R. McKay, and R.~K. Mallik, ``{MIMO} {Terahertz} quantum key distribution,'' \emph{IEEE Communications Letters}, vol.~25, no.~10, pp. 3345--3349, Oct. 2021.

\bibitem{KMC:23:TQE}
N.~K. Kundu, M.~R. McKay, A.~Conti, R.~K. Mallik, and M.~Z. Win, ``{MIMO} {Terahertz} quantum key distribution under restricted eavesdropping,'' \emph{IEEE Transactions on Quantum Engineering}, vol.~4, pp. 1--15, 2023.

\bibitem{GNR:23:Conf}
A.~Gokul, H.~Natarajan, and V.~Raghunathan, ``Experimental demonstration of coexistence of classical and quantum communication in quantum key distribution link,'' in \emph{2023 {IEEE} {Microwaves}, {Antennas}, and {Propagation} {Conference} ({MAPCON})}, Dec. 2023, pp. 1--5.

\bibitem{JQZ:24:Chip}
X.~Jing, C.~Qian, X.~Zheng, H.~Nian, C.~Wang, J.~Tang, X.~Gu, Y.~Kong, T.~Chen, Y.~Liu, C.~Sheng, D.~Jiang, B.~Niu, and L.~Lu, ``Coexistence of multiuser entanglement distribution and classical light in optical fiber network with a semiconductor chip,'' \emph{Chip}, p. 100083, Jan. 2024.

\bibitem{ATB:21:conf}
O.~Alia, R.~S. Tessinari, T.~D. Bradley, H.~Sakr, K.~Harrington, J.~Hayes, Y.~Chen, P.~Petropoulos, D.~Richardson, F.~Poletti, G.~T. Kanellos, R.~Nejabati, and D.~Simeonidou, ``1.6 {Tbps} classical channel coexistence with dv-qkd over hollow core nested antiresonant nodeless fibre ({HC}-{NANF}),'' in \emph{2021 {European} {Conference} on {Optical} {Communication} ({ECOC})}, Sep. 2021, pp. 1--4.

\bibitem{TAJ:21:conf}
R.~S. Tessinari, O.~Alia, S.~K. Joshi, D.~Aktas, M.~Clark, E.~Hugues-Salas, G.~T. Kanellos, J.~Rarity, R.~Nejabati, and D.~Simeonidou, ``Towards co-existence of 100 {Gbps} classical channel within a {WDM} quantum entanglement network,'' in \emph{2021 {Optical} {Fiber} {Communications} {Conference} and {Exhibition} ({OFC})}, Jun. 2021, pp. 1--3.

\bibitem{BRS:16:SI}
S.~Bahrani, M.~Razavi, and J.~A. Salehi, ``Crosstalk reduction in hybrid quantum-classical networks,'' \emph{Scientia Iranica}, vol.~23, no.~6, pp. 2898--2907, Dec. 2016.

\bibitem{BRS:18:SR}
------, ``Wavelength assignment in hybrid quantum-classical networks,'' \emph{Scientific Reports}, vol.~8, no.~1, p. 3456, Feb. 2018.

\bibitem{BRS:23:PRA}
M.~Bathaee, M.~Rezai, and J.~A. Salehi, ``Quantum wavelength-division-multiplexing and multiple-access communication systems and networks: {Global} and unified approach,'' \emph{Physical Review A}, vol. 107, no.~1, p. 012613, Jan. 2023.

\bibitem{he2023sensitive}
D.~He, X.~Feng, and L.~Wei, ``Sensitive enhancement of cat state quantum illumination,'' \emph{Optics Express}, vol.~31, no.~11, pp. 17\,709--17\,715, 2023.

\bibitem{bergmann2016quantum}
M.~Bergmann and P.~van Loock, ``Quantum error correction against photon loss using multicomponent cat states,'' \emph{Physical Review A}, vol.~94, no.~4, p. 042332, 2016.

\bibitem{vlastakis2013deterministically}
B.~Vlastakis, G.~Kirchmair, Z.~Leghtas, S.~E. Nigg, L.~Frunzio, S.~M. Girvin, M.~Mirrahimi, M.~H. Devoret, and R.~J. Schoelkopf, ``Deterministically encoding quantum information using 100-photon schr{\"o}dinger cat states,'' \emph{Science}, vol. 342, no. 6158, pp. 607--610, 2013.

\bibitem{laghaout2013amplification}
A.~Laghaout, J.~S. Neergaard-Nielsen, I.~Rigas, C.~Kragh, A.~Tipsmark, and U.~L. Andersen, ``Amplification of realistic schr{\"o}dinger-cat-state-like states by homodyne heralding,'' \emph{Physical Review A}, vol.~87, no.~4, p. 043826, 2013.

\bibitem{winnel2024deterministic}
M.~S. Winnel, J.~J. Guanzon, D.~Singh, and T.~C. Ralph, ``Deterministic preparation of optical squeezed cat and gottesman-kitaev-preskill states,'' \emph{Physical Review Letters}, vol. 132, no.~23, p. 230602, 2024.

\bibitem{ourjoumtsev2007generation}
A.~Ourjoumtsev, H.~Jeong, R.~Tualle-Brouri, and P.~Grangier, ``Generation of optical ‘schr{\"o}dinger cats’ from photon number states,'' \emph{Nature}, vol. 448, no. 7155, pp. 784--786, 2007.

\bibitem{takase2021generation}
K.~Takase, J.-i. Yoshikawa, W.~Asavanant, M.~Endo, and A.~Furusawa, ``Generation of optical schr{\"o}dinger cat states by generalized photon subtraction,'' \emph{Physical Review A}, vol. 103, no.~1, p. 013710, 2021.

\bibitem{brune1992manipulation}
M.~Brune, S.~Haroche, J.-M. Raimond, L.~Davidovich, and N.~Zagury, ``Manipulation of photons in a cavity by dispersive atom-field coupling: Quantum-nondemolition measurements and generation of ‘‘schr{\"o}dinger cat’’states,'' \emph{Physical Review A}, vol.~45, no.~7, p. 5193, 1992.

\bibitem{buvzek1992schrodinger}
V.~Bu{\v{z}}ek, H.~Moya-Cessa, P.~Knight, and S.~Phoenix, ``Schr{\"o}dinger-cat states in the resonant jaynes-cummings model: Collapse and revival of oscillations of the photon-number distribution,'' \emph{Physical review A}, vol.~45, no.~11, p. 8190, 1992.

\bibitem{liao2016generation}
J.-Q. Liao, J.-F. Huang, and L.~Tian, ``Generation of macroscopic schr{\"o}dinger-cat states in qubit-oscillator systems,'' \emph{Physical Review A}, vol.~93, no.~3, p. 033853, 2016.

\bibitem{zeng2020macroscopic}
Y.-X. Zeng, J.~Shen, M.-S. Ding, and C.~Li, ``Macroscopic schr{\"o}dinger cat state swapping in optomechanical system,'' \emph{Optics Express}, vol.~28, no.~7, pp. 9587--9602, 2020.

\bibitem{chamberland2022building}
C.~Chamberland, K.~Noh, P.~Arrangoiz-Arriola, E.~T. Campbell, C.~T. Hann, J.~Iverson, H.~Putterman, T.~C. Bohdanowicz, S.~T. Flammia, A.~Keller \emph{et~al.}, ``Building a fault-tolerant quantum computer using concatenated cat codes,'' \emph{PRX Quantum}, vol.~3, no.~1, p. 010329, 2022.

\bibitem{gottesman2001encoding}
D.~Gottesman, A.~Kitaev, and J.~Preskill, ``Encoding a qubit in an oscillator,'' \emph{Physical Review A}, vol.~64, no.~1, p. 012310, 2001.

\bibitem{campagne2020quantum}
P.~Campagne-Ibarcq, A.~Eickbusch, S.~Touzard, E.~Zalys-Geller, N.~E. Frattini, V.~V. Sivak, P.~Reinhold, S.~Puri, S.~Shankar, R.~J. Schoelkopf \emph{et~al.}, ``Quantum error correction of a qubit encoded in grid states of an oscillator,'' \emph{Nature}, vol. 584, no. 7821, pp. 368--372, 2020.

\bibitem{fluhmann2019encoding}
C.~Fl{\"u}hmann, T.~L. Nguyen, M.~Marinelli, V.~Negnevitsky, K.~Mehta, and J.~Home, ``Encoding a qubit in a trapped-ion mechanical oscillator,'' \emph{Nature}, vol. 566, no. 7745, pp. 513--517, 2019.

\bibitem{kendell2024deterministic}
H.~C. Kendell, G.~Ferranti, and C.~A. Weidner, ``Deterministic generation of highly squeezed gkp states in ultracold atoms,'' \emph{APL Quantum}, vol.~1, no.~2, 2024.

\bibitem{takase2023gottesman}
K.~Takase, K.~Fukui, A.~Kawasaki, W.~Asavanant, M.~Endo, J.-i. Yoshikawa, P.~van Loock, and A.~Furusawa, ``Gottesman-kitaev-preskill qubit synthesizer for propagating light,'' \emph{npj Quantum Information}, vol.~9, no.~1, p.~98, 2023.

\bibitem{vasconcelos2010all}
H.~M. Vasconcelos, L.~Sanz, and S.~Glancy, ``All-optical generation of states for “encoding a qubit in an oscillator”,'' \emph{Optics letters}, vol.~35, no.~19, pp. 3261--3263, 2010.

\bibitem{weigand2018generating}
D.~J. Weigand and B.~M. Terhal, ``Generating grid states from schr{\"o}dinger-cat states without postselection,'' \emph{Physical Review A}, vol.~97, no.~2, p. 022341, 2018.

\bibitem{eaton2022measurement}
M.~Eaton, C.~Gonz{\'a}lez-Arciniegas, R.~N. Alexander, N.~C. Menicucci, and O.~Pfister, ``Measurement-based generation and preservation of cat and grid states within a continuous-variable cluster state,'' \emph{Quantum}, vol.~6, p. 769, 2022.

\bibitem{yanagimoto2020engineering}
R.~Yanagimoto, T.~Onodera, E.~Ng, L.~G. Wright, P.~L. McMahon, and H.~Mabuchi, ``Engineering a kerr-based deterministic cubic phase gate via gaussian operations,'' \emph{Physical Review Letters}, vol. 124, no.~24, p. 240503, 2020.

\bibitem{PhysRevResearch.3.033118}
\BIBentryALTinterwordspacing
K.~Fukui, R.~N. Alexander, and P.~van Loock, ``All-optical long-distance quantum communication with gottesman-kitaev-preskill qubits,'' \emph{Phys. Rev. Res.}, vol.~3, p. 033118, Aug 2021. [Online]. Available: \url{https://link.aps.org/doi/10.1103/PhysRevResearch.3.033118}
\BIBentrySTDinterwordspacing

\bibitem{Azari2024QuantumSF}
\BIBentryALTinterwordspacing
M.~Azari, P.~Polakos, and K.~P. Seshadreesan, ``Quantum switches for gottesman-kitaev-preskill qubit-based all-photonic quantum networks,'' \emph{ArXiv}, vol. abs/2402.02721, 2024. [Online]. Available: \url{https://api.semanticscholar.org/CorpusID:267411868}
\BIBentrySTDinterwordspacing

\bibitem{Car:15:Book}
\BIBentryALTinterwordspacing
G.~Cariolaro, \emph{Quantum {Communications}}, ser. Signals and {Communication} {Technology}.\hskip 1em plus 0.5em minus 0.4em\relax Cham: Springer International Publishing, 2015. [Online]. Available: \url{https://link.springer.com/10.1007/978-3-319-15600-2}
\BIBentrySTDinterwordspacing

\bibitem{PLOB:17:NC}
S.~Pirandola, R.~Laurenza, C.~Ottaviani, and L.~Banchi, ``Fundamental limits of repeaterless quantum communications,'' \emph{Nature Communications}, vol.~8, no.~1, p. 15043, Apr. 2017.

\bibitem{Survey:21:TPI}
\BIBentryALTinterwordspacing
F.~Bouchard, A.~Sit, Y.~Zhang, R.~Fickler, F.~M. Miatto, Y.~Yao, F.~Sciarrino, and E.~Karimi, ``Two-photon interference: the hong–ou–mandel effect,'' \emph{Reports on Progress in Physics}, vol.~84, no.~1, p. 012402, dec 2020. [Online]. Available: \url{https://dx.doi.org/10.1088/1361-6633/abcd7a}
\BIBentrySTDinterwordspacing

\bibitem{HOM:87}
\BIBentryALTinterwordspacing
Z.~Y. Ou, C.~K. Hong, and L.~Mandel, ``Detection of squeezed states by cross correlation,'' \emph{Phys. Rev. A}, vol.~36, pp. 192--196, Jul 1987. [Online]. Available: \url{https://link.aps.org/doi/10.1103/PhysRevA.36.192}
\BIBentrySTDinterwordspacing

\bibitem{Heralded:12:86}
\BIBentryALTinterwordspacing
J.~Hofmann, M.~Krug, N.~Ortegel, L.~Gérard, M.~Weber, W.~Rosenfeld, and H.~Weinfurter, ``Heralded entanglement between widely separated atoms,'' \emph{Science}, vol. 337, no. 6090, pp. 72--75, 2012. [Online]. Available: \url{https://www.science.org/doi/abs/10.1126/science.1221856}
\BIBentrySTDinterwordspacing

\bibitem{RCRE:16:88}
\BIBentryALTinterwordspacing
A.~Narla, S.~Shankar, M.~Hatridge, Z.~Leghtas, K.~M. Sliwa, E.~Zalys-Geller, S.~O. Mundhada, W.~Pfaff, L.~Frunzio, R.~J. Schoelkopf, and M.~H. Devoret, ``Robust concurrent remote entanglement between two superconducting qubits,'' \emph{Phys. Rev. X}, vol.~6, p. 031036, Sep 2016. [Online]. Available: \url{https://link.aps.org/doi/10.1103/PhysRevX.6.031036}
\BIBentrySTDinterwordspacing

\bibitem{MDIQKD:12:x84}
\BIBentryALTinterwordspacing
H.-K. Lo, M.~Curty, and B.~Qi, ``Measurement-device-independent quantum key distribution,'' \emph{Phys. Rev. Lett.}, vol. 108, p. 130503, Mar 2012. [Online]. Available: \url{https://link.aps.org/doi/10.1103/PhysRevLett.108.130503}
\BIBentrySTDinterwordspacing

\bibitem{PQKDP:14:x102}
\BIBentryALTinterwordspacing
T.~Sasaki, Y.~Yamamoto, and M.~Koashi, ``Practical quantum key distribution protocol without monitoring signal disturbance,'' \emph{Nature}, vol. 509, no. 7501, pp. 475--478, 2014. [Online]. Available: \url{https://doi.org/10.1038/nature13303}
\BIBentrySTDinterwordspacing

\bibitem{PRRDPSQKD:15:x85}
\BIBentryALTinterwordspacing
J.-Y. Guan, Z.~Cao, Y.~Liu, G.-L. Shen-Tu, J.~S. Pelc, M.~M. Fejer, C.-Z. Peng, X.~Ma, Q.~Zhang, and J.-W. Pan, ``Experimental passive round-robin differential phase-shift quantum key distribution,'' \emph{Phys. Rev. Lett.}, vol. 114, p. 180502, May 2015. [Online]. Available: \url{https://link.aps.org/doi/10.1103/PhysRevLett.114.180502}
\BIBentrySTDinterwordspacing

\bibitem{IIPEBS:14:x146}
\BIBentryALTinterwordspacing
M.~C. Tichy, ``Interference of identical particles from entanglement to boson-sampling,'' \emph{Journal of Physics B: Atomic, Molecular and Optical Physics}, vol.~47, no.~10, p. 103001, may 2014. [Online]. Available: \url{https://dx.doi.org/10.1088/0953-4075/47/10/103001}
\BIBentrySTDinterwordspacing

\bibitem{CLTC:14:x147}
\BIBentryALTinterwordspacing
H.~de~Guise, S.-H. Tan, I.~P. Poulin, and B.~C. Sanders, ``Coincidence landscapes for three-channel linear optical networks,'' \emph{Phys. Rev. A}, vol.~89, p. 063819, Jun 2014. [Online]. Available: \url{https://link.aps.org/doi/10.1103/PhysRevA.89.063819}
\BIBentrySTDinterwordspacing

\bibitem{PITME:15:x148}
\BIBentryALTinterwordspacing
V.~S. Shchesnovich, ``Partial indistinguishability theory for multiphoton experiments in multiport devices,'' \emph{Phys. Rev. A}, vol.~91, p. 013844, Jan 2015. [Online]. Available: \url{https://link.aps.org/doi/10.1103/PhysRevA.91.013844}
\BIBentrySTDinterwordspacing

\bibitem{17:DMPI:mi144}
\BIBentryALTinterwordspacing
A.~J. Menssen, A.~E. Jones, B.~J. Metcalf, M.~C. Tichy, S.~Barz, W.~S. Kolthammer, and I.~A. Walmsley, ``Distinguishability and many-particle interference,'' \emph{Phys. Rev. Lett.}, vol. 118, p. 153603, Apr 2017. [Online]. Available: \url{https://link.aps.org/doi/10.1103/PhysRevLett.118.153603}
\BIBentrySTDinterwordspacing

\bibitem{12:QWTIB:mi153}
\BIBentryALTinterwordspacing
Y.~Lahini, M.~Verbin, S.~D. Huber, Y.~Bromberg, R.~Pugatch, and Y.~Silberberg, ``Quantum walk of two interacting bosons,'' \emph{Phys. Rev. A}, vol.~86, p. 011603, Jul 2012. [Online]. Available: \url{https://link.aps.org/doi/10.1103/PhysRevA.86.011603}
\BIBentrySTDinterwordspacing

\bibitem{12:TPPE:mi154}
\BIBentryALTinterwordspacing
E.~Poem, Y.~Gilead, and Y.~Silberberg, ``Two-photon path-entangled states in multimode waveguides,'' \emph{Phys. Rev. Lett.}, vol. 108, p. 153602, Apr 2012. [Online]. Available: \url{https://link.aps.org/doi/10.1103/PhysRevLett.108.153602}
\BIBentrySTDinterwordspacing

\bibitem{09:IQP:mi156}
A.~Politi, J.~C. Matthews, M.~G. Thompson, and J.~L. O'Brien, ``Integrated quantum photonics,'' \emph{IEEE Journal of Selected Topics in Quantum Electronics}, vol.~15, no.~6, pp. 1673--1684, 2009.

\bibitem{19:QTHD:mi164}
\BIBentryALTinterwordspacing
Y.-H. Luo, H.-S. Zhong, M.~Erhard, X.-L. Wang, L.-C. Peng, M.~Krenn, X.~Jiang, L.~Li, N.-L. Liu, C.-Y. Lu, A.~Zeilinger, and J.-W. Pan, ``Quantum teleportation in high dimensions,'' \emph{Phys. Rev. Lett.}, vol. 123, p. 070505, Aug 2019. [Online]. Available: \url{https://link.aps.org/doi/10.1103/PhysRevLett.123.070505}
\BIBentrySTDinterwordspacing

\bibitem{13:LMEG:mi165}
\BIBentryALTinterwordspacing
M.~C. Tichy, F.~Mintert, and A.~Buchleitner, ``Limits to multipartite entanglement generation with bosons and fermions,'' \emph{Phys. Rev. A}, vol.~87, p. 022319, Feb 2013. [Online]. Available: \url{https://link.aps.org/doi/10.1103/PhysRevA.87.022319}
\BIBentrySTDinterwordspacing

\bibitem{11:CCLO:mi166}
\BIBentryALTinterwordspacing
S.~Aaronson and A.~Arkhipov, ``The computational complexity of linear optics,'' in \emph{Proceedings of the Forty-Third Annual ACM Symposium on Theory of Computing}, ser. STOC '11.\hskip 1em plus 0.5em minus 0.4em\relax New York, NY, USA: Association for Computing Machinery, 2011, p. 333–342. [Online]. Available: \url{https://doi.org/10.1145/1993636.1993682}
\BIBentrySTDinterwordspacing

\bibitem{RN77}
\BIBentryALTinterwordspacing
B.~Ndagano, B.~Perez-Garcia, F.~S. Roux, M.~McLaren, C.~Rosales-Guzman, Y.~Zhang, O.~Mouane, R.~I. Hernandez-Aranda, T.~Konrad, and A.~Forbes, ``Characterizing quantum channels with non-separable states of classical light,'' \emph{Nature Physics}, vol.~13, no.~4, pp. 397--402, 2017. [Online]. Available: \url{https://doi.org/10.1038/nphys4003}
\BIBentrySTDinterwordspacing

\bibitem{PhysRevLett.84.4737}
\BIBentryALTinterwordspacing
W.~Tittel, J.~Brendel, H.~Zbinden, and N.~Gisin, ``Quantum cryptography using entangled photons in energy-time bell states,'' \emph{Phys. Rev. Lett.}, vol.~84, pp. 4737--4740, May 2000. [Online]. Available: \url{https://link.aps.org/doi/10.1103/PhysRevLett.84.4737}
\BIBentrySTDinterwordspacing

\bibitem{RN78}
G.~L. Roberts, M.~Lucamarini, J.~F. Dynes, S.~J. Savory, Z.~Yuan, and A.~J. Shields, ``Modulator‐free coherent‐one‐way quantum key distribution,'' \emph{Laser \& Photonics Reviews}, vol.~11, no.~4, p. 1700067, 2017.

\bibitem{Survey:CLQS:24}
\BIBentryALTinterwordspacing
A.~Anshu and S.~Arunachalam, ``A survey on the complexity of learning quantum states,'' \emph{Nature Reviews Physics}, vol.~6, no.~1, pp. 59--69, 2024. [Online]. Available: \url{https://doi.org/10.1038/s42254-023-00662-4}
\BIBentrySTDinterwordspacing

\bibitem{Mele2024LearningQS}
F.~A. Mele, A.~A. Mele, L.~Bittel, J.~Eisert, V.~Giovannetti, L.~Lami, L.~Leone, and S.~F.~E. Oliviero, ``Learning quantum states of continuous variable systems,'' \emph{arXiv}, May 2024, arXiv:2405.01431 [quant-ph].

\bibitem{QCB:20:survey-cer}
\BIBentryALTinterwordspacing
J.~Eisert, D.~Hangleiter, N.~Walk, I.~Roth, D.~Markham, R.~Parekh, U.~Chabaud, and E.~Kashefi, ``Quantum certification and benchmarking,'' \emph{Nature Reviews Physics}, vol.~2, no.~7, pp. 382--390, 2020. [Online]. Available: \url{https://doi.org/10.1038/s42254-020-0186-4}
\BIBentrySTDinterwordspacing

\bibitem{QSTCS:10:c53}
\BIBentryALTinterwordspacing
D.~Gross, Y.-K. Liu, S.~T. Flammia, S.~Becker, and J.~Eisert, ``Quantum state tomography via compressed sensing,'' \emph{Phys. Rev. Lett.}, vol. 105, p. 150401, Oct 2010. [Online]. Available: \url{https://link.aps.org/doi/10.1103/PhysRevLett.105.150401}
\BIBentrySTDinterwordspacing

\bibitem{FST:20:c56}
\BIBentryALTinterwordspacing
M.~Guţă, J.~Kahn, R.~Kueng, and J.~A. Tropp, ``Fast state tomography with optimal error bounds,'' \emph{Journal of Physics A: Mathematical and Theoretical}, vol.~53, no.~20, p. 204001, apr 2020. [Online]. Available: \url{https://dx.doi.org/10.1088/1751-8121/ab8111}
\BIBentrySTDinterwordspacing

\bibitem{QTPPCSP:15:c69}
\BIBentryALTinterwordspacing
A.~Kalev, R.~L. Kosut, and I.~H. Deutsch, ``Quantum tomography protocols with positivity are compressed sensing protocols,'' \emph{npj Quantum Information}, vol.~1, no.~1, p. 15018, 2015. [Online]. Available: \url{https://doi.org/10.1038/npjqi.2015.18}
\BIBentrySTDinterwordspacing

\bibitem{SRDM:2013:c10}
\BIBentryALTinterwordspacing
T.~Baumgratz, D.~Gross, M.~Cramer, and M.~B. Plenio, ``Scalable reconstruction of density matrices,'' \emph{Phys. Rev. Lett.}, vol. 111, p. 020401, Jul 2013. [Online]. Available: \url{https://link.aps.org/doi/10.1103/PhysRevLett.111.020401}
\BIBentrySTDinterwordspacing

\bibitem{EQST:10:Nature:c31}
\BIBentryALTinterwordspacing
M.~Cramer, M.~B. Plenio, S.~T. Flammia, R.~Somma, D.~Gross, S.~D. Bartlett, O.~Landon-Cardinal, D.~Poulin, and Y.-K. Liu, ``Efficient quantum state tomography,'' \emph{Nature Communications}, vol.~1, no.~1, p. 149, 2010. [Online]. Available: \url{https://doi.org/10.1038/ncomms1147}
\BIBentrySTDinterwordspacing

\bibitem{WTMPS:13:c66}
\BIBentryALTinterwordspacing
R.~H\"ubener, A.~Mari, and J.~Eisert, ``Wick's theorem for matrix product states,'' \emph{Phys. Rev. Lett.}, vol. 110, p. 040401, Jan 2013. [Online]. Available: \url{https://link.aps.org/doi/10.1103/PhysRevLett.110.040401}
\BIBentrySTDinterwordspacing

\bibitem{ELQS:19:c101}
\BIBentryALTinterwordspacing
A.~Rocchetto, S.~Aaronson, S.~Severini, G.~Carvacho, D.~Poderini, I.~Agresti, M.~Bentivegna, and F.~Sciarrino, ``Experimental learning of quantum states,'' \emph{Science Advances}, vol.~5, no.~3, p. eaau1946, 2019. [Online]. Available: \url{https://www.science.org/doi/abs/10.1126/sciadv.aau1946}
\BIBentrySTDinterwordspacing

\bibitem{EOHL:12:c52}
\BIBentryALTinterwordspacing
C.~E. Granade, C.~Ferrie, N.~Wiebe, and D.~G. Cory, ``Robust online hamiltonian learning,'' \emph{New Journal of Physics}, vol.~14, no.~10, p. 103013, oct 2012. [Online]. Available: \url{https://dx.doi.org/10.1088/1367-2630/14/10/103013}
\BIBentrySTDinterwordspacing

\bibitem{SRUPH:15:c63}
\BIBentryALTinterwordspacing
M.~Holz\"apfel, T.~Baumgratz, M.~Cramer, and M.~B. Plenio, ``Scalable reconstruction of unitary processes and hamiltonians,'' \emph{Phys. Rev. A}, vol.~91, p. 042129, Apr 2015. [Online]. Available: \url{https://link.aps.org/doi/10.1103/PhysRevA.91.042129}
\BIBentrySTDinterwordspacing

\bibitem{DFEP:2011:c46}
\BIBentryALTinterwordspacing
S.~T. Flammia and Y.-K. Liu, ``Direct fidelity estimation from few pauli measurements,'' \emph{Phys. Rev. Lett.}, vol. 106, p. 230501, Jun 2011. [Online]. Available: \url{https://link.aps.org/doi/10.1103/PhysRevLett.106.230501}
\BIBentrySTDinterwordspacing

\bibitem{OSEAF:13:c98}
\BIBentryALTinterwordspacing
D.~M. Reich, G.~Gualdi, and C.~P. Koch, ``Optimal strategies for estimating the average fidelity of quantum gates,'' \emph{Phys. Rev. Lett.}, vol. 111, p. 200401, Nov 2013. [Online]. Available: \url{https://link.aps.org/doi/10.1103/PhysRevLett.111.200401}
\BIBentrySTDinterwordspacing

\bibitem{VEOSS:24:mine}
L.~Oleynik, J.~ur~Rehman, H.~Al-Hraishawi, and S.~Chatzinotas, ``Variational estimation of optimal signal states for quantum channels,'' \emph{IEEE Transactions on Quantum Engineering}, vol.~5, pp. 1--8, 2024.

\bibitem{cariolaro2010theory:m10}
G.~Cariolaro and G.~Pierobon, ``Theory of quantum pulse position modulation and related numerical problems,'' \emph{IEEE Transactions on Communications}, vol.~58, no.~4, pp. 1213--1222, Apr. 2010.

\bibitem{2010_PQDTS:m9}
------, ``Performance of quantum data transmission systems in the presence of thermal noise,'' \emph{IEEE Transactions on Communications}, vol.~58, no.~2, pp. 623--630, Feb. 2010.

\bibitem{vasquez2023quantum:m15}
A.~V{\'a}zquez-Castro and B.~Samandarov, ``Quantum advantage of binary discrete modulations for space channels,'' \emph{IEEE Wireless Communications Letters}, vol.~12, no.~5, pp. 903--906, 2023.

\bibitem{2015_QSD_survey:m21}
J.~Bae and L.-C. Kwek, ``Quantum state discrimination and its applications,'' \emph{J. Phys. A: Math. Theor.}, vol.~48, no.~8, p. 083001, Jan. 2015.

\bibitem{QLTA:19:fs12}
\BIBentryALTinterwordspacing
A.~A. Semenov and W.~Vogel, ``Quantum light in the turbulent atmosphere,'' \emph{Phys. Rev. A}, vol.~80, p. 021802, Aug 2009. [Online]. Available: \url{https://link.aps.org/doi/10.1103/PhysRevA.80.021802}
\BIBentrySTDinterwordspacing

\bibitem{17:HONEFLC}
\BIBentryALTinterwordspacing
M.~Bohmann, J.~Sperling, A.~A. Semenov, and W.~Vogel, ``Higher-order nonclassical effects in fluctuating-loss channels,'' \emph{Phys. Rev. A}, vol.~95, p. 012324, Jan 2017. [Online]. Available: \url{https://link.aps.org/doi/10.1103/PhysRevA.95.012324}
\BIBentrySTDinterwordspacing

\bibitem{Usenko_2012}
\BIBentryALTinterwordspacing
V.~C. Usenko, B.~Heim, C.~Peuntinger, C.~Wittmann, C.~Marquardt, G.~Leuchs, and R.~Filip, ``Entanglement of gaussian states and the applicability to quantum key distribution over fading channels,'' \emph{New Journal of Physics}, vol.~14, no.~9, p. 093048, Sep. 2012. [Online]. Available: \url{http://dx.doi.org/10.1088/1367-2630/14/9/093048}
\BIBentrySTDinterwordspacing

\bibitem{09:FFSQKD:fs16}
D.~Elser, T.~Bartley, B.~Heim, C.~Wittmann, D.~Sych, and G.~Leuchs, ``Feasibility of free space quantum key distribution with coherent polarization states,'' \emph{New Journal of Physics}, vol.~11, no.~4, p. 045014, 2009.

\bibitem{16:FSQSUH:fs18}
\BIBentryALTinterwordspacing
C.~Croal, C.~Peuntinger, B.~Heim, I.~Khan, C.~Marquardt, G.~Leuchs, P.~Wallden, E.~Andersson, and N.~Korolkova, ``Free-space quantum signatures using heterodyne measurements,'' \emph{Phys. Rev. Lett.}, vol. 117, p. 100503, Sep 2016. [Online]. Available: \url{https://link.aps.org/doi/10.1103/PhysRevLett.117.100503}
\BIBentrySTDinterwordspacing

\bibitem{12:QTED100:fs5}
\BIBentryALTinterwordspacing
J.~Yin, J.-G. Ren, H.~Lu, Y.~Cao, H.-L. Yong, Y.-P. Wu, C.~Liu, S.-K. Liao, F.~Zhou, Y.~Jiang, X.-D. Cai, P.~Xu, G.-S. Pan, J.-J. Jia, Y.-M. Huang, H.~Yin, J.-Y. Wang, Y.-A. Chen, C.-Z. Peng, and J.-W. Pan, ``Quantum teleportation and entanglement distribution over 100-kilometre free-space channels,'' \emph{Nature}, vol. 488, no. 7410, pp. 185--188, 2012. [Online]. Available: \url{https://doi.org/10.1038/nature11332}
\BIBentrySTDinterwordspacing

\bibitem{10:ETTTA:fs19}
\BIBentryALTinterwordspacing
A.~A. Semenov and W.~Vogel, ``Entanglement transfer through the turbulent atmosphere,'' \emph{Phys. Rev. A}, vol.~81, p. 023835, Feb 2010. [Online]. Available: \url{https://link.aps.org/doi/10.1103/PhysRevA.81.023835}
\BIBentrySTDinterwordspacing

\bibitem{16:GETA:fs22}
\BIBentryALTinterwordspacing
M.~Bohmann, A.~A. Semenov, J.~Sperling, and W.~Vogel, ``Gaussian entanglement in the turbulent atmosphere,'' \emph{Phys. Rev. A}, vol.~94, p. 010302, Jul 2016. [Online]. Available: \url{https://link.aps.org/doi/10.1103/PhysRevA.94.010302}
\BIBentrySTDinterwordspacing

\bibitem{OGH:16:PRL}
M.~Oszmaniec, A.~Grudka, M.~Horodecki, and A.~W\'ojcik, ``Creating a superposition of unknown quantum states,'' \emph{Phys. Rev. Lett.}, vol. 116, p. 110403, Mar 2016.

\bibitem{Ban:20:PRA}
S.~Bandyopadhyay, ``Impossibility of creating a superposition of unknown quantum states,'' \emph{Phys. Rev. A}, vol. 102, p. 050202, Nov 2020.

\bibitem{walshe2021streamlined}
B.~W. Walshe, R.~N. Alexander, N.~C. Menicucci, and B.~Q. Baragiola, ``Streamlined quantum computing with macronode cluster states,'' \emph{Physical Review A}, vol. 104, no.~6, p. 062427, 2021.

\end{thebibliography}

\end{document}